\documentclass[%
    aps,
    prd,
    letterpaper,
    superscriptaddress,
    preprintnumbers,
    floatfix,
    twocolumn,
    nofootinbib,
    showpacs,
    hyphens
]{revtex4-1}  

\usepackage{amsmath}
\usepackage{amsfonts}
\usepackage{mathtools}
\usepackage{cases}
\usepackage{lmodern}
\usepackage[T1]{fontenc}
\usepackage{upgreek}
\usepackage{microtype}
\usepackage{graphicx} 
\usepackage{comment}    
\usepackage{dcolumn}   
\usepackage{bm}       
\usepackage[caption=false,justification=justified]{subfig}
\usepackage{ragged2e} % for the \justifying macro
\DeclareCaptionJustification{justified}{\justifying}
\usepackage{multirow}
\usepackage{slashed}
\usepackage{placeins}
\usepackage{epstopdf}
\usepackage{color}
\usepackage{tikz,pgfplots}
\pgfplotsset{compat=newest}
\usepackage{enumerate,comment}
\usepackage{setspace}
\usepackage{dsfont}
\usepackage{nicefrac}
\usepackage{siunitx}
\usepackage{physics}
\AtBeginDocument{\RenewCommandCopy\qty\SI}
\ExplSyntaxOn
\msg_redirect_name:nnn { siunitx } { physics-pkg } { none }
\ExplSyntaxOff
\usepackage[normalem]{ulem}
\usepackage{tensor}
\usepackage{tabularx}
\usepackage{listings}
\usepackage{adjustbox}
\usepackage{hyperref}  
\hypersetup{breaklinks=true}
\usepackage[capitalize]{cleveref}

\definecolor{c1}{RGB}{206,0,0}
\definecolor{c2}{RGB}{249,149,0}
\definecolor{c3}{RGB}{153,0,210}
\definecolor{c4}{RGB}{0,109,219}
\definecolor{c5}{RGB}{0,146,146}
\definecolor{c6}{RGB}{255,109,182}
\pgfplotscreateplotcyclelist{cycle1}{
{c1}, {c2}, {c3}, {c4}, {c5}, {c6}
}

\begin{document}
\title{Collins-Soper kernel from lattice QCD at the physical pion mass}
\author{Artur Avkhadiev}
\affiliation{Center for Theoretical Physics, Massachusetts Institute of Technology, Cambridge, MA 02139, U.S.A.}
\author{Phiala E. Shanahan}
\affiliation{Center for Theoretical Physics, Massachusetts Institute of Technology, Cambridge, MA 02139, U.S.A.}
\author{Michael L. Wagman}
 \affiliation{Fermi National Accelerator Laboratory, Batavia, IL 60510, USA}  
\author{Yong Zhao}%
\affiliation{Physics Division, Argonne National Laboratory, Lemont, IL 60439, USA}
\preprint{FERMILAB-PUB-23-375-T,\,MIT-CTP/5587}
\renewcommand{\vec}[1]{\boldsymbol{#1}}
\newcommand{\mom}[1]{\mathbf{#1}}
\newcommand{\pos}[1]{\mathbf{#1}}
\newcommand{\tran}{T} % transverse
\newcommand{\wf}{\tilde{\phi}}
\newcommand{\unexp}{{\mathrm{u}}} 
\newcommand{\ren}{{\small \mathrm{ren.}}} % renormalized
\newcommand{\nn}{\nonumber} % nonumber
\DeclareRobustCommand{\Eq}[1]{Eq.~\eqref{eq:#1}}
\DeclareRobustCommand{\Eqs}[2]{Eqs.~\eqref{eq:#1} and \eqref{eq:#2}}
\DeclareRobustCommand{\fig}[1]{Fig.~\ref{fig:#1}}
\DeclareRobustCommand{\figs}[2]{Figs.~\ref{fig:#1} and \ref{fig:#2}}
\DeclareRobustCommand{\app}[1]{App.~\ref{app:#1}}
\DeclareRobustCommand{\sec}[1]{Sec.~\ref{sec:#1}}
\DeclareRobustCommand{\secs}[2]{Secs.~\ref{sec:#1} and \ref{sec:#2}}
\DeclareRobustCommand{\tbl}[1]{Table~\ref{tbl:#1}}
\DeclareRobustCommand{\refcite}[1]{Ref.~\cite{#1}}
\DeclareRobustCommand{\refcites}[1]{Refs.~\cite{#1}}
\newcommand{\MSbar}{
\overline{\rm MS}    
} % renormalized
 \newcommand{\xMOM}{ 
 \rm RI/xMOM
 } % renormalized
\newcommand{\MOM}{
    \rm RI^\prime/MOM
 } % renormalized
\begin{abstract}
    This work presents a determination of the quark Collins-Soper kernel, which relates transverse-momentum-dependent parton distributions (TMDs) at different rapidity scales, using lattice quantum chromodynamics (QCD). 
    This is the first lattice QCD calculation of the kernel at quark masses corresponding to a close-to-physical value of the pion mass, with next-to-next-to-leading logarithmic matching to TMDs from the corresponding lattice-calculable distributions, and includes a complete analysis of systematic uncertainties arising from operator mixing. 
    The kernel is extracted at transverse momentum scales $\SI{240}{\MeV}\lesssim q_{T}\lesssim\SI{1.6}{\GeV}$ with a precision sufficient to begin to discriminate between different phenomenological models in the nonperturbative region.
\end{abstract}

\maketitle
\section{Introduction}
    \par Since the 1970s it has been understood that the intrinsic motion of partons inside hadrons in the direction transverse to the hadron's momentum plays an important role in experimentally observed processes, beginning historically with Drell-Yan scattering (DY)~\cite{Feynman:1977yr,Feynman:1978dt,Berger:1982vs}. 
    The effect of this motion on the DY cross section has been rigorously derived in QCD in the form of a factorization theorem~\cite{Collins:1981uk,Collins:1981va,Collins:1984kg} and thereby described in terms of transverse-momentum-dependent parton distribution functions (TMDs). 
    TMDs are universal, appearing in the factorization of cross-sections for processes including also semi-inclusive deep inelastic scattering (SIDIS) and di-hadron production in $e^+ e^-$ collisions.
    Constraints on TMDs, particularly for the nucleon, have thus been the target of experimental programs since the 2000s (see Refs.~\cite{Avakian:2016rst,Boussarie:2023izj} for a review) and remain key targets of current and future experiments at facilities including the Thomas Jefferson National Accelerator Facility~\cite{Burkert:2008rj,Dudek:2012vr}, the Large Hadron Collider~\cite{Kikola:2017hnp,Feng:2022inv}, and the Electron-Ion Collider~\cite{Boer:2010zf,Boer:2011fh,Zheng:2018ssm,xue:2021svd,AbdulKhalek:2022hcn,Burkert:2022hjz,Abir:2023fpo}. 
    Simultaneously, significant efforts are being made from the theoretical perspective to constrain TMDs, including through lattice QCD calculations~\cite{Musch:2010ka,Musch:2011er,Engelhardt:2015xja,Yoon:2017qzo,Shanahan:2019zcq,Shanahan:2020zxr,LatticeParton:2020uhz,Li:2021wvl,Shanahan:2021tst,Schlemmer:2021aij,Engelhardt:2021kdo,Zhang:2022xuw,LPC:2022ibr,LPC:2022zci,Shu:2023cot,Chu:2023jia,Alexandrou:2023ucc,Chu:2023flm}.
    \par TMDs have a functional dependence on two scales: a virtuality scale $\mu$ and a rapidity scale $\zeta$, which is related to the hadron momentum in a scattering process. While the renormalization group (RG) evolution of TMDs with $\mu$ is perturbative for perturbative scales $\mu$ and $\zeta$, the evolution with $\zeta$ is inherently nonperturbative in certain regions of parameter space, even for perturbative $\mu$. 
    The $\zeta$-evolution of TMDs is encoded in the Collins-Soper (CS) kernel~\cite{Collins:1981uk,Collins:1981va,Collins:1984kg}, which can be defined as the rapidity anomalous dimension entering the relevant RG evolution equations (up to a conventional factor):
    \begin{equation}
    \label{eq:cs-kernel}
        \gamma_{p}(b_\tran, \mu) = 2 \dv{}{\ln\zeta} \ln\phi_p(b_T,\mu, x, \zeta),
    \end{equation}
    where $\phi_p(b_\tran, x, \mu, \zeta)$ is a TMD, chosen here as a TMD wavefunction (TMD WF) encoding the transverse motion of a parton $p \in \{ q, g \}$ in a meson state~\cite{Farrar:1979aw, Lepage:1979zb, Li:1992nu, Ji:2019sxk}. 
    The TMD WF is defined in a factorization formula valid in the limit of ultra-relativistic hadron momentum $\mom{P}$ and depends on the fraction $x$ of the parton's momentum collinear with $\mom{P}$, as well as the parton's momentum transverse to $\mom{P}$ as given by its Fourier conjugate $b_{\tran}$, the transverse displacement.
    The CS kernel depends on $\mu$, $b_{\tran}$ and parton type $p$, but is independent of $x$ and the hadronic state.
    \par Experimental DY and SIDIS data has been used to constrain phenomenological parameterizations of the quark CS kernel~\cite{Davies:1984sp,Ladinsky:1993zn,Landry:2002ix,Konychev:2005iy,Sun:2014dqm,DAlesio:2014mrz,Bacchetta:2017gcc,Scimemi:2017etj,Bertone:2019nxa,Scimemi:2019cmh,Bacchetta:2019sam,Hautmann:2020cyp,Bury:2022czx,Bacchetta:2022awv,Moos:2023yfa}.
    A number of parameterizations are in some tension in the region $b_\tran \gtrsim \SI{0.2}{\femto\meter}$ (at $\mu = \SI{2}{\GeV}$), which may be partially understood to arise from different approaches to modelling nonpertubative effects.
    In the more recent analyses~\cite{Bacchetta:2022awv,Moos:2023yfa}, the tensions have been reduced as larger sets of experimental data sensitive to the CS kernel in the nonperturbative regime~\cite{Grewal:2020hoc, Hautmann:2020cyp} were included. 
    Further improvements are expected with future data from the LHC~\cite{Feng:2022inv} and the Electron-Ion Collider~\cite{AbdulKhalek:2022hcn, Burkert:2022hjz}.
    A direct way of constraining the kernel from cross-section ratios has also been proposed and demonstrated on synthetic data~\cite{BermudezMartinez:2022ctj} and could be applied to experimental data in the future.
    A more precise determination of the nonperturbative CS kernel is important in particular for measurements of electroweak observables such as the $W^{\pm}$-boson mass~\cite{Bozzi:2019vnl} and especially for studies of nucleon and nuclear structure via deep inelastic scattering~\cite{AbdulKhalek:2022hcn}.
    \par Complementing phenomenological approaches, lattice QCD offers a pathway towards first-principles constraints of the CS kernel in the nonperturbative regime.
    One approach to such calculations is provided by Large-Momentum Effective Theory (LaMET)~\cite{Ji:2013dva,Ji:2014gla,Ji:2020ect}, in which physical TMDs, defined by matrix elements of lightlike-separated operators, and quasi-distributions, defined by the matrix elements of the corresponding spacelike-separated operators which are computable in lattice QCD, are perturbatively matched at large hadron momentum $\lvert \mom{P} \rvert \gg \Lambda_{\mathrm{QCD}}$~\cite{Ji:2014hxa,Ji:2018hvs,Ebert:2018gzl,Ebert:2019okf,Ebert:2019tvc,Ji:2019sxk,Ji:2019ewn,Ebert:2020gxr,Ji:2020jeb,Ji:2021znw,Ebert:2022fmh,Schindler:2022eva,Zhu:2022bja}.
    For example, a TMD WF $\phi_p(b_T, \mu, x, \zeta)$ is matched to a quasi-TMD WF ${\wf_p(b_\tran, \mu, x, \zeta)}$ with matching coefficients computed perturbatively in LaMET~\cite{Ji:2021znw,Deng:2022gzi} up to a nonperturbative soft factor independent of $x$ and $\zeta$ and power corrections that vanish in the limit of infinite boost.
    To date, several lattice QCD calculations have been carried out using quasi-TMD WFs and other quasi-distributions to extract the quark CS kernel~\cite{Shanahan:2019zcq,Shanahan:2020zxr,LatticeParton:2020uhz,Li:2021wvl,Schlemmer:2021aij,Shanahan:2021tst,LPC:2022ibr,Shu:2023cot,Chu:2023flm} and the soft function~\cite{LatticeParton:2020uhz,Li:2021wvl,Chu:2023flm}, as well as the full kinematic dependence of TMDs~\cite{LPC:2022zci,Chu:2023jia}.
    \par Using quasi-TMD WFs and LaMET, this work presents the first lattice QCD calculation of the quark CS kernel at valence quark masses corresponding to a close-to-physical value of the pion mass, $m_\pi = \SI{148.8\pm 0.1}{\MeV}$, thereby addressing the systematic uncertainty arising from the sensitivity of the kernel to the QCD vacuum structure~\cite{Vladimirov:2020umg} and reducing those arising from perturbative LaMET matching and proportional to $m^2_\pi/(x \lvert \mom{P} \rvert)^2$ and $m^2_\pi/((1-x) \lvert \mom{P} \rvert)^2$.
    Other $b_T$-dependent systematic uncertainties associated with matching are better quantified relative to previous calculations.
    The matching is performed at next-to-next-to-leading order (NNLO) and next-to-next-to-leading logarithmic (NNLL) accuracies for the first time in a calculation of the CS kernel, using recent results of Refs.~\cite{delRio:2023pse,Ji:2023pba}.
    Moreover, previously dominant~\cite{Shanahan:2021tst} systematic uncertainties from the Fourier transformation of quasi-TMDs are reduced in this work, and the associated model dependence is eliminated.
    Finally, renormalization-induced mixing effects for the nonlocal operators associated with quasi-TMDs are fully quantified for the first time in the $\xMOM$ renormalization scheme~\cite{Ji:2017oey,Green:2017xeu,Green:2020xco}.
    Taken together, this work achieves sufficient control and precision to begin to discriminate in the nonperturbative region between phenomenological parameterizations~\cite{Landry:2002ix,Scimemi:2019cmh,Bacchetta:2019sam,Bacchetta:2022awv,Moos:2023yfa} of the quark CS kernel
    and provides a better understanding of perturbative convergence in LaMET matching and the associated power corrections.
\section{The Collins-Soper kernel from quasi-TMD wavefunctions}
\label{sec:theory}
    \par The quark CS kernel can be computed in lattice QCD from ratios of matrix elements of nonlocal staple-shaped Wilson line operators in hadron states at different finite boost momenta $P_1^z$, $P_2^z$~\cite{Ji:2014hxa,Ebert:2018gzl,Ji:2019sxk}:
    \begin{widetext}
    \begin{equation}
    \begin{aligned}
    \label{eq:kernel-wf-lattice}
    \gamma_q^{\MSbar}(b_{\tran}, \mu)
                  &=
                    %\lim_{a\to 0}
                    \lim_{\ell \to \infty}
                    \frac{1}{\ln(P_{1}^{z}/P_{2}^{z})} 
                    \ln\frac
                    {\displaystyle
                        \int_{-\infty}^\infty \frac{\dd{b^z}}{2\pi}
                        e^{i \left(x-\frac{1}{2}\right) P_1^z b^z}
                        P_1^{z}
                        N_\Gamma(P_1^z)
                        \sum_{\Gamma^\prime} Z^{\MSbar}_{\Gamma \Gamma^\prime}(\mu)
                        W^{(0)}_{\Gamma^\prime}(b_\tran, b^{z}, P_1^{z}, \ell)
                    }
                    {\displaystyle
                        \int_{-\infty}^\infty \frac{\dd{b^z}}{2\pi}
                        e^{i \left(x-\frac{1}{2}\right) P_{2}^z b^z}
                        P_2^{z}
                        N_\Gamma(P_2^z)
                        \sum_{\Gamma^\prime} Z^{\MSbar}_{\Gamma \Gamma^\prime}(\mu)
                        W^{(0)}_{\Gamma^\prime}(b_\tran, b^z, P_2^{z}, \ell)
                    } 
                    \\ &\quad\quad+
                    {\delta \gamma}^{\MSbar}_q(\mu, x, P_1^z, P_2^z)
                     + \mathrm{p.c.}
    \end{aligned}
    \end{equation}
    \end{widetext}
    Here the dependence on the lattice spacing, $a$, is suppressed.
    $\delta \gamma^{\MSbar}_q(\mu, x, P_1^z, P_2^z)$ denotes the perturbative matching correction defined at the end of this section, and
    $\mathrm{p.c.}$ denotes the associated power corrections that are power series in $1/(b_\tran (xP^z))^2$, $\Lambda^2_{\mathrm{QCD}}/(x P^z)^2$, $m^2_h/(x P^z)^2$, where $m_h$ is the meson mass and $P^z \in \{ P_1^z, P_2^z\}$, and analogous forms with $x$ replaced by $1-x$.
    $W^{(0)}_\Gamma(b_\tran, b^z, P^{z}, \ell)$ denote ratios of bare quark quasi-TMD WFs (defined further below), such that
    \begin{align}
    \label{eq:quasi-wf-ratio}
        W^{(0)}_{\Gamma}(b_\tran, b^z, P^{z}, \ell) 
            &= \frac
                {\wf_{\Gamma}(b_\tran, b^{z}, P^z, \ell)}
                {\wf_{\gamma_4\gamma_5}(b_\tran, 0, 0, \ell)}.
    \end{align}
    As only quark quasi-TMD WFs are studied in this work, parton labels on WFs and WF ratios are omitted. Subscripts $\Gamma^{(')}$ denote Dirac structures; in the limit of infinite boosts $P_1^z,P_2^z\rightarrow \infty$, quasi-TMD WFs with $\Gamma \in \lbrace \gamma_3 \gamma_5, \gamma_4 \gamma_5 \rbrace$ approach $\gamma_{+}\gamma_5$.
    Renormalization factors $Z^{\MSbar}_{\Gamma \Gamma^\prime}(\mu)$ are $16\times 16$ matrices, detailed further below, and the normalization factors $N_{\Gamma}(P^z)$ correspond to 
    \begin{equation}
    \label{eq:tmd-normalization}
    N_{\Gamma}(P^z)=
    {\displaystyle
    \begin{cases}
         \dfrac{-i m_{h}}{P^{z}}, 
             & \Gamma = \gamma_3\gamma_5, \vspace{2mm} \\
        \dfrac{m_h}{E_h(P^z\hat{\mom{z}})}, 
             & \Gamma = \gamma_4 \gamma_5,
    \end{cases}
    }
    \end{equation}
    where $E_h(P^z\hat{\mom{z}})$ and $m_h$ are the meson energy and mass, respectively.
    \begin{figure}[t]
        \centering
        \includegraphics[scale=0.83]{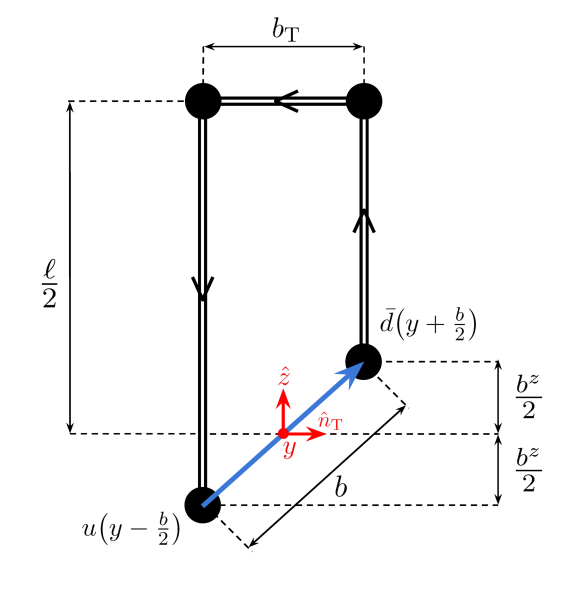}
        \caption{\label{fig:staple-shaped-operator}
        Diagrammatic representation of the nonlocal operator $\mathcal{O}^{ \Gamma}_{u\bar{d}}(b_\tran, b^z, y, \ell)$ defined in \cref{eq:staple-shaped-op}.
        The operator comprises a staple-shaped Wilson line of length $\ell + b_\tran$ connecting a quark-antiquark pair $u \bar{d}$ separated by $b = (\pos{b}_{\tran}, b^{z}, 0)$ (blue).
        The origin is defined at the midpoint between the quark and the antiquark (red).
        }
    \end{figure}
    \par Bare quark quasi-TMD WFs in position space are given by Euclidean equal-time correlation functions
    \begin{equation}
    \label{eq:wf-bare}
    \begin{aligned}
    \wf_{\Gamma}(b_\tran, b^z, P^{z}, \ell)
                &= \mel{0} {\mathcal{O}^{\Gamma}_{u\bar{d}}(b_\tran, b^z, 0, \ell)}{h(P^z)},
    \end{aligned} 
    \end{equation}
    where $\ket{0}$ and $\ket{h(P^z)}$ denote the QCD vacuum and a pseudoscalar meson state, respectively.
    The meson is taken to contain the isovector $u \bar{d}$ valence quark-antiquark pair, and the operator $\mathcal{O}^{\Gamma}_{u\bar{d}}(b_\tran, b^z, y, \ell)$ is depicted in Fig.~\ref{fig:staple-shaped-operator} and defined as  
    \begin{align}
       \mathcal{O}&_{u \bar{d}}^{ \Gamma }(b_\tran, b^z, y, \ell) \nonumber \\
    &\begin{aligned}
    \label{eq:staple-shaped-op}
            &\!= \bar{d}\bigg(\!y+\frac{b}{2}\bigg) \frac{\Gamma}{2} \mathcal{W}_{\hat{z}}\bigg(
                    y + \frac{b}{2};\, \frac{\ell - b^z}{2}
                    \bigg) 
                      \\
                &\quad\times 
                \mathcal{W}_{{-\hat{n}_\tran}}\bigg(
                    y + \frac{\ell}{2} \hat{z} + \frac{b_\tran}{2};\,
                    b_\tran\! 
                \bigg)   \\
                &\quad\times 
                \mathcal{W}_{- \hat{z}}\bigg(
                    y + \frac{\ell}{2} \hat{z} - \frac{b_\tran}{2};\, \frac{\ell + b^z}{2}
                    \bigg)
                u\bigg(\!y-\frac{b}{2}\bigg) \\
            &\!\equiv \bar{d}\bigg(\!y+\frac{b}{2}\bigg) \frac{\Gamma}{2} 
                \mathcal{W}_{\sqsupset}\bigg(\!y+\frac{b}{2}, y-\frac{b}{2}, \ell\bigg) 
                u\bigg(\!y-\frac{b}{2}\bigg),
    \end{aligned}
    \end{align}
    where $b = (\pos{b}_{\tran},\,b^{z},\,0)$, $u(y)$ and $d(y)$ denote up- and down-quark fields, respectively, $\mathcal{W}_{\hat{n}}(x;\,\xi)$ denotes a Wilson line of length $\xi$ starting at $x$ directed along $\hat{n}$,  $\hat{n}_\tran$ denotes a unit four-vector along $b_\tran$, and $\ell$ denotes the total collinear length of the staple-shaped Wilson line.
    The transformation properties of these operators and quasi-TMD WFs under sign changes of $b_{\tran}$ and $b^z$ as well as other discrete symmetries are presented in Appendix~\ref{sec:app:discrete-transformations}.
    Forming ratios in Eq.~\eqref{eq:quasi-wf-ratio} cancels divergences logarithmic in $a$, as well as power divergences linear in $\ell/a$ and $b_\tran/a$, in the quasi-TMD WFs~\cite{Ji:2017oey,Ishikawa:2017faj,Green:2017xeu}.
    Furthermore, forming the ratios eliminates $\ell$ dependence up to discretization artifacts and power corrections of order $1/(P^{z} \ell)$ and $b_\tran / \ell$.
    This leads to finite $\ell \to \infty$ limits of infinite collinear staple length for the ratios $W^{(0)}_\Gamma(b_\tran, b^z, P^z, \ell)$.
    \par The $16 \times 16$ renormalization matrices $Z^{\MSbar}_{\Gamma \Gamma^\prime}(\mu)$ appearing in \cref{eq:kernel-wf-lattice} may be computed as
    \begin{equation}
    \label{eq:mod-quasi-wf-renorm}
    \begin{aligned}
    Z&^{\MSbar}_{\Gamma\Gamma^\prime}(\mu)  
            \\&=  C^{\MSbar}_{\xMOM}(\mu, p_\mathrm{R}, \xi_{\mathrm{R}})
            Z^{\xMOM}_{\Gamma\Gamma^\prime}(p_\mathrm{R}, \xi_{\mathrm{R}}),
    \end{aligned}
    \end{equation}
    where
    \begin{equation}
    \label{eq:renorm-factor}
    \begin{aligned}
    Z&^{\xMOM}_{\Gamma\Gamma^\prime}(p_\mathrm{R}, \xi_{\mathrm{R}}) \\&= \Tr\bigg\lbrack
            \big[Z_{\Lambda_{d,-z}}^{\xMOM}\big]^\dagger(p_\mathrm{R}, \xi_{\mathrm{R}})\,\Gamma\, \\
            &\qquad\qquad\qquad\times
            Z^{\xMOM}_{\Lambda_{u,+z}}
            (p_\mathrm{R}, \xi_{\mathrm{R}})
            \left\lbrack
                \Gamma^\prime
            \right\rbrack^\dagger
            \bigg\rbrack.
    \end{aligned}
    \end{equation}
    Here $\mathrm{Tr}$ denotes a spinor trace and $\Gamma^\prime$ runs over the $16$ Dirac matrices.
    Conversion from the $\xMOM$ renormalization scheme~\cite{Ji:2017oey,Green:2017xeu,Green:2020xco} at the scale defined by $p^\mu_\mathrm{R}$ and $\xi^\mu_{\mathrm{R}}$ to the $\MSbar$ scheme at the scale $\mu$ is achieved with the conversion coefficient $C^{\MSbar}_{\xMOM}(\mu, p_\mathrm{R}, \xi_{\mathrm{R}})$ computed in continuum perturbation theory~\cite{Green:2020xco}. 
    $Z^{\xMOM}_{\Lambda_{q, \pm z}}$, where $q \in \{ u, d \}$, are $4\times 4$ matrices in spinor space 
    renormalizing the corresponding Green's functions $\Lambda_{q, \pm z}(p, \xi)$ defined as 
    \begin{equation}
    \label{eq:nonamp}
    \begin{aligned}
    \Lambda&_{q, \pm z}(p, \xi) \\
    &= \expval{0 | \mathcal{W}_{\mp z}(\pm\xi;\xi) | 0} \\ 
    &\qquad\times\expval{0|
          \mathcal{W}_{\pm z}(0; \xi) q(0) \bar{q}(p) | 0}
    \expval{0 | S^{-1}_{q}(p) | 0},
    \end{aligned}
    \end{equation}
    where $S_{q}(p)$ denotes the momentum-space quark propagator.
    $Z^{\xMOM}_{\Lambda_{q,\pm z}}(p_{\mathrm{R}}, \xi_{\mathrm{R}})$ is computed from the renormalization condition
    \begin{equation}
    \label{eq:rixmom}
    \begin{aligned}
        Z&^{\xMOM}_{\Lambda_{q,\pm z}}(p_{\mathrm{R}}, \xi_{\mathrm{R}})
        \Lambda_{q,\pm z}(p_{\mathrm{R}}, \xi_{\mathrm{R}})
                =
                \Lambda^{\mathrm{tree}},
    \end{aligned}
    \end{equation}
    in a fixed gauge, where $\Lambda^{ \mathrm{tree}}$ is the tree-level Green's function corresponding to $\Lambda_{q,\pm z}(p_{\mathrm{R}}, \xi_{\mathrm{R}})$.
    Further details are provided in \cref{sec:app:xmom}.
    \par Since $C^{\MSbar}_{\xMOM}$ has no Dirac structure, it cannot change the mixing patterns encoded by $Z^{\xMOM}_{\Lambda_{q,\pm z}}(p_{\mathrm{R}}, \xi_{\mathrm{R}})$ and their dependence on the auxiliary renormalization scales $p^\mu_{\mathrm{R}}$ and $\xi^\mu_{\mathrm{R}}$.
    Moreover, if determined for any given $p^\mu_{\mathrm{R}}$ and $\xi^\mu_{\mathrm{R}}$,
    $C^{\MSbar}_{\xMOM}$ simply cancels in the ratio of \cref{eq:kernel-wf-lattice}.
    However, in practice, if a calculation of $Z^{\MSbar}_{\Gamma\Gamma^\prime}(\mu)$ is realized as an average over multiple auxiliary scales, conversion (in both the numerator and the denominator in \cref{eq:kernel-wf-lattice} before averaging) may affect the value and systematic uncertainties in the lattice QCD determination of the CS kernel.
    \par The matching correction $\delta\gamma^{\MSbar}_q(\mu, x,P_1^z, P_2^z)$ appearing in \cref{eq:kernel-wf-lattice} is perturbative and given by
    \begin{equation}
    \label{eq:matching-correction}
    \begin{aligned}
	\delta\gamma&_q^{\text{N}^k\text{LO}}(\mu, x, P_1^z, P_2^z) \\ 
	&\equiv - \dfrac{1}{\ln(P_{1}^{z}/P_{2}^{z})} \Bigg(\!\ln\frac{C^{\text{N}^k\text{LO}}_\phi(\mu, xP_1^z)}{C^{\text{N}^k\text{LO}}_\phi(\mu, xP_2^z)} 
    \\&\quad+ (x \leftrightarrow \bar{x})\!
    \Bigg),
    \end{aligned}
    \end{equation}
    where $\mathrm{N}^k \mathrm{LO}$ denotes a fixed-order accuracy, $\bar{x} \equiv 1-x$, the renormalization scheme dependence is omitted for brevity, and $C^{\mathrm{N}^k \mathrm{LO}}_\phi(\mu, p^z)$, with $p^z\in\lbrace x P_1^z, x P_2^z, \bar{x} P_1^z, \bar{x} P_2^z \rbrace$ denote the TMD WF matching coefficients.
    The corresponding matching formula between physical and quasi-TMD WF receives power corrections as discussed around \cref{eq:kernel-wf-lattice}.
    \par The $C^{\mathrm{N}^k \mathrm{LO}}_\phi(\mu, p^z)$ are computed perturbatively in the strong coupling $\alpha_s(\mu)$, with
    $C^{\mathrm{LO}}_\phi = 1$. 
    The NLO contribution has been computed in Refs.~\cite{Ji:2021znw,Deng:2022gzi}; the NNLO contribution may be inferred from the matching formula for quasi-TMD PDFs~\cite{delRio:2023pse,Ji:2023pba}.
    For further discussion, see \cref{app:FO}.
    \par Fixed-order coefficients $C^{\mathrm{N}^k \mathrm{LO}}_\phi(\mu, p^z)$ may be resummed from initial scales $(\mu_0, p^z_0) = (2 p^z, p^z)$ as~\cite{Ji:2019ewn,Ebert:2022fmh} 
    \begin{equation}
    \label{eq:matching-resummation}
    \begin{aligned}
           C&^{\mathrm{N}^k \mathrm{LL}}_\phi(\mu, p^z)  
           \\ &= C^{\mathrm{N}^{k-1} \mathrm{LO}}_\phi\!\left(2 p^z,p^z\right) \exp\!\left[-K^{\mathrm{N}^k \mathrm{LL}}_\phi\big(2 p^z, \mu\big)\right]\!, 
    \end{aligned}
    \end{equation}
    where $\mathrm{N}^k \mathrm{LL}$ denotes a logarithmic accuracy and $K^{\mathrm{N}^k \mathrm{LL}}_\phi\big( \mu_0, \mu\big)$ is a resummation kernel.
    Since the $\mu$ dependence cancels in the ratio of quasi-TMD WFs (excluding the effects of conversion to the $\MSbar$ scheme which may arise in practice as discussed above), the CS kernel in \cref{eq:kernel-wf-lattice} is dependent on $\mu$ only through perturbative corrections, and the above choice of $\mu_0 = 2p^z$ further isolates the $\mu$ dependence to the resummation kernel.
    Resummations are independent of initial scale at infinite order but differ by higher-order terms at finite order. 
    For any choice of $\mu_0$, variations around $\mu_0$ provide a measure of the associated perturbative uncertainties.
    The resummed matching correction to the CS kernel is given by \cref{eq:matching-correction,eq:matching-resummation} as
    \begin{equation}
    \label{eq:matching-correction-NNLO}
    \begin{aligned}
	\delta&\gamma_q^{\text{N}^k\text{LL}}(\mu, x,P_1^z, P_2^z) \\
	   &= - \dfrac{1}{\ln(P_{1}^{z}/P_{2}^{z})}  \bigg(\!\ln\frac{C^{\mathrm{N}^{k-1} \mathrm{LO}}_\phi(2 p_1^z, p_1^z)}{C^{\mathrm{N}^{k-1} \mathrm{LO}}_\phi(2 p_2^z, p_2^z)}
         \\
        &\quad- \left(\!K^{\text{N}^k\text{LL}}_\phi\!\left(2 p_1^z, \mu\right) - K^{\text{N}^k\text{LL}}_\phi\!\left(2 p_2^z, \mu\right)\!\right) 
        \\ &\quad+ (x \leftrightarrow \bar{x})\!
        \bigg),
    \end{aligned}
    \end{equation}
    where the logarithmic ratio is expanded perturbatively in $\alpha_s(2p_1^z)$ and $\alpha_s(2p_2^z)$.
    For further discussion, see \cref{app:resum}.
    \par To partially account for the $b_T$-dependent power corrections, a practical choice is to replace $\delta\gamma^{\text{N}^k\text{LO}}_q(\mu,x, P_1^z, P_2^z)$ in \cref{eq:kernel-wf-lattice} with a $b_\tran$-unexpanded correction:
    \begin{equation}
    \label{eq:unexpanded-matching}
        \begin{aligned}
        &\delta\gamma^{\unexp \text{N}^k\text{LO}}_q(b_T, \mu, x, P_1^z, P_2^z) \\
            &\,\equiv - \dfrac{1}{\ln(P_{1}^{z}/P_{2}^{z})} \Bigg(\!\ln\frac{C^{\unexp \text{N}^k\text{LO}}_\phi(b_\tran, \mu, xP_1^z)}{C^{\unexp \text{N}^k\text{LO}}_\phi(b_\tran, \mu, xP_2^z)} 
            \\ &\quad\;+
            (x \leftrightarrow \bar{x})\!
    \Bigg) \\
            &\,= \delta\gamma^{ \text{N}^k\text{LO}}_q(\mu, x, P_1^z, P_2^z) + \ldots, \\
    \end{aligned}    
    \end{equation}
    where the ellipsis denotes terms that are power- and exponentially suppressed in $b_\tran (x P_1^z)$, $b_\tran (x P_2^z)$, and the analogous terms with $x$ replaced by $\bar{x}$. 
    $C^{\unexp \text{N}^k\text{LO}}_\phi(b_\tran, \mu, p^z)$ are the $b_\tran$-unexpanded TMD WF coefficients such that $C^{\unexp \text{N}^k\text{LO}}_\phi(b_\tran, \mu, p^z)$ is equal to $C^{\text{N}^k\text{LO}}_{\phi}( \mu, p^z)$ in the limit $b_\tran \gg 1/p^z$.
    They are computed perturbatively, with
    $C^{\unexp\mathrm{LO}}_\phi = C^{\mathrm{LO}}_\phi$. 
    The NLO contribution may be inferred from the corresponding TMD PDF coefficients \cite{Ebert:2019okf,Deng:2022gzi}.
    \par $C^{\unexp \text{N}^k\text{LO}}_\phi(b_\tran, \mu, p^z)$ may be resummed as in \cref{eq:matching-resummation}, using the same kernel $K^{\mathrm{N}^{k}\mathrm{LL}}_\phi(\mu,\mu_0)$.
    Both $C^{\unexp \text{N}^k\text{LO}}_\phi(b_\tran, \mu, p^z)$ and the corresponding resummed unexpanded correction $\delta\gamma^{\unexp \text{N}^k\text{LL}}_q(b_T, \mu, x, P_1^z, P_2^z)$ are conjectured in this work to reduce the $b_\tran$-dependent power corrections relative to the resummed matching correction in \cref{eq:matching-correction-NNLO} at the same accuracy, as is investigated numerically in the following section and in \cref{app:finite};  further study and a more systematic treatment of power corrections is left to future work.
\section{Numerical Investigation
\label{sec:numerical-investigation}}
    \par The quark CS kernel is computed numerically using an ensemble of lattice gauge-field configurations produced by the MILC collaboration~\cite{MILC:2012znn} with $2+1+1$ dynamical quark flavors and four-volume $V = L^3 \times T = (48a)^3 \times 64a$ with $a = \SI{0.12}{\femto\meter}$. 
    The one-loop Symanzik improved gauge action~\cite{Symanzik:1983dc,Curci:1983an,Luscher:1984xn,Luscher:1985zq} and the highly improved staggered quark action with sea quark masses tuned to produce a close-to-physical pion mass~\cite{Follana:2003fe,Follana:2003ji,Follana:2006rc} are used for gauge field generation.
    Gauge field configurations are subjected to Wilson flow with flow-time $\mathfrak{t}=\num{1.0}$~\cite{Luscher:2010iy} to enhance the signal-to-noise ratio in numerical results, and are gauge fixed to Landau gauge. 
    Calculations are performed in a mixed-action setup with the tree-level $\mathcal{O}(a)$-improved Wilson clover fermion action~\cite{Sheikholeslami:1985ij,Luscher:1996ug,Jansen:1998mx} used for propagator computation, with hopping parameter $\kappa = \num{.12547}$ and clover term coefficient $c_{\mathrm{sw}}=\num{1.0}$, resulting in a pion mass of $m_\pi = \SI{148.8
    \pm 0.1}{\MeV}$. 
    \par The following subsections detail the steps of the calculation of the quark CS kernel, including calculations of the bare quasi-TMD WF ratios and renormalization matrices, the Fourier transform to $b_\tran$-space, and finally the extraction of the CS kernel from ratios of quasi-TMD WF ratios with perturbative matching corrections.
    \subsection{Bare quasi-TMD WF ratios
    \label{sec:numerical-investigation-bare}
    }
        \par The CS kernel is computed according to \cref{eq:kernel-wf-lattice}, using quasi-TMD WF ratios with a pion $\ket{h(P^z)} = \ket{\pi(P^z)}$  chosen as the hadronic state. 
        The ratios $W_\Gamma^{(0)}(b_\tran, b^{z}, P^{z}, \ell)$ in \cref{eq:quasi-wf-ratio} are extracted from fits to pion two-point correlation functions. In particular, Euclidean correlation functions both with and without staple-shaped operators are constructed as
        \begin{equation}
        \begin{aligned}
                C&_{\mathrm{2pt}}^{\pi}(t, \mom{P})  \equiv a^6 \sum_{\mathclap{\pos{y}}}     
                    e^{i \mom{P}\cdot \pos{y}} 
                \left<
                            \chi_{\mathbf{P}}(y)
                            \chi^\dagger_{\mathbf{P}}(0) 
                \right>,
        \end{aligned}
        \end{equation}
        and
        \begin{equation}
        \begin{aligned}
            C&^{\Gamma}_{\mathrm{2pt}}(t, b_\tran, b^z, \mom{P}, \ell)  \\
                &\equiv a^6 \sum_{\mathclap{\pos{y}}}    
                e^{i \mom{P}\cdot \pos{y}} 
                \left<
                    \mathcal{O}^{\Gamma}_{u\bar{d}}(b_\tran, b^z, y, \ell)
                        \chi^\dagger_{\mom{P}}(0)
                \right>,
        \end{aligned}
        \end{equation}
        where $\mom{P} = P^{z} \hat{\mom{z}}$, $t= y_4$, and pion states are created with momentum-smeared interpolating fields 
        \begin{equation}
            \chi^\dagger_{\mom{P}}(x) = \bar{u}_{F_{\mom{P}/2}}(x) \gamma_5 d_{ F_{-\mom{P}/2}}(x),
        \end{equation}
        where the quasi-local quark fields are constructed using a Gaussian momentum smearing kernel $F_{\mom{K}}$ with $\mom{K}=\pm \mom{P}/2$ realized iteratively with $n_{\mathrm{smear}} = \num{50}$ smearing steps and a smearing kernel width defined by $\varepsilon = \num{0.2}$~\cite{Bali:2016lva}. 
        These correlation functions have spectral representations
        \begin{equation}
        \begin{aligned}
        \label{eq:pion-2pt-spectral}
            C&_{\mathrm{2pt}}^{\pi}(t, \mom{P}) \\ 
            &= a^3 \sum_{\mathfrak{n}=0}^\infty \dfrac
            { 
                \left| Z^{S}_{\mathfrak{n} \pi}(\mom{P})
                \right|^2
            }
            {2 E_{\mathfrak{n} \pi}(\mom{P})} \\
            &\hspace{30pt} \times
                \left[ e^{-E_{\mathfrak{n}\pi}(\mom{P}) t} + e^{-E_{\mathfrak{n}\pi}(\mom{P}) (T-t)} \right] + \ldots,
        \end{aligned}
        \end{equation}
        and
        \begin{equation}
        \label{eq:wf-2pt}
        \begin{aligned}
            C&^{\Gamma}_{\mathrm{2pt}}(t, b_\tran, b^z, \mom{P}, \ell) \\
            &= a^3 \sum_{\mathfrak{n}=0}^\infty \dfrac{
                  Z_{\mathfrak{n}\pi}^{S}(\mom{P})
                }
                {2 E_{\mathfrak{n}\pi}(\mom{P})}
                \wf_{\mathfrak{n}\Gamma}(b_\tran, b^{z}, \mom{P}, \ell) \\
                &\hspace{30pt} \times \left[ e^{- E_{\mathfrak{n}\pi}(\mom{P}) t}  \pm e^{- E_{\mathfrak{n}\pi}(\mom{P}) (T-t)}  \right] + \ldots, 
        \end{aligned}
        \end{equation}
        where $E_{\mathfrak{n}\pi}$ denotes the energy of the $\mathfrak{n}$-th eigenstate of the LQCD transfer matrix with quantum number of the pion, denoted $\ket{\pi_{\mathfrak{n}}}$, and in particular $E_\pi(\mom{P}) \equiv E_{0\pi}(\mom{P})$.
        Staple-shaped operator matrix elements are defined as
        \begin{equation}
            \label{eq:overlap-wf}
            \wf_{\mathfrak{n}\Gamma}(b_\tran, b^{z}, P^{z}, \ell)
              \equiv \mel{0}{\mathcal{O}^{\Gamma}_{u\bar{d}}(b_\tran, b^z, 0, \ell)}{\pi_{\mathfrak{n}}({\mom{P})}},
        \end{equation} 
        where $\wf_{\Gamma}(b_\tran, b^{z}, P^{z}, \ell) \equiv  \wf_{0\Gamma}(b_\tran, b^{z}, P^{z}, \ell)$.
        The overlap factors of the pion interpolating field between $\ket{\pi_{\mathfrak{n}}}$ and the vacuum state are defined as
        \begin{equation}
                \label{eq:overlap}
                Z^{S}_{\mathfrak{n}\pi}(\mom{P})
                  \equiv \mel{0}{\chi_{\mom{P}}(0)}{\pi_{\mathfrak{n}}({\mom{P})}}.
        \end{equation} 
        In \cref{eq:pion-2pt-spectral,eq:wf-2pt}, $T$ denotes the temporal extent of the lattice, and the ellipses denote additional contributions where the vacuum state is replaced by finite-temperature excited states.
        These contributions are suppressed by factors of order $e^{-2m_\pi (T/2)}$ or smaller in comparison with the terms shown and are therefore neglected below.
        \par The ground-state overlap $Z^{S}_{\pi}(\mom{P}) \equiv Z^{S}_{0\pi}(\mom{P})$ is guaranteed to be real-valued and positive up to discretization artifacts\footnote{
            A combination of the nonsinglet axial Ward identity in the isospin limit and the partially conserved axial current relation (PCAC) guarantee that
            \begin{equation*}
                    2 m_{q} \bra{0} P(0) \ket{\pi}
                    = \bra{0} \partial_\mu J^{\mu5}(0) \ket{\pi} = m_{\pi}^2 f_\pi,
            \end{equation*} 
            where $m_{q}$ is the renormalized light quark mass, $P(x)$ is a local pseudoscalar interpolating field for an isovector pion, $J^{\mu 5}(x)$ is the corresponding axial vector current, and $f_\pi$ is the pion's decay constant~\cite{Maris:1997hd}.
            The above applies to renormalized fields\thinspace---\thinspace for the bare pseudoscalar interpolating field, the pion overlap factor is therefore real and positive up to discretization artifacts from possible mixing with higher-dimension operators.
            This continues to hold for boosted pion states and if the quark fields in $P(x)$ are smeared with a self-adjoint smearing kernel.
            }.
        This ensures that $Z^{S}_{\pi}(\mom{P}) = \sqrt{ \left| Z^{S}_{\pi}(\mom{P})\right|^2 }$ can be extracted from fits to Eq.~\eqref{eq:pion-2pt-spectral} and therefore that both the magnitude and phase of the complex-valued TMD WF $\wf_{\Gamma}(b_\tran, b^{z}, P^{z}, \ell)$ can be extracted from joint fits to \cref{eq:pion-2pt-spectral,eq:wf-2pt}. 
        The $\pm$ sign appearing in Eq.~\eqref{eq:wf-2pt} depends on $\Gamma$ as detailed in App.~\ref{sec:app:discrete-transformations} and in particular is negative for $\gamma_4 \gamma_5$ and positive for $\gamma_3 \gamma_5$.
        \begin{ruledtabular}
        \begin{table}[t]
            \begin{tabular}{ccccc}
                $n^z$ & $P^z$ [GeV] & $\ell/a$ & $N_\text{cfg}$\\\hline
                0  & 0    & \{11,\,14,\,17,\,20,\,26,\,32\}  & 79     \\
                4  & 0.86 & \{26,\,32\} & 469 \\
                6  & 1.29 & \{17,\,20\} & 472 \\
                8  & 1.72 & \{14,\,17\} & 523 \\
                10 & 2.15 & \{11,\,14\} & 481 
            \end{tabular}
            \caption{\label{tab:measurements}
            Momenta $P^z = \frac{2\pi}{L} n^z$, number of configurations $N_\text{cfg}$, and operator extents $\ell/a$, used in the computation of two-point correlation functions in \cref{eq:wf-2pt}. For a given extent $\ell/a$, geometries with all of the 16 Dirac structures, asymmetries $-\ell/a \leq b^z \leq \ell/a$ and transverse displacements $0 \leq b_\tran/a \leq 7$ along $\hat{n}_T \in \lbrace \pm \hat{x}, \pm \hat{y} \rbrace$ are computed. 
            } 
        \end{table}
        \end{ruledtabular}
        \par The operator geometries and number of configurations $N_{\mathrm{cfg}}(P^{z})$ used to compute the two-point correlation functions for each choice of pion momentum $P^{z} = \frac{2\pi}{L} n^z$ are summarized in Table~\ref{tab:measurements}. Correlation functions are computed with propagators calculated from sources on a $2^4$ grid bisecting the lattice along each dimension for all of the $16$ Dirac structures.\footnote{For $n^z = 10$ measurements were performed on slightly fewer configurations (corresponding to at least $80\%$ of the the $N_{\rm cfg}(P^z)$ shown in Table~\ref{tab:measurements}) for some Dirac structures that are found to make negligible contributions to the renormalized quantities studied here.}
        The operator geometries used, illustrated in Fig.~\ref{fig:staple-shaped-operator}, are such that for each $b_\tran/a \in \{0,\ldots 7\}$ along $\hat{n}_T \in \lbrace \pm \hat{x}, \pm \hat{y} \rbrace$, all possible staple asymmetries $b_z$ are constructed with the fixed values of $\ell/a$ specified, i.e., $-\ell/a \leq b^z \leq \ell/a$, which are by construction restricted to be either even or odd integers for any fixed $\ell/a$. This choice is convenient as power divergences are proportional to the total length of the Wilson line~\cite{Ji:2017oey,Ishikawa:2017faj,Green:2017xeu} in the operator, so all operator geometries computed for a given $\ell$ and $b_\tran$ have equal power divergences across all $b^{z}$, simplifying renormalization. This is in contrast to the staple geometries chosen in the work of Refs.~\cite{Shanahan:2019zcq,Shanahan:2020zxr,Shanahan:2021tst} where various geometries with a given $b_\tran$ were constructed with fixed values of $\frac{1}{2}(\ell + b^z)$, leading to $b^{z}$-dependent renormalization factors.
        \par Correlation functions computed on each gauge-field configuration are averaged over sources, forward and backwards propagation in time, and operator structures with $\hat{n}_\tran \in \lbrace \pm \hat{x}, \pm\hat{y}\rbrace$ for $C^{\Gamma}_{\mathrm{2pt}}$. 
        The bare quasi-TMD WF ratios $W_\Gamma^{(0)}(b_\tran, b^{z}, P^{z}, \ell)$ in \cref{eq:quasi-wf-ratio} are then determined using a multi-step fitting procedure:
        \begin{enumerate}
            \item Determination of $E_\pi(\mom{P})$ and $Z_\pi^S(\mom{P})$ from a simultaneous fit to $C_{\rm{2pt}}^\pi$ and the statistically most precise $C_{\rm{2pt}}^\Gamma$ for a given $\mom{P}$;
            \item Determination of $\tilde{\phi}_\Gamma$ from fits to the $t$-dependence of combinations of $C_{\rm{2pt}}^\Gamma$ using the results for $E_\pi(\mom{P})$ and $Z_\pi^S(\mom{P})$, accounting for correlations between quasi-TMD WF ratios with different staple geometries using bootstrap resampling;
            \item Construction of $W^{(0)}_\Gamma(b_\tran, b^{z}, P^{z}, \ell)$ from ratios of $\tilde{\phi}_{\Gamma}$ as in \cref{eq:quasi-wf-ratio} for each bootstrap sample.
        \end{enumerate}
        Each of these steps is detailed in the following subsections.
        \subsubsection{Determination of \texorpdfstring{$E_\pi(\mom{P})$}{pion energies} and \texorpdfstring{$Z_\pi^S(\mom{P})$}{overlap factors}}
        \label{subsec:EandZ}
            \par As the exponential $t$-dependencies of both $C_{\rm{2pt}}^\pi$ and $C_{\rm{2pt}}^\Gamma$ are governed by $E_{\mathfrak{n}\pi}(\mom{P})$, these correlation functions may be fit simultaneously to extract $E_{\mathfrak{n}\pi}(\mom{P})$ and $Z_{\mathfrak{n}\pi}^S(\mom{P})$. 
            In practice, only the statistically most precise $C_{\rm{2pt}}^\Gamma$ for a given $\mom{P}$ is used, corresponding to the two-point function constructed with the operator geometry with the minimum value of $\ell/ a$ computed, $b_\tran/a=1$, $b^z=0$ (even $\ell/a$), or an average of $b^z/a=\pm 1$ (odd $\ell/a$), and $\Gamma = \gamma_4 \gamma_5$.
            For each $\mom{P}$, the two correlation functions $C_{\rm{2pt}}^\pi$ and $C_{\rm{2pt}}^\Gamma$ are jointly fit to the spectral representations of Eq.~\eqref{eq:pion-2pt-spectral} and Eq.~\eqref{eq:wf-2pt} for a variety of fit ranges using correlated $\chi^2$-minimization with the fitting procedures detailed in Refs.~\cite{Shanahan:2020zxr,NPLQCD:2020ozd} and summarized here.
            \par Results using $t \leq t_{\text{max}}$ are used for fitting, where $t_{\text{max}}$ is chosen to be the largest $t$ for which a given correlation function has signal-to-noise ratio $\geq\!\!1/3$.
            Fits are performed with all possible fit windows $[t_\text{min},t_\text{max}]$ such that $t_\text{min} \ge 2$ and $t_\text{max} - t_\text{min} \ge 3$, where $t_{\text{min}}$ is chosen independently for $C_{\rm{2pt}}^\pi$ and $C_{\rm{2pt}}^\Gamma$.\footnote{The final results are insensitive to changes in the smallest allowed $t_{\text{max}} - t_{\text{min}}$ and other numerical tolerances included in the fitting procedure, as verified by performing analyses with a range of alternative choices.}
            For each fit range, the covariance matrix is estimated using bootstrap resampling~\cite{davison_hinkley_1997} with optimal linear shrinkage~\cite{stein1956,Ledoit:2004}.
            First, fits using one-state truncations of \cref{eq:pion-2pt-spectral} and \cref{eq:wf-2pt} are performed.
            For $P^z > 0$, VarPro methods~\cite{varpro0,varpro1} are used in which the best-fit $Z^S_{0\pi}(P^z \hat{\mom{z}})$, which enters $\chi^2$ linearly, is determined using linear methods during each step of nonlinear optimization for $E_{0\pi}(P^z \hat{\mom{z}})$.
            For $P^z = 0$, where there is negligible signal-to-noise degradation, VarPro methods lead to less efficient $\chi^2$-minimization and are not employed. 
            Fits to two-state truncations of Eq.~\eqref{eq:pion-2pt-spectral} and Eq.~\eqref{eq:wf-2pt} are then performed analogously.
            \par The Akaike Information Criterion (AIC)~\cite{AkaikeAIC} is used to select whether one- or two-state fits are preferred for each fit range. 
            To penalize overfitting, two-state fits are only accepted if they improve the AIC by at least 2 times the number of degrees of freedom and if excited-state contributions do not severely dominate over ground-state contributions\thinspace---\thinspace in particular $Z^S_{0\pi}(P^z \hat{\mom{z}})>0.2\,Z^S_{1\pi}(P^z \hat{\mom{z}})$ and $\wf_{0\Gamma}(b_\tran, b^{z}, \mom{P}, \ell)>0.2\,\wf_{1\Gamma}(b_\tran, b^{z}, \mom{P}, \ell)$ is required.
            In cases where two-state fits are preferred, three-state fits are also performed but are not found to be preferred by the AIC in any case. 
            Further selection cuts are then applied as described in Ref.~\cite{NPLQCD:2020ozd}: fits are discarded for which two nonlinear optimizers disagree on the ground state by more than $10^{-5}$, the bootstrap median and mean disagree by more than $2\sigma$, or correlated and uncorrelated fits disagree by more than $5\sigma$.
            \par Weighted averages of all results from fits passing these cuts are then used to determine the final results for $Z^S_{\pi}(P^z \hat{\mom{z}})$ and $E_{\pi}(P^z \hat{\mom{z}})$.  
            The same weights are used as in Refs.~\cite{Rinaldi:2019thf,Shanahan:2020zxr,NPLQCD:2020ozd}, which for each fit parameter ($Z^{S}_{\mathfrak{n}\pi}$ and $E_{\mathfrak{n}\pi}$) correspond to the $p$-value of each fit divided by the variance of the fitted parameter.
            For each momentum, at least 6 fit ranges are found to lead to fits passing the cuts described above and are therefore included in these weighted averages.
            The same weights are also used to perform averages of the bootstrap samples of $Z^S_\pi(P^z \hat{\mom{z}})$ and $E_\pi(P^z \hat{\mom{z}})$ generated using a common set of bootstrap ensembles for each fit range, which are used below to enable correlated determinations of $\tilde{\phi}^\Gamma(b_T,b^z,P^z,\ell)$ for different $\Gamma$, $b_T$, $b^z$, and $\ell$.
            \begin{figure}[t]
            \centering
                \includegraphics[width=0.46\textwidth]{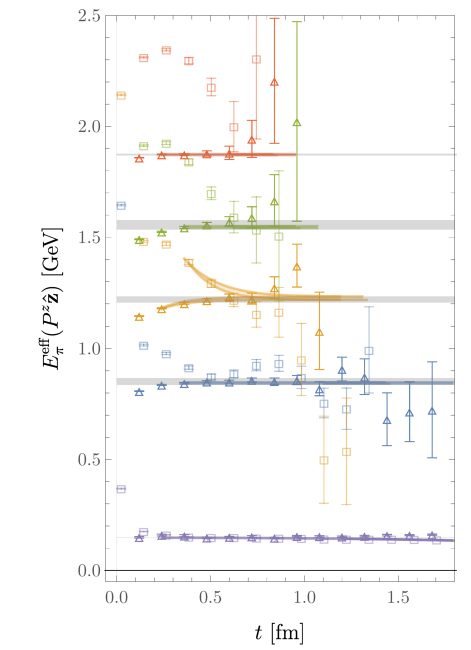}   
                \caption{\label{fig:EMP} Effective energies defined in \cref{eq:eff-energy} and constructed using $C^{\pi}_{2\mathrm{pt}}$ (squares, offset slightly on the horizontal axis) and the most statistically precise $C^{\Gamma }_{2\mathrm{pt}}$ (triangles) for each choice of $P^z$.
                The gray bands show the weighted average of $E_\pi(P^z \hat{\mom{z}})$ from fits with all possible $t_{\rm min}$ using the number of excited states preferred by the AIC, as described in the main text. 
                The colored bands show the corresponding highest-weight fit included in the average.
                }
            \end{figure}
            \par Fig.~\ref{fig:EMP} shows a comparison of the fit results for $E_\pi(P^z \hat{\mom{z}})$ with effective energy functions constructed from each correlation function as
            \begin{equation}
            \label{eq:eff-energy}
            \begin{aligned}
                a E_{\pi}^{\mathrm{eff}}\left(t + \frac{1}{2}a, \mom{P}=P^z\hat{\mom{z}}\right)
                        &=  \mathrm{log}
                                \left[ \frac{
                                     C^{\pi/\Gamma}_{2\mathrm{pt}}(t, \mom{P})
                                }
                                {C^{\pi/\Gamma}_{2\mathrm{pt}}(t+a, \mom{P})} \right] \\
                        &\xrightarrow{T \gg t \gg 0} a E_{\pi}(\mom{P}) + \ldots,
                    \end{aligned}
            \end{equation}
            where the ellipsis denotes exponentially-suppressed corrections from excited states and the finite temporal extent of the lattice geometry.
            The momentum dependence of the choice of $N_{\mathrm{cfg}}(P^z)$ and $\ell(P^z)$ leads to a complicated dependence of the statistical uncertainties of the determination of $E_\pi(P^z \hat{\mom{z}})$ on $P^z$.
            \begin{figure}[t]
            \centering
                \centering    \includegraphics[width=0.46\textwidth]{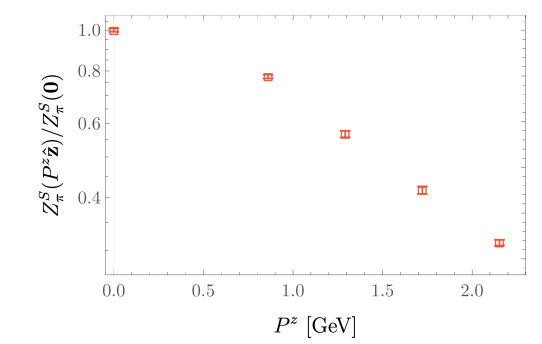}
                \includegraphics[width=0.46\textwidth]{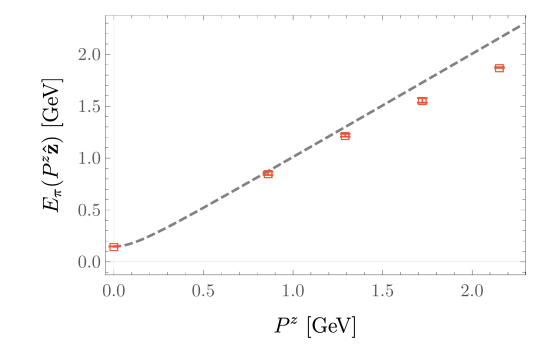}
                \caption{
                Fitted momentum-smeared overlap factors $Z_\pi^S(P^z \vec{\mom{z}})$ normalized by $Z_\pi^S(\mom{0})$ (top panel) and energies $E_\pi(P^z\hat{\mom{z}})$ (bottom panel, red data points) as functions of $P^z$.
                The gray dashed line represents the energy determined from the continuum dispersion relation.
                \label{fig:overlap-and-energy}}
            \end{figure}
            The momentum dependence of the extracted values of $E_\pi(P^z \hat{\mom{z}})$ and $Z_\pi^S(P^z \hat{\mom{z}})$ is shown in \cref{fig:overlap-and-energy}. 
            The continuum dispersion relation $E_{\pi}(P^z \hat{\mom{z}}) = \sqrt{E_\pi(0)^2 + \left\lvert P^{z} \hat{\mom{z}} \right\rvert^2}$ is also shown for comparison.
            The relative differences between $E_\pi(P^z \hat{\mom{z}})$ and the continuum dispersion relation in order of increasing $P^z$ are $\num{0.03(2)}$, $\num{0.06(1)}$, $\num{0.10(1)}$, and $\num{0.13(1)}$ for the four nonzero $a P^z$ values studied.
            The increase in these differences with $a P^z$ is observed to be approximately linear, which is consistent with the expected form of lattice artifacts since the clover term has not been nonpertubatively tuned to remove $O(a)$ chiral symmetry breaking effects.
            Further calculations at other values of the lattice spacing are required to study these lattice artifacts in more detail.
        \subsubsection{Determination of bare quasi-TMD WFs \texorpdfstring{$\tilde{\phi}^{\lbrack\Gamma\rbrack}$}{} }
            \par The results for $E_{\mathfrak{n}\pi}(\mom{P})$ and $Z_{\mathfrak{n}\pi}^S(\mom{P})$, detailed in the previous section, are subsequently used to determine $\wf_{\mathfrak{n}\Gamma}(b_\tran, b^{z}, P^{z}, \ell)$ from fits of $C^{\Gamma }_{2\mathrm{pt}}(t, b_\tran, b^z, P^{z} \mom{\hat{z}},\ell)$ to Eq.~\eqref{eq:wf-2pt} with all operator geometries.
            Combinations of $C^{\Gamma }_{2\mathrm{pt}}$, $E_\pi(\mom{P})$ and $Z_\pi^S(\mom{P})$ are formed at the bootstrap level:
            \begin{equation}
                \begin{aligned}
                \label{eq:ratio}
                \mathcal{R}&^\Gamma(t, b_\tran, b^z, P^z, \ell) \\
                            &= {C^{\Gamma }_{2\mathrm{pt}}(t, b_\tran, b^z, P^{z} \mom{\hat{z}},\ell)} \\
                            &\hspace{20pt} \times \frac{2 E_\pi(P^{z} \hat{\mom{z}}) }{Z_\pi^S(P^{z} \hat{\mom{z}})\left[ e^{- E_{\pi}(\mom{P}) t}  \pm e^{- E_{\pi}(\mom{P}) (T-t)}  \right] } \\
                        &\xrightarrow{T \gg t \gg 0}
                        \wf_\Gamma(b_\tran, b^z, P^{z}, \ell) + \ldots,
                \end{aligned}
                \end{equation}
                which are fit to the appropriate spectral representations obtained by multiplying Eq.~\eqref{eq:wf-2pt} by $\mathcal{R}^\Gamma(t, b_\tran, b^z, P^z, \ell) / C^{\Gamma }_{2\mathrm{pt}}(t, b_\tran, b^z, P^{z} \vec{\hat{z}},\ell)$.
                \par
\begin{figure}[t]
                    \centering
                        \includegraphics[width=0.46\textwidth]{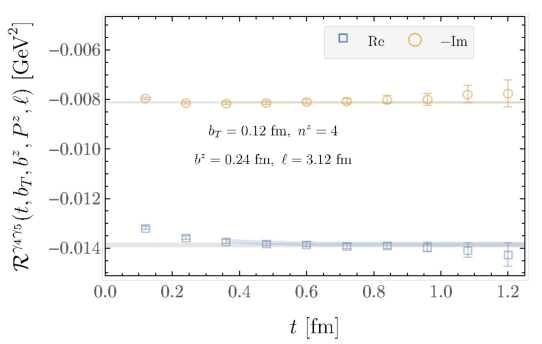}
                        \includegraphics[width=0.46\textwidth]{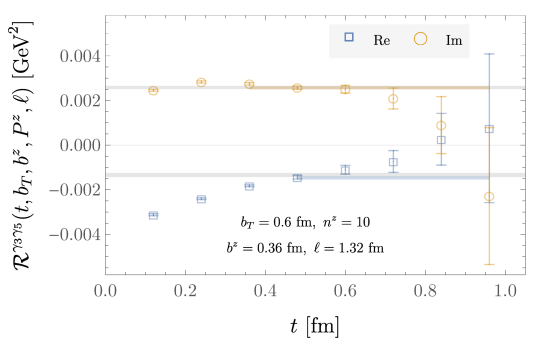} 
                        \caption{
                        Examples of fits of real and imaginary parts of the bare quasi-TMD WFs $\wf_\Gamma(b_\tran, b^z, P^z, \ell)$ defined in \cref{eq:wf-bare} to combinations  $\mathcal{R}^\Gamma(t, b_T, b^z, P^z, \ell)$ defined in \cref{eq:ratio}. The gray bands show the weighted average of $\wf_\Gamma(b_\tran, b^z, P^z, \ell)$ from fits with all possible $t_{\rm min}$, as described in the main text. 
                        The colored bands show the corresponding single highest-weight fit included in the average.
                        The imaginary part in the top panel is multiplied by \num{-1} for visual clarity. \label{fig:phi_fits}
                        }
                \end{figure}

 The same procedure described in Sec.~\ref{subsec:EandZ} is used to choose $t_{\text{max}}$ for these fits; however, for some staple-shaped operator geometries $C^{\Gamma }_{2\mathrm{pt}}$ is consistent with zero within the statistical precision of this work.
                Therefore, $t_{\text{max}} \geq 9$ is imposed in cases where the signal-to-noise criterion described above would lead to a smaller $t_{\text{max}}$.
                The same procedure described above is then used to sample over possible values of $t_{\text{min}}$, construct the bootstrap covariance matrix with optimal linear shrinkage for each choice of fit range, and determine weighted averages of the fit parameter $\wf_\Gamma(b_\tran, b^z, P^{z}, \ell)$ for each operator geometry.
                Examples of the resulting fits are shown in Fig.~\ref{fig:phi_fits}.
        
        \subsubsection{Construction of bare quasi-TMD WF ratios \texorpdfstring{$\tilde{W}^{\lbrack\Gamma\rbrack}$}{}}

        Bare quasi-TMD WF ratios are obtained at the boostrap level from bare quasi-TMD WFs via \cref{eq:quasi-wf-ratio} for each $\Gamma$, $P^{z}$, $\ell$, $b^z$, and $b_\tran$ combination considered.
        For the symmetric staple geometries used here, $b^z/a$ is necessarily odd (even) for odd (even) $\ell/a$.
        For the geometries where $\ell/a$ and therefore $b^z/a$ are odd, $b^z=0$ matrix elements are replaced by averaging over those with $b^z/a = \pm 1$. 
        The replacement leads to differences in the normalization of even and odd $\ell/a$ matrix elements at nonzero lattice spacing; however, these differences vanish in the continuum limit and can be analyzed in conjunction with other lattice artifacts when the continuum limit is performed.
        $W^{(0)}_\Gamma(b_\mathrm{T}, b^z, P^z, \ell)$ are shown as a function of $P^{z} b^{z}$ at different $b_\tran$ for each $\Gamma$ and $\ell$, with examples for particular choices of $P^{z}$ and $b_\tran$, in \cref{fig:analysis-bare}.
        Additional examples are provided in Appendix~\ref{sec:app:plots}.
        \par The statistical precision of the quasi-TMD WF ratios diminishes with increasing $b_\tran$, with the smallest signal-to-noise ratio observed for the largest computed $b_\tran/a = 7$ for quasi-TMD WF ratios with the largest collinear length, $\ell/a = 32$ at $n^{z} = 4$ ($P^z = \SI{0.86}{\GeV}$).
        Signal-to-noise ratio also decreases with increasing $P^z$; however, the use of smaller $\ell$ at larger $P^z$ with roughly constant $P^z \ell$ leads to relatively mild signal-to-noise scaling with $P^z$ in the quasi-TMD WF results.
        \par Both real and imaginary parts of $W_\Gamma^{(0)}(b_\tran, b^z, P^{z}, \ell)$ are symmetric in $P^{z} b^{z}$ within uncertainties, which is consistent with expectations given the symmetric definition of the staple-shaped operator's origin (shown in Fig.~\ref{fig:staple-shaped-operator}) placed at the midpoint between the quarks.
        For further discussion, see Appendix~\ref{sec:app:discrete-transformations}.
    \subsection{Renormalized quasi-TMD WF ratios
    \label{sec:numerical-investigation-renorm}
    }
      
    \par The renormalization factors $Z^{\MSbar}_{\Gamma\Gamma^\prime}(\mu)$ are determined from $C^{\MSbar}_{\xMOM}(\mu, p_\mathrm{R}, \xi_{\mathrm{R}})$ and $Z^{\xMOM}_{\Lambda_{\pm z}}(p_\mathrm{R},\xi_\mathrm{R})$ via \cref{eq:mod-quasi-wf-renorm} for Wilson lines along $\pm \hat{z}$ and the set of renormalization scales defined by Wilson line lengths $\xi_\mathrm{R}/a \in \lbrace 2, 3, 4 \rbrace$ and off-shell quark momenta $p^\mu_\mathrm{R} = \frac{2\pi}{L}(0,0,n^z,0)$, with $n^z \in \lbrace 8, 10, 12 \rbrace$.
    The coefficients $C^{\MSbar}_{\xMOM}(\mu, p_\mathrm{R}, \xi_{\mathrm{R}})$ are computed perturbatively, with $\alpha_{s}(\mu)$ determined as prescribed in Ref.~\cite{Bethke:2009jm}, setting $\mu = \SI{2}{\GeV}$, and neglecting the running from the scale set by $\mu^2 = p_{\mathrm{R}}^2$.
    The factors $Z^{\xMOM}_{\Lambda_{\pm z}}(p_\mathrm{R},\xi_\mathrm{R})$ are calculated numerically from the $\xMOM$ renormalization condition in \cref{eq:rixmom} using $\num{32}$ gauge-field configurations. %fixed to Landau gauge.
    Quark propagators are computed using wall sources with fixed four-momentum $p_R$.
    Means and standard errors are estimated with bootstrap resampling with $\num{200}$ bootstrap ensembles.
      \begin{figure*}[t]
        \centering
            \captionsetup[subfigure]{format=hang}
            \subfloat[The mixing matrix at the renormalization scale used in the analysis of the CS kernel.
            White disks denote off-diagonal elements with contributions expected at one-loop order in lattice perturbation theory~\cite{Constantinou:2019vyb}.
            Examples for other renormalization scales are provided in \cref{fig:app-mixing} of \cref{sec:app:xmom}.
            \label{fig:mixing-full}
            ]{
            \includegraphics[width=0.46\textwidth]{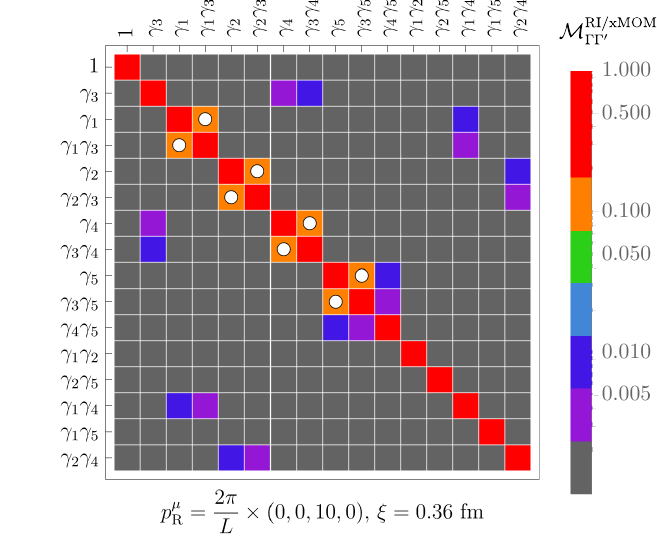}
            }
            \hfill
            \subfloat[Dominant off-diagonal $\Gamma, \Gamma^\prime$ elements of the mixing matrix for $\Gamma \in \lbrace \gamma_3\gamma_5, \gamma_4 \gamma_5 \rbrace$ as a function of $p_{\mathrm{R}}$.
            Data corresponding to
            $\xi_\mathrm{R}/a \in \{2, 3, 4\}$ are denoted by squares, circles and triangles, respectively.
            Statistical and systematic uncertainties, denoted by
            error bars and color bands, respectively, are computed as described in text.
            \label{fig:mixing-sub}
            ]{
            \includegraphics[width=0.46\textwidth]{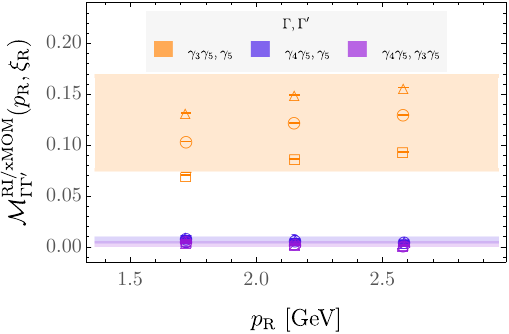} 
            }
            \caption{An example of renormalization-induced mixing effects as defined by the mixing matrix $\mathcal{M}^{\xMOM}_{\Gamma\Gamma^\prime}(p_{\mathrm{R}}, \xi_{\mathrm{R}})$ in \cref{eq:mixing-matrix}. 
            \label{fig:mixing}
            }
        \end{figure*}
    \par Renormalization affects the determination of the CS kernel only via mixing induced between staple-shaped operators.
    The mixing is characterized by off-diagonal elements of $Z^{\xMOM}_{\Gamma \Gamma^\prime}$ normalized relative to their diagonal components:
    \begin{equation}
    \label{eq:mixing-matrix}
    \begin{aligned}
        \mathcal{M}&^{\xMOM}_{\Gamma \Gamma^\prime}(p_{\mathrm{R}},\, \xi_{\mathrm{R}}) \\&\equiv \frac{\mathrm{Abs}\lbrack Z^{\xMOM}_{\Gamma \Gamma^\prime}(p_{\mathrm{R}}, \xi_{\mathrm{R}}) \rbrack}{\frac{1}{16} \sum_{\Gamma}\mathrm{Abs}\lbrack Z^{\xMOM}_{\Gamma \Gamma} (p_{\mathrm{R}}, \xi_{\mathrm{R}})\rbrack}.
    \end{aligned}
    \end{equation}
    The central values of $Z^{\xMOM}_{\Gamma \Gamma^\prime}(p_{\mathrm{R}},\, \xi_{\mathrm{R}})$ are calculated at $\xi_\mathrm{R}/a = 3$ and $p_{\mathrm{R}} = \SI{2.15}{\GeV}$, and systematic uncertainty for each pair $\Gamma, \Gamma^\prime$ is estimated as half the difference between the maximum and the minimum of $Z^{\xMOM}_{\Gamma \Gamma^\prime}(p_{\mathrm{R}},\, \xi_{\mathrm{R}})$ over all values of $p_\mathrm{R}$ and $\xi_\mathrm{R}$ studied.
    Systematic and statistical uncertainties are added in quadrature.
    \cref{fig:mixing-full} illustrates $\mathcal{M}^{\xMOM}_{\Gamma \Gamma^\prime}(p_{\mathrm{R}},\, \xi_{\mathrm{R}})$ computed from the central values of $Z^{\xMOM}_{\Gamma \Gamma^\prime}(p_{\mathrm{R}},\, \xi_{\mathrm{R}})$, and \cref{fig:mixing-sub} illustrates the auxiliary-scale dependence of dominant off-diagonal contributions to $\mathcal{M}^{\xMOM}_{\Gamma \Gamma^\prime}(p_{\mathrm{R}},\, \xi_{\mathrm{R}})$ for $\Gamma \in \{\gamma_3 \gamma_5, \gamma_4 \gamma_5\}$, which are given by $\Gamma^\prime \in \{\gamma_3 \gamma_5, \gamma_4 \gamma_5, \gamma_5\}$.
    \par At the level of renormalization constants as defined by \cref{eq:mixing-matrix}, mixing effects for collinear configurations of $p^\mu_{\mathrm{R}}$ and $\xi^\mu_{\mathrm{R}}$ are consistent with constraints on staple-shaped operator mixing from $\mathcal{C}$, $\mathcal{P}$ and $\mathcal{T}$ transformations~\cite{Ji:2021uvr,Alexandrou:2023ucc}, and the dominant contributions are as expected from lattice perturbation theory at one-loop order~\cite{Constantinou:2019vyb}.
    While noncollinear momentum configurations are not used in the determination of the kernel, an investigation of mixing effects using such a definition of the associated renormalization scales, summarized in Appendix~\ref{sec:app:xmom}, reveals contributions to mixing in addition to those expected in lattice perturbation theory at one-loop order.
    The additional contributions may be understood as artifacts of an off-shell renormalization scheme.
    \par The ratios of the $\MSbar$-renormalized quasi-TMD WFs, $W^{\MSbar}_{\Gamma}(b_\tran, \mu, b^z, P^z, \ell)$, are computed according to
    \begin{equation}
    \begin{aligned}
    \label{eq:quasi-wf-pos-ren}
    W^{\MSbar}_{\Gamma}(b_\tran, \mu, b^z, P^z, \ell) = \sum_{\Gamma^\prime}
            Z^{\MSbar}_{\Gamma \Gamma^\prime}(\mu) \,
            W^{(0)}_{\Gamma^\prime}(b_\tran, b^z, P^z, \ell),
    \end{aligned}
    \end{equation}
    using $W^{(0)}_{\Gamma^\prime}(b_\tran, b^z, P^z, \ell)$ and $Z^{\MSbar}_{\Gamma \Gamma^\prime}(\mu)$ for all of the 16 $\Gamma^\prime$ structures; the uncertainties are combined in quadrature. 
    \begin{figure*}[t]
            \subfloat[][{Examples of real and imaginary parts of the bare quasi-TMD WF ratios $W^{(0)}_{\Gamma}(b_\tran, \mu, b^z, P^z, \ell)$, computed as described in \cref{sec:numerical-investigation-bare}.
            \label{fig:analysis-bare}
            }]
            {   
                \centering
                \includegraphics[width=0.46\textwidth]{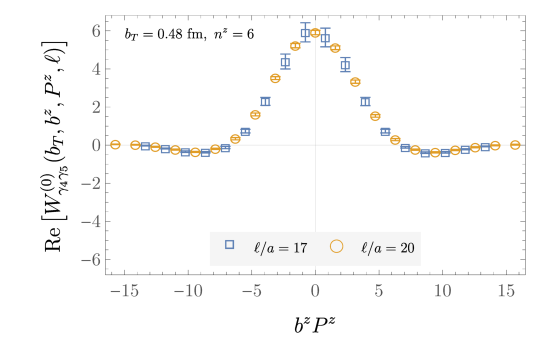}
                \hspace{20pt}
                \includegraphics[width=0.46\textwidth]{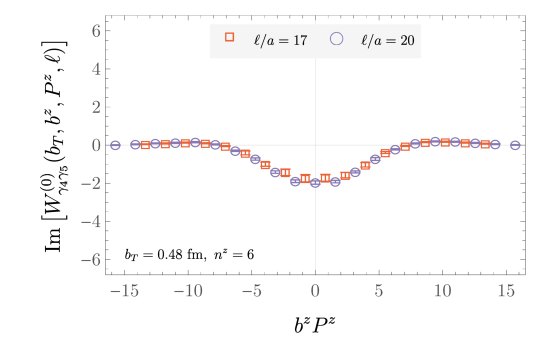} 
            }
            \hfill
            \subfloat[][{Examples of real and imaginary parts of the renormalized quasi-TMD WF ratios  $W^{\MSbar}_{\Gamma}(b_\tran, \mu, b^z, P^z, \ell)$, $\mu = \SI{2.0}{\GeV}$, computed as described in \cref{sec:numerical-investigation-renorm}.
            Further examples are shown \cref{fig:wf_ms_pz4_gamma7_a,fig:wf_ms_pz4_gamma7_b,fig:wf_ms_pz4_gamma11_a,fig:wf_ms_pz4_gamma11_b,fig:wf_ms_pz6_gamma7_a,fig:wf_ms_pz6_gamma7_b,fig:wf_ms_pz6_gamma11_a,fig:wf_ms_pz6_gamma11_b,fig:wf_ms_pz8_gamma7_a,fig:wf_ms_pz8_gamma7_b,fig:wf_ms_pz8_gamma11_a,fig:wf_ms_pz8_gamma11_b,fig:wf_ms_pz10_gamma7_a,fig:wf_ms_pz10_gamma7_b,fig:wf_ms_pz10_gamma11_a,fig:wf_ms_pz10_gamma11_b} of \cref{sec:app:plots}.
            \label{fig:analysis-ren}}]
            {
                \centering
                \includegraphics[width=0.46\textwidth]{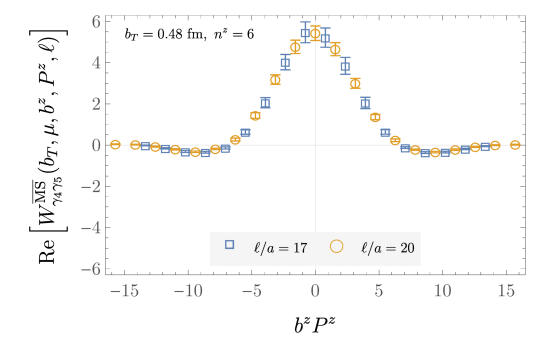} 
                \hspace{20pt}
                \includegraphics[width=0.46\textwidth]
                {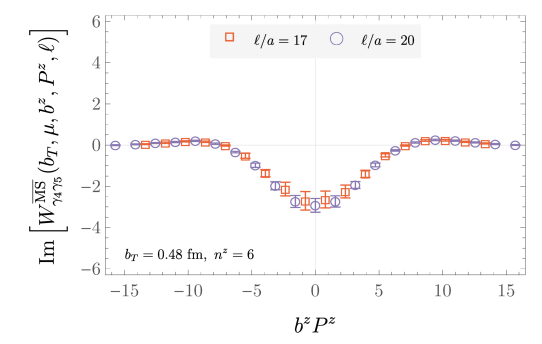} 
            }
            \hfill
            \caption{\label{fig:analysis-1}
                Example comparison of quasi-TMD WF ratios before and after accounting for renormalization-induced mixing effects.
                }
        \end{figure*}
    The effects of mixing on quasi-TMD WF ratios are illustrated in \cref{fig:analysis-ren,fig:analysis-bare}.
    
    \subsection{Fourier-transformed quasi-TMD WF ratios
        \label{sec:numerical-investigation-ft}
        }
        \par The Fourier transform of the $\MSbar$-renormalized position-space quasi-TMD WF ratios is realized as a Discrete Fourier Transform (DFT), i.e.,
        \begin{equation}
        \label{eq:wf-ratio-fourier}
        \begin{aligned}
        W^{\MSbar}_{\Gamma}(b_\tran, \mu, x, P^z)
                &= \frac{P^{z}}{2\pi} N_\Gamma(P)  \sum_{\mathclap{|b^z|\leq 
                b^z_{\mathrm{max}}}}
                    e^{i \left(x - \frac{1}{2}\right) P^z b^z}
                \\ &\quad\times 
                \bar{W}_\Gamma^{\MSbar}(b_\tran, \mu, b^z, P^z),
        \end{aligned}
        \end{equation}
        where $b^{\mathrm{max}}_z$ denotes the truncation point in position space and $\bar{W}_\Gamma^{\MSbar}(b_\tran, \mu, b^z, P^z)$ denotes a position-space quasi-TMD WF ratio whose real and imaginary parts have been averaged at each $P^z$ over $\pm b^z$ and all values of $\ell(P^z)$ relevant for a given $b^z$  with weights proportional to the inverse variance of each contribution.
        \begin{figure*}[t]
        \centering
        \includegraphics[width=0.46\textwidth]{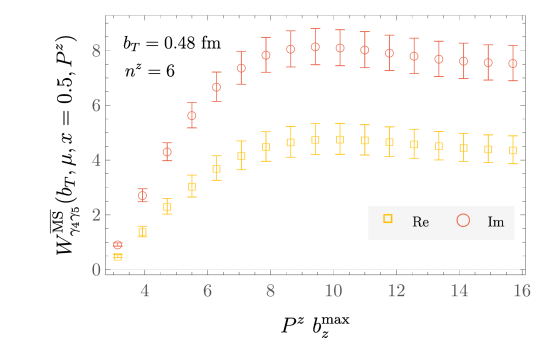}
        \hspace{20pt}
        \includegraphics[width=0.46\textwidth]{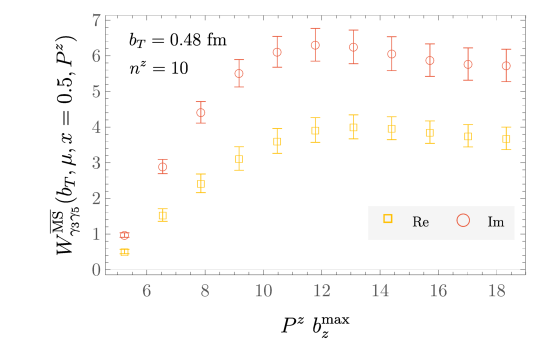}  
        \caption{Examples of $\MSbar$-renormalized quasi-TMD WF ratios in $x$-space, defined in \cref{eq:wf-ratio-fourier}, evaluated at $x=0.5$ and $\mu = \SI{2.0}{\GeV}$, as a function of the truncation of the position-space data at $ b^z_{\mathrm{max}}$ for different momenta and Dirac structures.
        Further examples are shown in \cref{fig:wf_dft_trunc} of \cref{sec:app:plots}.
        \label{fig:ft-truncation-robust}
        }
        \end{figure*}
        As can be seen in \cref{sec:app:plots}, the values that are averaged are in all cases consistent within $\approx 1\sigma$.
        As demonstrated in Fig.~\ref{fig:ft-truncation-robust}, with additional examples provided in \cref{fig:wf_dft_trunc} of \cref{sec:app:plots}, the values of quasi-TMD WF ratios are robust to decreasing $b_z^\mathrm{max}/a$ from the largest computed values, remaining constant within uncertainties for $P^z b_z^\mathrm{max} \gtrsim 12$ for all $b_\tran$ and momenta studied. 
    \begin{figure*}[t]
                \centering
                \includegraphics[width=0.46\textwidth]{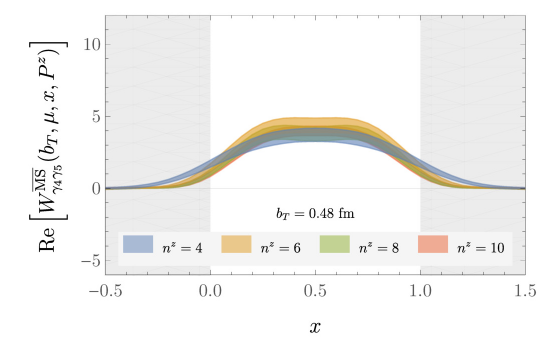}
                \hspace{20pt}
                \includegraphics[width=0.46\textwidth]{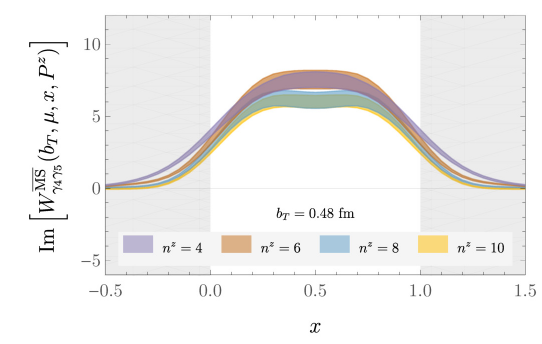}
            \caption{Examples of real and imaginary parts of the Fourier-transformed $\MSbar$-renormalized quasi-TMD WF ratios $W^{\MSbar}_{\Gamma}(b_\tran, \mu, x, P^z)$, $\mu = \SI{2.0}{\GeV}$, computed as described in \cref{sec:numerical-investigation-ft}.
            Further examples are shown in \cref{fig:wf_ms_x_gamma7_a,fig:wf_ms_x_gamma7_b,fig:wf_ms_x_gamma11_a,fig:wf_ms_x_gamma11_b} of \cref{sec:app:plots}.
            \label{fig:analysis-2}
            }
        \end{figure*}
        \par Selected $x$-space quasi-TMD WF ratios obtained via DFT are shown in \cref{fig:analysis-2} (with further examples provided in \cref{sec:app:plots}). 
        Consistent with their symmetry properties in $b^z P^z$ space, $W^{\MSbar}_\Gamma(b_\tran, x, P^{z}, \ell)$ are generally complex distributions, with a vanishing imaginary part as $b_\tran \to 0$ or $P^z \to 0$, where $W^{\MSbar}_\Gamma(b_\tran, x, P^{z}, \ell)$ is expected to be real.
        Finally, since the LaMET matching coefficients to NLO are independent of Dirac structure, $W^{\MSbar}_\Gamma(b_\tran, b^z, P^{z}, \ell)$ for $\Gamma \in \lbrace \gamma_3\gamma_5, \gamma_4\gamma_5 \rbrace$ are expected to agree up to power corrections.
        The magnitude of both real and imaginary parts of the quasi-TMD WFs are reduced outside of the physical region $x \in [0,1]$ as $P^z$ increases, which is consistent with expectations from the factorization formula~\cite{Ebert:2018gzl,Ebert:2019okf,Ji:2019sxk,Ji:2019ewn,Ji:2021znw,Ebert:2022fmh,Deng:2022gzi}. 
        Since the factorization scales are proportional to the hard parton momenta $xP^z$ and $(1-x)P^z$, the power corrections are always enhanced near the end-point regions $x\to0$ and $x\to1$, and lead to nonvanishing tails when $P^z$ is finite.
    \subsection{Perturbative matching \label{sec:numerical-investigation-matching}} 
        \par The final determination of the CS kernel in this work employs the $b_\tran$-unexpanded resummed perturbative correction at NNLL accuracy, denoted uNNLL,
        \begin{equation}
        \label{eq:matching-correction-uNNLL}
        \begin{aligned}
    	 \delta&\gamma_q^{\MSbar,\,\unexp\mathrm{NNLL}}(b_T, \mu, x,P_1^z, P_2^z) \\
    	   &= - \dfrac{1}{\ln(P_{1}^{z}/P_{2}^{z})}  \bigg(\!\ln\frac{C^{\MSbar,\,\unexp\mathrm{NLO}}_\phi(b_\tran, 2 p_1^z, p_1^z)}{C^{\MSbar,\,\unexp\mathrm{NLO}}_\phi(b_\tran, 2 p_2^z,  p_2^z)}
             \\
            &\quad- \left(\!K^{\MSbar,\,\mathrm{NNLL}}_\phi\!\left(\mu, 2 p_1^z\right) - K^{\MSbar,\,\mathrm{NNLL}}_\phi\!\left(\mu, 2 p_2^z\right)\!\right) \\
            &\quad+ (x \leftrightarrow \bar{x})\!
            \bigg),
        \end{aligned}
        \end{equation}
        which is derived from \cref{eq:unexpanded-matching} by resumming the $b_\tran$-unexpanded coefficients $C^{\MSbar,\,\unexp\mathrm{NLO}}_\phi(b_\tran, \mu_0,  p^z)$ with the kernel $K^{\MSbar,\,\mathrm{NNLL}}_\phi\!\left(\mu, \mu_0\right)$ for $\mu_0 = 2 p^z$. 
        The logarithmic ratio of the uNLO coefficients is expanded in $\alpha_s(2p_1^z)$ and $\alpha_s(2p_2^z)$ analogously to that of the $b_\tran$-independent coefficients in \cref{eq:matching-nnlo-2}.
        \begin{figure*}[t]
        \centering    \includegraphics[width=0.46\textwidth]{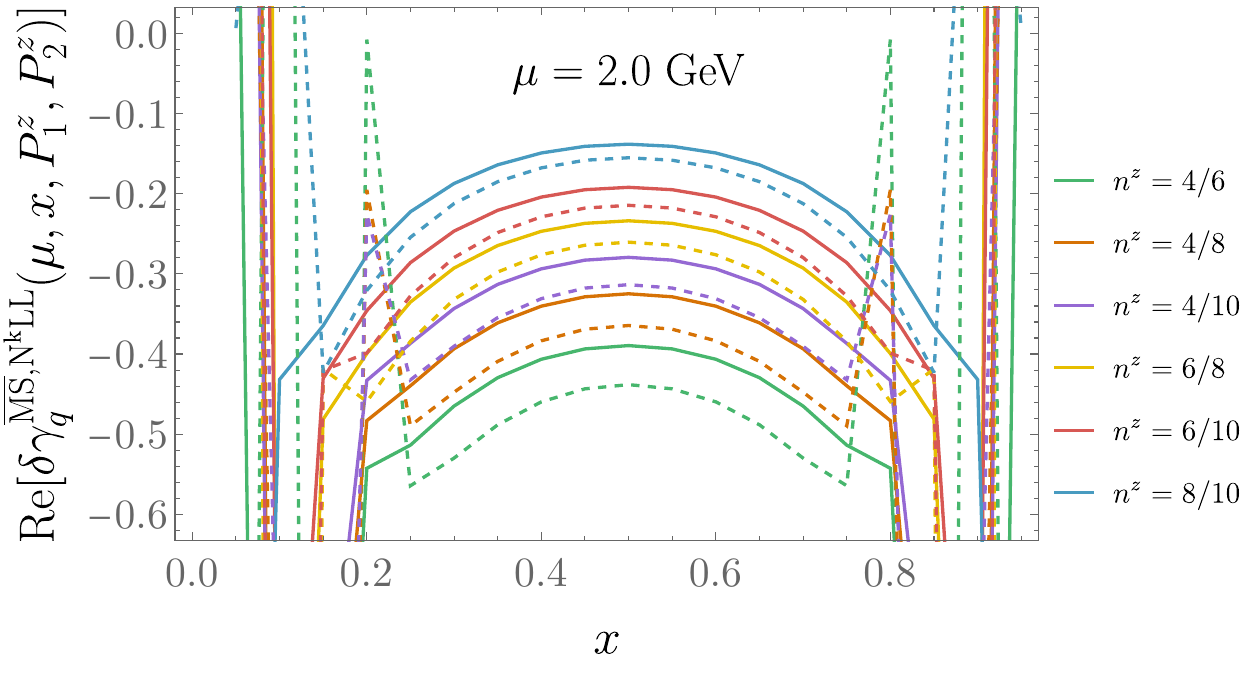}
        \includegraphics[width=0.46\textwidth]{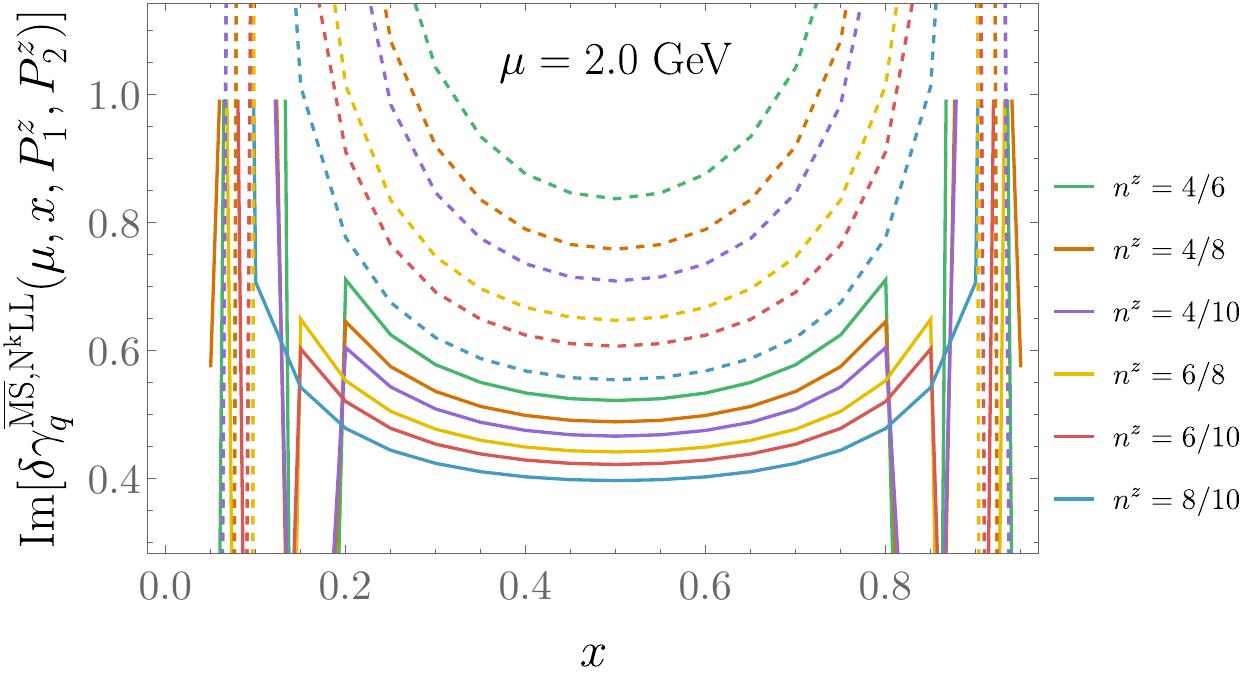}
             \caption{Examples of real and imaginary parts of resummed matching corrections to the CS kernel, defined in \cref{eq:matching-correction-NNLO}, at NLL (solid)
             and NNLL (dashed) for each momentum pair $n_1^z/n_2^z$.
             Analogous comparison of fixed-order corrections at NLO and NNLO is illustrated in \cref{fig:nlovsnnlo} of \cref{app:FO}.
             \label{fig:nllvsnnll}
             }
        \end{figure*}
    \par In addition to $b_T$-unexpanded NNLL (uNNLL), corrections at several other accuracies are computed to study perturbative convergence: fixed-order NLO and NNLO corrections computed according to \cref{eq:matching-correction}, $b_T$-unexpanded NLO (uNLO) corrections computed analogously, and NLL and NNLL resummations computed according to \cref{eq:matching-correction-NNLO}.
    In all comparisons beyond LO, for example that of NNLL and NLL illustrated in \cref{fig:nllvsnnll}, the
    $\mathrm{Re}[\delta\gamma^{\MSbar}_q(\mu,x, P_1^z, P_2^z)]$ exhibit qualitative agreement between different accuracies for $x \in [0.3,\,0.7]$ at each pair $(P_1^z, P_2^z)$, with better agreement at larger momenta.
    When compared analogously, the $\mathrm{Im}[\delta\gamma^{\MSbar}_q(\mu,x, P_1^z, P_2^z)]$ exhibit worse agreement and are larger in magnitude than the real parts.
    This indicates different rates of perturbative convergence in real and imaginary parts of matching corrections.
    The same qualitative picture is observed for fixed-order corrections in \cref{fig:nlovsnnlo} of \cref{app:FO}.
    \begin{figure*}[t]
    \centering 
    \includegraphics[width=0.46\textwidth]{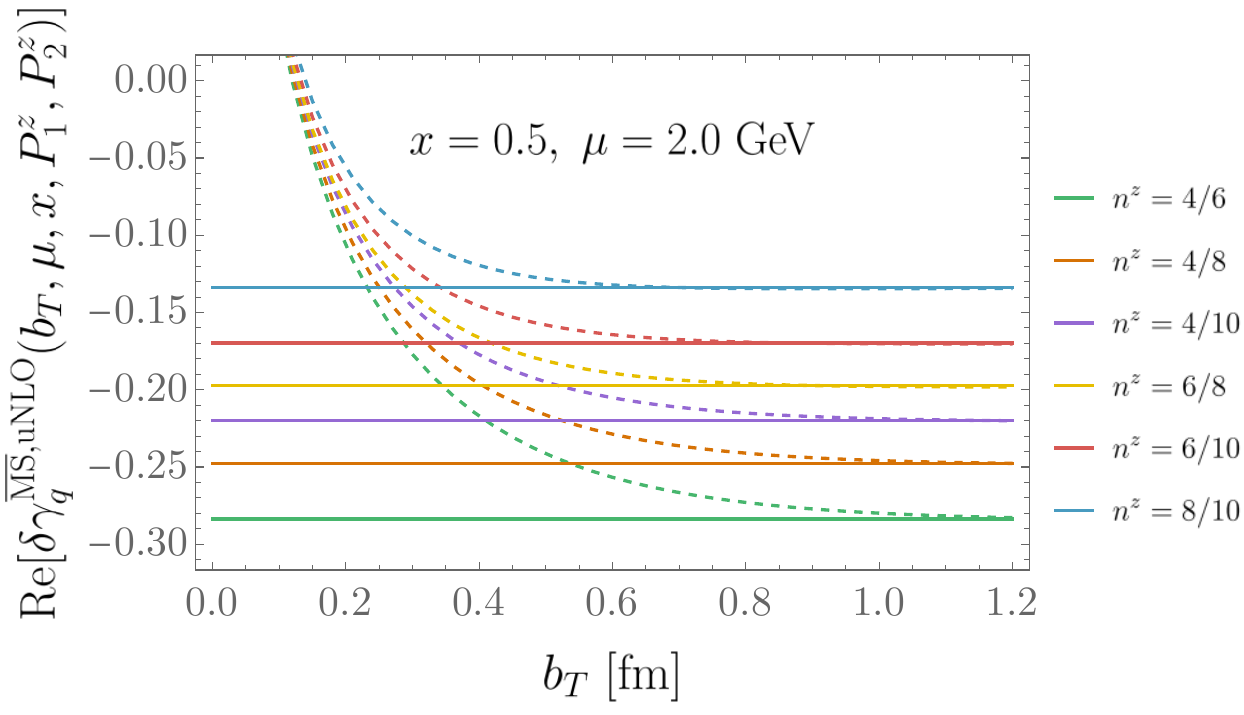} 
    \includegraphics[width=0.46\textwidth]{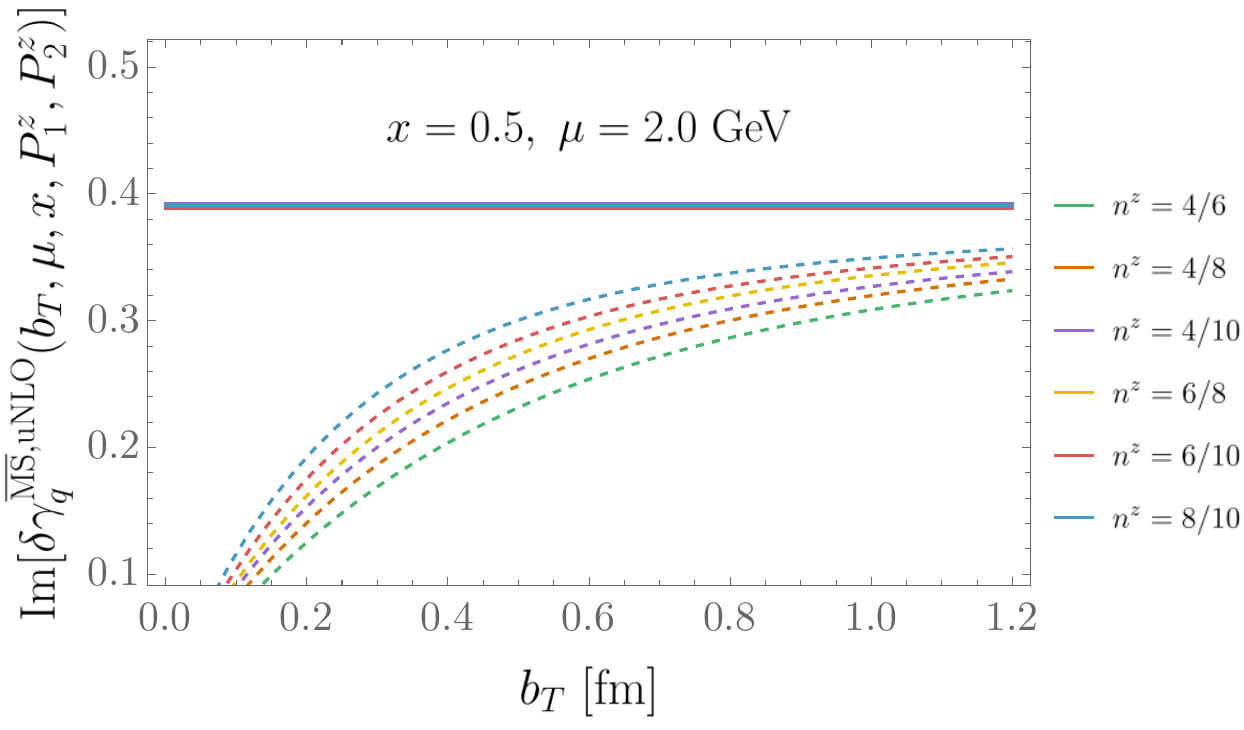}
    \caption{Examples of real and imaginary parts of $b_\tran$-dependent uNLO matching corrections to the CS kernel (dashed), defined in \cref{app:finite}, compared with those of the $b_\tran$-independent corrections at NLO (solid).
    Further examples of $b_\tran$-unexpanded matching are illustrated in \cref{fig:unexp,fig:CSunexp} of \cref{app:finite}.
    \label{fig:CSunexp_bT}
    }
    \end{figure*}
    Sensitivity to $b_\tran$-dependent power corrections is also different between real and imaginary parts, as may be seen by comparing corrections expanded and unexpanded in $b_\tran$, such as the comparison of NLO and uNLO illustrated in \cref{fig:CSunexp_bT} and further examples provided in \cref{app:finite}.
    These comparisons reveal a $b_\tran$-dependent sensitivity to power corrections which, for momenta used in this work, is significant for $b_\tran/a \lesssim 3$ in the real part and across the entire range in $b_\tran/a$ in the imaginary part.
    \subsection{The Collins-Soper kernel}
    \label{sec:numerical-investigation-cs}
    Using \cref{eq:kernel-wf-lattice} and replacing integral Fourier transforms of quasi-TMD WF ratios with the DFTs defined in \cref{eq:wf-ratio-fourier}, the $\MSbar$-renormalized quark CS kernel is determined via the estimator
    \begin{equation}
    \begin{aligned}
    \label{eq:kernel-wf-estimator}
    \hat{\gamma}&_{\Gamma}^{\MSbar}(b_{\tran}, \mu, x, P_1^z, P_2^z) \\
    &= \dfrac{1}{\ln(P_{1}^{z}/P_{2}^{z})}
                  \ln\Bigg\lbrack
                   \dfrac
                    {\displaystyle
                        W_\Gamma^{\MSbar}(b_\tran,  \mu, x, P^{z}_1)
                    }
                    {\displaystyle
                        W_\Gamma^{\MSbar}(b_\tran,  \mu, x, P^{z}_2)
                    } 
                \Bigg\rbrack
        \\&\quad+ \delta \gamma_q^{\MSbar}(b_\tran,  \mu, x, P_1^z, P_2^z),
    \end{aligned}
    \end{equation}
    for each chosen perturbative accuracy in the correction $\delta \gamma_q^{\MSbar}$.
    The estimator coincides with the kernel up to power corrections and discretization artifacts, whereby the dependence on $x$, $P_{1}^z$, $P_{2}^z$, $\Gamma$ and the implicit dependence on $a$ is introduced.
    Examples of $\mathrm{Re}[ \hat{\gamma}_\Gamma^{\MSbar}]$ with LO and uNNLL matching are illustrated in \cref{fig:analysis-2}, with additional examples illustrated in \cref{fig:wf_cs_x_a1,fig:wf_cs_x_a2,fig:wf_cs_x_b1,fig:wf_cs_x_b2} of \cref{sec:app:plots}.
    \begin{figure}[t]
                \centering
                \includegraphics[width=0.46\textwidth]
                {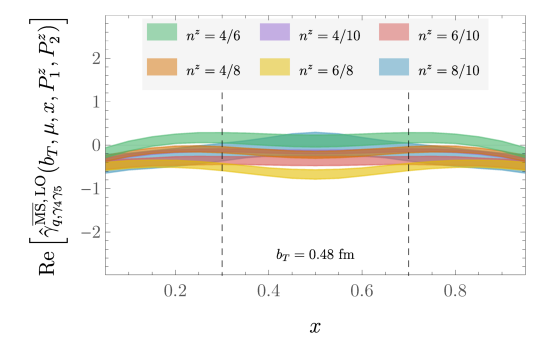}
                \hspace{10pt}
                \includegraphics[width=0.46\textwidth]
                {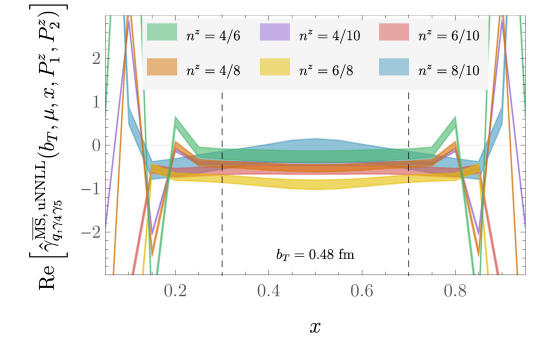}
                \hspace{10pt}
            \caption{Examples of real parts of CS kernel estimators $\hat{\gamma}_{\Gamma}^{\MSbar}(b_{\tran}, x, P_1^z, P_2^z, \mu)$, computed with matching corrections at LO (top panel) and uNNLL (bottom panel) accuracies as described in Section~\ref{sec:numerical-investigation-cs}, using $b_\tran = \SI{0.48}{\femto\meter}$ and  $\Gamma = \gamma_4\gamma_5$. 
            The black dashed lines enclose the region in $x$ used to determine the CS kernel.
            The notation $n^z = P_1^z/P_2^z$ displays momenta in lattice units.
            Further examples are shown in \cref{fig:wf_cs_x_a1,fig:wf_cs_x_a2,fig:wf_cs_x_b1,fig:wf_cs_x_b2} of \cref{sec:app:plots}.
            \label{fig:analysis-3}
            }
        \end{figure}
    The contribution of $a P^z$-dependent discretization artifacts to $\hat{\gamma}^{\MSbar}_\Gamma$ can be expected to be comparable to that of $x P^z$-dependent power corrections in the intermediate $x$ region. Since both effects are $P^z$-dependent, they can not be disentangled and it is left to future work to quantify their separate contributions to systematic uncertainty in the CS kernel determination. Here, the overall systematic uncertainty arising from these effects is estimated from the variation of $\hat{\gamma}^{\MSbar}_\Gamma$ over the choices of $x$, $P^1_z$, $P^2_z$, and $\Gamma$ for each choice of matching. 
    \par
\begin{figure}[t]
    \centering
        \includegraphics[width=0.46\textwidth]{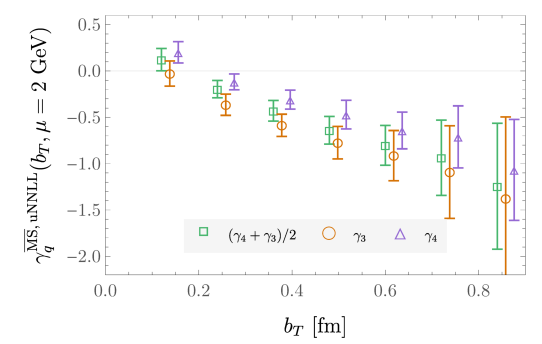}   
        \hspace{20pt}
        \includegraphics[width=0.46\textwidth]{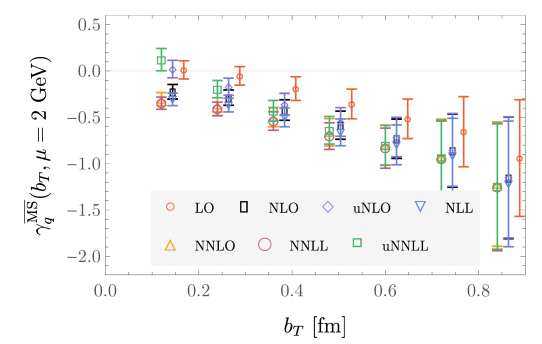}   
        \caption{\label{fig:cs_results}CS kernel in $b_T$ space for different choices of Dirac structure $\Gamma$ with uNNLL matching (top panel) and for all computed accuracies of the matching correction $\delta \gamma_q^{\MSbar}(b_\tran, \mu, x, P_1^z, P_2^z)$ (bottom panel). 
        }
    \end{figure}

 Precisely, the CS kernel is determined from an average of $\mathrm{Re}\big\lbrack\hat{\gamma}_{\Gamma}^{\MSbar}(b_{\tran}, x, P_1^z, P_2^z, \mu)\big\rbrack$ over $\Gamma \in \{\gamma_4 \gamma_5, \gamma_3\gamma_5\}$, computed pairs $\{P_1^z, P_2^z\}$, and a range of $x$.
    In particular, $x \in [0.3, 0.7]$ is taken to be the largest range of intermediate $x$ for which perturbative matching corrections including resummation avoid significant effects from singularities near $x=0$ and $x=1$.
    Weighted averages of $\mathrm{Re}\big\lbrack\hat{\gamma}_{\Gamma}^{\MSbar}(b_{\tran}, x, P_1^z, P_2^z, \mu)\big\rbrack$ are computed at the bootstrap level with weights taken to be proportional to the inverse variance of $\mathrm{Re}\big\lbrack\hat{\gamma}_{\Gamma}^{\MSbar}(b_{\tran}, x, P_1^z, P_2^z, \mu)\big\rbrack$.
    The estimator $\mathrm{Re}\big\lbrack\hat{\gamma}^{\MSbar}_{q,\Gamma}\big\rbrack$ is computed for a uniform grid of points in $x$ with spacing $\Delta x = 0.05$; a wide range of different choices of $\Delta x$ lead to indistinguishable results as long as correlations between $\mathrm{Re}\big\lbrack\hat{\gamma}_{\Gamma}^{\MSbar}(b_{\tran}, x, P_1^z, P_2^z, \mu)\big\rbrack$ with different $x$ are accounted for.
    \begin{figure}[t]
        \centering
        \includegraphics[width=0.46\textwidth]{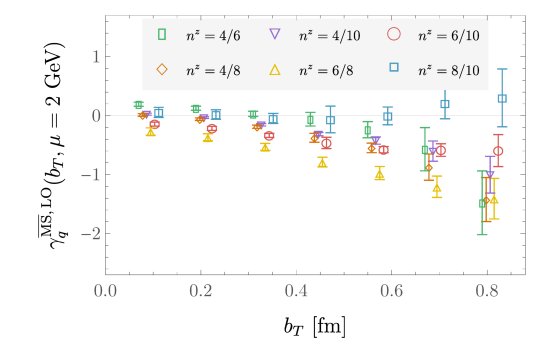}
        \hspace{20pt}   
        \includegraphics[width=0.46\textwidth]{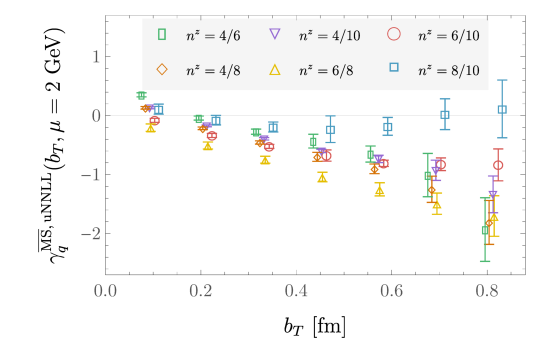}   
        \caption{\label{fig:cs_mtm} CS kernel in $b_T$ space evaluated sepatately for each momentum pair with LO (top panel) and uNNLL (bottom panel) matching.}
    \end{figure}
    Comparisons of these averaged estimators with different choices of $\Gamma$, perturbative matching accuracy, and momentum pairs, are shown in \cref{fig:cs_results,fig:cs_mtm}.
    The fitting procedure for $\mathrm{Im}\big\lbrack\hat{\gamma}_{\Gamma}^{\MSbar}\big\rbrack$ is identical. 
    \begin{figure}[t]
    \centering    
    \includegraphics[width=0.46\textwidth]{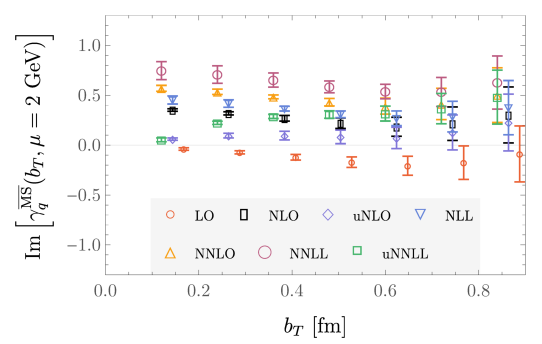} 
    \caption{Imaginary part of the CS kernel estimator shown for various accuracies of the perturbative matching correction $\delta \gamma_q^{\MSbar}(b_\tran, \mu, x, P_1^z, P_2^z)$. 
    \label{fig:cs_imag_part}
    }
    \end{figure}
    \par Whereas the CS kernel is a real quantity, averages of $\mathrm{Im}\big\lbrack\hat{\gamma}^{\MSbar}_{q,\Gamma}\big\rbrack$ at different perturbative accuracies indicate a nonzero imaginary part as illustrated in \cref{fig:cs_imag_part}.
    By comparing to the LO estimate, where the matching correction vanishes, it is clear that matching is a dominant source of the imaginary part.
    As discussed in \cref{sec:numerical-investigation-matching}, the imaginary part from the matching is attributed to $b_\tran$-dependent power corrections enhanced at small $b_\tran$ and mitigated by uNLO and uNNLL corrections.
    Consistent with this explanation, for small $b_\tran$, $\mathrm{Im}\big\lbrack\hat{\gamma}^{\MSbar}_{q,\Gamma}\big\rbrack$ at uNLO and uNNLL are reduced relative to all other orders of matching; as $b_\tran$ increases and power corrections are suppressed, they approach NLO and NNLL results, respectively.
    However, uNLO and uNNLL accuracies still do not lead to values of $\mathrm{Im}\big\lbrack\hat{\gamma}^{\MSbar}_{q,\Gamma}\big\rbrack$ that are consistent with zero within the accessible range of $b_\tran P^z$.
    This suggests that power corrections beyond those that have been accounted for by the unexpanded matching are relevant at the level of precision of this calculation.
    \par Since matching corrections with smallest expected power corrections are given by uNNLL, this accuracy is used for the final estimate of the CS kernel.
    Furthermore, considering both the larger qualitative difference between $\mathrm{Im}\big\lbrack\hat{\gamma}^{\MSbar}_{q,\Gamma}\big\rbrack$ for different accuracies and momenta, as well as the parametrically larger estimates of $b_\tran$-dependent power corrections compared to $\mathrm{Re}\big\lbrack\hat{\gamma}^{\MSbar}_{q,\Gamma}\big\rbrack$, the central value of the CS kernel is determined from fits to $\mathrm{Re}\big\lbrack\hat{\gamma}^{\MSbar,\,\mathrm{uNNLL}}_{q,\Gamma}\big\rbrack$ while $\mathrm{Im}\big\lbrack\hat{\gamma}^{\MSbar}_{q,\Gamma}\big\rbrack$ is not treated as a direct source of systematic uncertainty.
    Finally, scale variation in resummed corrections around $\mu_0 = 2 p^z$, with $p ^z \in \{ x P^z, (1-x) P^z \}$, is not used to estimate the associated perturbative uncertainties. 
    This choice is motivated by the range of $p^z$ used to determine the CS kernel,  and in particular because results at scales $\mu_0/2$ are sensitive to nonperturbative effects.
    The significance of higher-order perturbative effects may instead be judged by comparing the final uNNLL CS kernel determination to those obtained with other accuracies, as shown in \cref{fig:cs_results}.
    \begin{ruledtabular}
    \begin{table*}[t]
        \centering
        \begin{tabular}{ccccccccc}
            $b_T$ [fm] & 0.12 & 0.24 & 0.36 & 0.48 & 0.60 & 0.72 & 0.84 \\\hline
            $\gamma_q^{\MSbar,\,\mathrm{uNNLL}}$ 
                & 0.12(12)
                & -0.20(9)
                & -0.43(11)
                & -0.64(15)
                & -0.80(15)
                & -0.94(41)
                & -1.24(68)
        \end{tabular}
        \caption{Quark Collins-Soper kernel $\gamma_q^{\MSbar}(b_{\tran}, \mu=\SI{2}{\GeV})$ as a function of $b_T$.}
        \label{tab:CS}
    \end{table*}
    \end{ruledtabular}
     
    \par The final CS kernel results of this work are summarized in Table~\ref{tab:CS}. These results are shown as a function of $b_\tran$ and compared with phenomenological determinations of the CS kernel in Fig.~\ref{fig:cs_outlook}.
\section{Outlook}
    \begin{figure}[t]
    \centering    
    \includegraphics[width=0.46\textwidth]{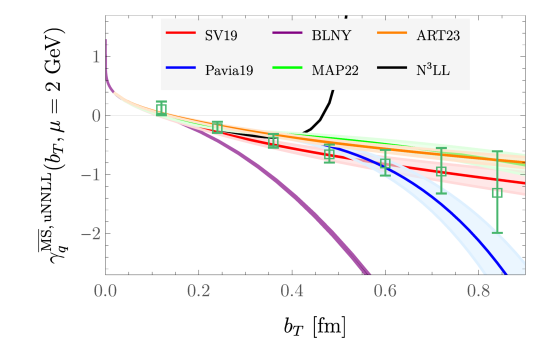}
    \caption{\label{fig:cs_outlook} CS kernel with uNNLL matching in $b_T$ space (green squares) compared to phenomenological parameterizations of experimental data in Refs.~\cite{Landry:2002ix,Scimemi:2019cmh,Bacchetta:2019sam,Bacchetta:2022awv,Moos:2023yfa} labeled BLNY, SV19, Pavia19, MAP22, and ART23, respectively,
    as well as perturbative results from Refs.~\cite{Collins:2014jpa,Li:2016ctv,Vladimirov:2016dll} labeled $\mathrm{N}^3\mathrm{LL}$.
    }
    \end{figure}
    \par This work presents a numerical determination of the quark Collins-Soper kernel $\gamma_q^{\MSbar}(b_{\tran}, \mu=\SI{2}{\GeV})$ in the nonperturbative range of $b_\tran$ corresponding to transverse momentum scales $\SI{240}{\MeV}\lesssim q_{\mathrm{T}}\lesssim\SI{1.6}{\GeV}$, through a lattice QCD calculation at a fixed lattice spacing and volume, quark masses corresponding to an approximately physical value of the pion mass $m_\pi = \SI{148.8
    \pm 0.1}{\MeV}$, and uNNLL perturbative matching power corrections in LaMET.
    Additionally, this work
    presents improved estimates of systematic uncertainties associated with perturbative matching from LaMET, the associated power corrections, and mixing effects in staple-shaped operators using the $\xMOM$ renormalization scheme. 
    \par While a complete quantification of systematic uncertainties would require performing lattice QCD calculations at multiple lattice spacings and at larger boosts or higher-order perturbative matching, the precision and control over systematic uncertainties achieved in this work is sufficient to preliminarily compare the CS kernel determination with phenomenological parameterizations of the kernel fit to experimental data. 
    In Fig.~\ref{fig:cs_outlook} the final determination is compared with the following parameterizations: Scimemi and Vladimirov (SV19)~\cite{Scimemi:2019cmh}, Bachetta et al. (Pavia19)~\cite{Bacchetta:2019sam}, the MAP Collaboration (MAPTMD22)~\cite{Bacchetta:2022awv}, Moos et al. (ART23)~\cite{Moos:2023yfa}, as well as an older parameterization based on the work of Brock, Landry, Nadolsky and Yuan (BLNY)~\cite{Landry:2002ix} and employed in recent code packages for resummation calculations relevant to precision electroweak measurements~\cite{Isaacson:2017hgb, Isaacson:2022rts}. 
    Within quantified uncertainties, the data agrees with all models in the range $\SI{.12}{\femto\meter} \lesssim b_\tran \lesssim \SI{.24}{\femto\meter}$, with all but BLNY for $\SI{.24}{\femto\meter} \lesssim b_\tran \lesssim \SI{0.6}{\femto\meter}$, and with SV19, MAPTMD22 and ART23 for $b_\tran \gtrsim \SI{0.6}{\femto\meter}$.
    Finally, for $b_\tran \geq \SI{0.6}{\femto\meter}$, the results are consistent with a constant, as suggested for the large-$b_\tran$ behavior in Ref.~\cite{Collins:2014jpa}.
    Discretization artifacts and power corrections, both enhanced at small $b_\tran$, will be studied in more detail in future work. More refined comparisons would also take into account the differences in the number of quark flavors and their masses between the lattice QCD determination and the global analyses, which lead to perturbative corrections described in Ref.~\cite{Pietrulewicz:2017gxc}.
    \par The first-principles QCD calculations achieved in this work provide new constraints on the quark CS kernel, with better control of the associated systematic uncertainties. The results are complementary to those achieved experimentally and, once the continuum limit is taken, can be rigorously compared to phenomenological parameterizations of the CS kernel from current global analyses. Moreover, in future analyses, lattice QCD constraints could be used to constrain the parameterizations in joint fits with experimental data.

\begin{acknowledgments}
\par We thank Feng Yuan for helpful discussions on the resummation in perturbative matching and Johannes Michel for valuable comments on the manuscript.
\par This work is supported in part by the U.S. Department of Energy, Office of Science, Office of Nuclear Physics, under grant Contract Numbers DE-SC0011090, DE-AC02-06CH11357, and by Early Career Award DE-SC0021006. PES is supported in part by Simons Foundation grant 994314 (Simons Collaboration on Confinement and QCD Strings). YZ is also supported in part by the 2023 Physical Sciences and Engineering (PSE) Early Investigator Named Award program at Argonne National Laboratory.
This manuscript has been authored by Fermi Research Alliance, LLC under Contract No.\ DE-AC02-07CH11359 with the U.S.\ Department of Energy, Office of Science, Office of High Energy Physics.
\par This research used resources of the National Energy Research Scientific Computing Center (NERSC), a U.S. Department of Energy Office of Science User Facility operated under Contract No. DE-AC02-05CH11231, the Extreme Science and Engineering Discovery Environment (XSEDE) Bridges-2 at the Pittsburgh Supercomputing Center (PSC) through allocation TG-PHY200036, which is supported by National Science Foundation grant number ACI-1548562, facilities of the USQCD Collaboration, which are funded by the Office of Science of the U.S. Department of Energy. 
We gratefully acknowledge the computing resources provided on Bebop, a high-performance computing cluster operated by the Laboratory Computing Resource Center at Argonne National Laboratory, and the computing resources at the MIT SuperCloud and Lincoln Laboratory Supercomputing Center~\cite{reuther2018interactive}.
The Chroma~\cite{Edwards:2004sx}, QLua \cite{qlua},  QUDA \cite{Clark:2009wm,Babich:2011np,Clark:2016rdz}, QDP-JIT \cite{6877336}, and QPhiX~\cite{10.1007/978-3-319-46079-6_30} software libraries were used in this work.
Data analysis used NumPy~\cite{harris2020array} and Julia~\cite{Julia-2017,mogensen2018optim}, and figures were produced using Mathematica \cite{Mathematica}.

\end{acknowledgments}
\appendix

\section{Constraints on quasi-TMD WF from discrete Lorentz transormations
\label{sec:app:discrete-transformations}}
    \par The properties of quasi-TMD WFs under charge conjugation $\mathcal{C}$, a product of reflections $\mathcal{R}_{\tran} \equiv \mathcal{R}_1\mathcal{R}_2$ in the transverse directions,  and time reversal $\mathcal{T}$, follow from the properties of the relevant staple-shaped operators $\mathcal{O}_{u\bar{d}}^{ \Gamma }(b_\tran, b^z, y, \ell, a)$  defined in \cref{eq:staple-shaped-op}. 
    These operators transform as
    \begin{align}
      \mathcal{C} \mathcal{O}^{\Gamma}_{u\bar{d}}(&b_\tran, b^z, y, \ell) \mathcal{C}^{-1} \nonumber \\  
            \label{eq:app-discrete-c-op}
            &=
            \mathcal{O}^{M_C \Gamma^T {M_C}^{-1}}_{d\bar{u}}\!(-b_\tran, -b^z, y, \ell), \\
       \mathcal{R}_\tran \mathcal{O}^{\Gamma}_{u\bar{d}}(&b_\tran, b^{z}, y, \ell) \mathcal{R}_\tran^{-1}  \nonumber \\ 
       \label{eq:app-discrete-p-op}
            &=
            \mathcal{O}^{M_R \Gamma {M_R}^{-1} }_{u\bar{d}}\!(-b_\tran, b^{z}, \mathbb{R}_{\tran}(y), \ell), \\
       \mathcal{T} \mathcal{O}^{\Gamma}_{u\bar{d}}(&b_\tran, b^{z}, y, \ell) \mathcal{T}^{-1} \nonumber \\ 
       \label{eq:app-discrete-t-op}
            &=
            \mathcal{O}^{M_T \Gamma {M_T}^{-1} }_{u\bar{d}}\!(b_\tran, b^{z}, \mathbb{T}(y), \ell),
    \end{align}
    where $\mathbb{R}_{\tran}(y) = (-y^1, -y^2, y^3, y^4)$, $\mathbb{T}(y) = (\pos{y}, -y^4)$, and the Dirac representation matrices $M_C$, $M_R$, and $M_T$ are defined by 
    \begin{align}
    \label{eq:m_c}
    M_C  &= \gamma_2 \gamma_4, \\
    \label{eq:m_p}
      M_R  &= (\gamma_1 \gamma_5)(\gamma_2 \gamma_5) = \gamma_2\gamma_1, \\
    \label{eq:m_t}
    M_T  &= \gamma_4 \gamma_5.
    \end{align}
    For further discussion of discrete transformations of staple-shaped operators, see Ref.~\cite{Ji:2021uvr}.
    \par These operator transformation properties constrain the unsubtracted bare quasi-TMD WFs $\phi^{\Gamma}(b_{\tran}, b^z, P^z, \ell)$.
    Using \cref{eq:app-discrete-c-op} in the isospin limit, charge conjugation invariance of pion states, and $u \leftrightarrow d$ exchange symmetry in the isospin limit gives
    \begin{align}
    \label{eq:app-discrete-c} \nonumber
        &\wf_{\Gamma}(b_{\tran}, b^z, P^z, \ell) \\\nonumber
        &= \mel{0}{\mathcal{C}^{-1}\mathcal{C} \mathcal{O}^{\Gamma}_{u\bar{d}}(b_\tran, b^{z}, 0, \ell) \mathcal{C}^{-1}\mathcal{C}}{\pi(P^z)} \\
        &= \wf_{M_C \Gamma^T M_C^{-1}}(-b_{\tran}, -b^z, P^z, \ell).
    \end{align}
    Next considering transverse reflections, pion states are pseudoscalar and are therefore invariant under the product of reflections $\mathcal{R}_{\tran}$.
    \cref{eq:app-discrete-p-op} can therefore be used to obtain
    \begin{align}
    \label{eq:app-discrete-p} \nonumber
        &\wf_{\Gamma}(b_{\tran}, b^z, P^z, \ell) \\\nonumber
        &= \mel{0}{\mathcal{R}_\tran^{-1}\mathcal{R}_\tran \mathcal{O}^{\Gamma}_{u\bar{d}}(b_\tran, b^{z}, 0, \ell) \mathcal{R}_\tran^{-1}\mathcal{R}_\tran}{\pi(P^z)} \\
        &= \wf_{M_R \Gamma M_R^{-1}}(-b_{\tran}, b^z, P^z, \ell),
    \end{align}
    which provides the $\Gamma$-dependent signs with which correlation functions can be averaged over different staple orientations.
    Combining these results gives
    \begin{align}
    \label{eq:app-discrete-cp} \nonumber
      &\wf_{\Gamma}(b_{\tran}, b^z, P^z, \ell) \\
      &= \wf_{M_R M_C \Gamma^T M_C^{-1} M_R^{-1}}(b_{\tran}, -b^z, P^z, \ell),
    \end{align}
    which establishes the symmetry properties of $\phi_{\Gamma}(b_{\tran}, b^z, P^z, \ell)$ under sign changes of $b^z$.
    In particular, it follows from \cref{eq:app-discrete-cp} that $\phi_{\gamma_4 \gamma_5}(b_{\tran}, b^z, P^z, \ell)$ and $\phi_{\gamma_3 \gamma_5}(b_{\tran}, b^z, P^z, \ell)$ are both symmetric in $b^z$.
   
    Finally, \cref{eq:app-discrete-t-op} and the $\mathcal{T}$-odd transformations of pion interpolating operators $\chi_{\mathbf{P}}^\dagger(0)$ can be used to obtain
    \begin{align}
    \label{eq:app-discrete-t} \nonumber
        &\wf_{\Gamma}(b_{\tran}, b^z, P^z, \ell) \\\nonumber
        &= \mel{0}{\mathcal{T}^{-1}\mathcal{T} \mathcal{O}^{\Gamma}_{u\bar{d}}(b_\tran, b^{z}, 0, \ell) \mathcal{T}^{-1}\mathcal{T}}{\pi(P^z)} \\
        &= -\wf_{M_T \Gamma M_T^{-1}}(b_{\tran}, b^z, P^z, \ell),
    \end{align}
    which provides the $\Gamma$-dependent signs with which correlation functions can be averaged over forward and backward propagation in time.

    Discrete transformation properties for renormalization factors can be derived analogously and ensure that renormalized quasi-TMD WFs share the same transformation properties as the bare quasi-TMD WFs with the corresponding $\Gamma$.

\section{\texorpdfstring{$\xMOM$}{RI/xMOM} renormalization scheme
\label{sec:app:xmom}}
    \par As discussed in \cref{sec:theory}, the renormalization condition of \cref{eq:rixmom} includes a Green's function containing a Wilson line and gives all the mixing effects of the staple-shaped operator in the $\xMOM$ scheme.
    This simplifies renormalization compared to other $\mathrm{RI}$-type schemes, which involve Green's functions of the operator itself and depend on the geometry of the Wilson-line staple. 
    Encoding all the mixing effects in \cref{eq:rixmom} is possible by interpreting the Wilson lines in QCD as originating from propagators of free auxiliary fields $\zeta_{n}(x)$~\cite{Ji:2017oey,Green:2017xeu,Green:2020xco}, 
    \begin{equation}
        \label{eq:app-zeta-prop}
        \begin{aligned}
            S_{\zeta_n}(\xi) &\equiv \expval{\zeta_{n}(x+\xi \hat{n}) \bar{\zeta}_{n}(x)}_{\zeta} \\
             &= \theta(\xi) \expval{\mathcal{W}_{-n}(x+\xi; \xi)},
        \end{aligned}
    \end{equation}
    where $\zeta_{n}(x)$ denote auxiliary fields of scalar particles moving along straight space-like directions $n^\mu$ and carrying color charge in the fundamental representation~\cite{Ji:2017oey,Green:2017xeu}.
    That is, the QCD action is augmented by $\zeta_{n}(x)$ in a way that returns the original action when the field is integrated out and \cref{eq:app-zeta-prop} holds.
    \par The staple-shaped operator in \cref{eq:staple-shaped-op}, nonlocal in QCD due to Wilson lines, may be recast in terms of local fields in the extended theory:
    \begin{equation}
    \label{eq:app-op-xmom}
        \begin{aligned}
        \mathcal{O}&^{\Gamma}_{u \bar{d}}(b_\tran, b^z, y) \\
        &=  \Bigg\langle  
                \bar{Q}_{d,-\hat{z}}\bigg(\!y+\frac{b}{2}\bigg)  
            \frac{\Gamma}{2}
                    C_{-z,n_\tran}
                        \bigg(y + \ell z + \frac{b_\tran}{2} \bigg) \\
            &\qquad \times C_{n_\tran, z}
                       \bigg(y + \ell z - \frac{b_\tran}{2}  \bigg)
                    Q_{u,z}
                    \bigg(
                         y - \frac{b}{2} 
                    \bigg)
        \Bigg\rangle_\zeta, 
        \end{aligned}
    \end{equation}
    where $C_{n, n^\prime}(x) \equiv \bar{\zeta}_{n}(x) \zeta_{n^\prime}(x)$ denote cusp operators, and $Q_{q, n}(x) \equiv \bar{\zeta}_{n}(x) q(x)$ denote composite spin-$1/2$ fields.
    The renormalization constant of the operator is thereby factorized into $Z_{C_{n,n^\prime}}$, $Z_{Q_{q,n}}$, $Z_q$ and $Z_{\zeta_n}$, renormalizing $C_{\hat{n},\hat{n}^\prime}$, $Q_{q,\hat{n}}$, quark, and $\zeta$ fields, respectively, as well as a factor of $e^{-\delta m (\ell + b_\tran)}$ where $\delta m$ denotes the mass of $\xi$ fields induced by loop effects~\cite{Ji:2017oey,Ishikawa:2017faj,Green:2017xeu}.
    \begin{figure*}[tp]
                \subfloat[$p_{\mathrm{R}} = \SI{2.15}{\GeV}$, $z=1$ (collinear $p^\mu_{\mathrm{R}}$ and $\xi^\mu_{\mathrm{R}}$).\label{fig:app-mixing-col}]{   
                    \centering
                    \includegraphics[width=0.46\textwidth]{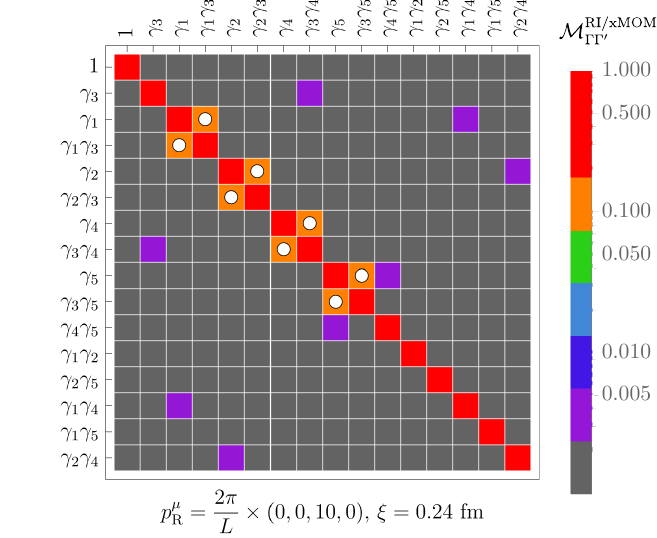}
                    \hspace{20pt}
                    \includegraphics[width=0.46\textwidth]{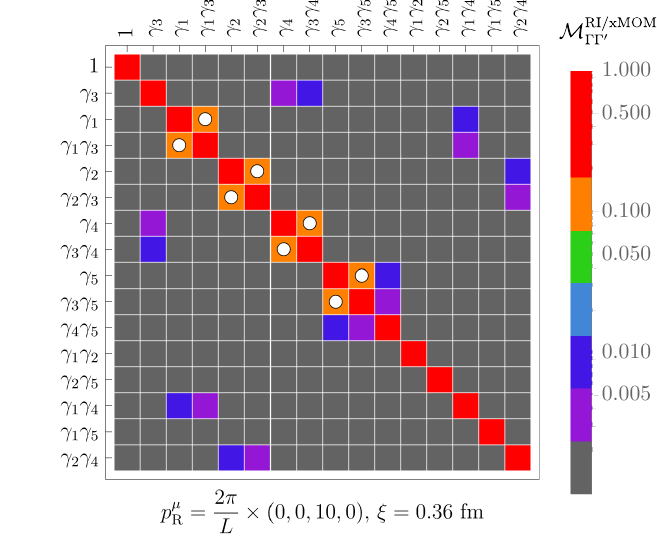} 
                }
                \hfill
                \subfloat[$p_{\mathrm{R}} = \SI{2.15}{\GeV}$ and $z=0$ (perpendicular $p^\mu_{\mathrm{R}}$ and $\xi^\mu_{\mathrm{R}}$).\label{mixing-app-1-2}]{
                    \centering
                    \includegraphics[width=0.46\textwidth]{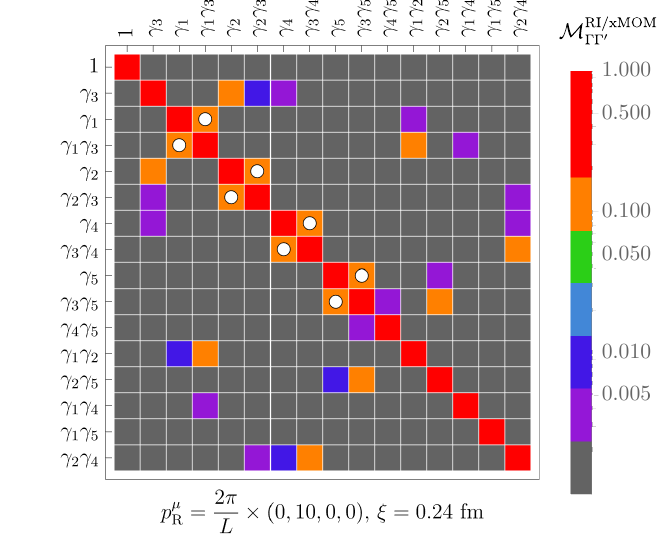} 
                    \hspace{20pt}
                    \includegraphics[width=0.46\textwidth]{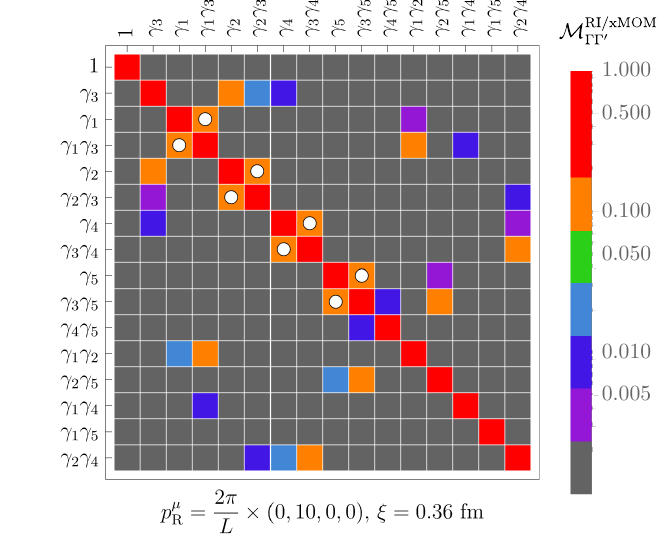} 
                }
                \hfill
                \subfloat[$p_{\mathrm{R}} = \SI{2.44}{\GeV}$ and $0<|z|<1$.\label{fig:app-mixing-mixed}]{
                    \centering
                    \includegraphics[width=0.46\textwidth]{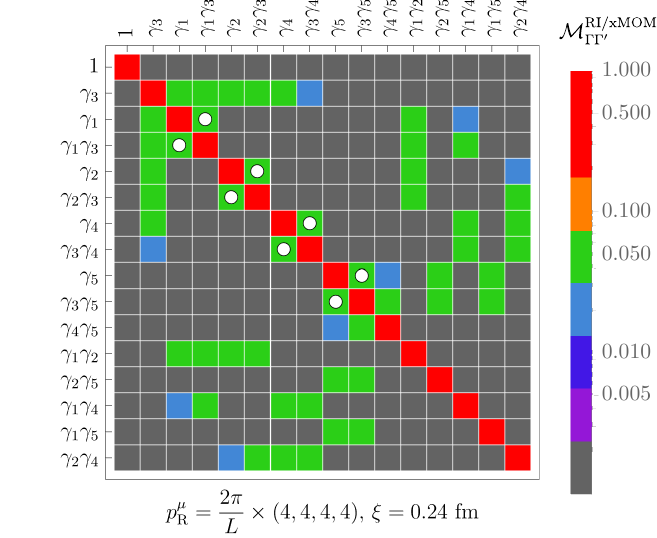} 
                    \hspace{20pt}
                    \includegraphics[width=0.46\textwidth]{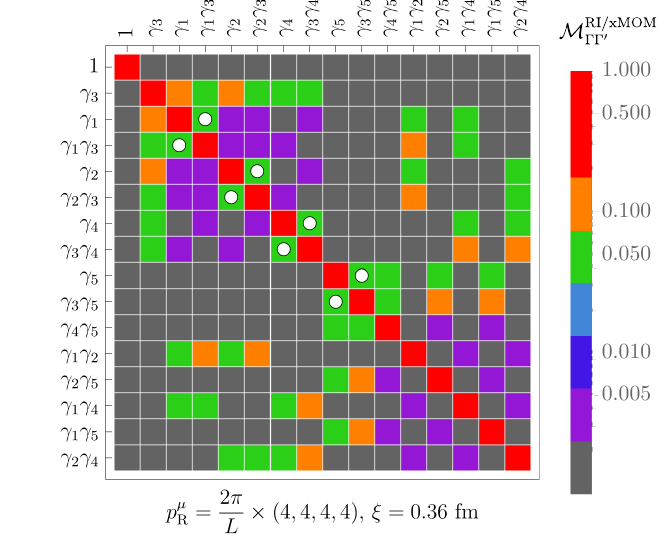} 
                }
                \hfill
                \caption{\label{fig:app-mixing}
                As in \cref{fig:mixing-full} in the main text, for a set of renormalization scales defined by $z \equiv p_{\mathrm{R}} \cdot \xi_{\mathrm{R}} / (|p_{\mathrm{R}}| |\xi_{\mathrm{R}}|)$ and $\xi_{\mathrm{R}}/a \in \lbrace 2, 3\rbrace$.
                }
            \end{figure*}
    \par In practice, the corresponding renormalization conditions can be solved in QCD by integrating out the auxiliary fields. 
    For example, while the Green's function in \cref{eq:nonamp} may be written as
    \begin{equation}
    \label{eq:app-amp-xmom}
    \begin{aligned}
        \Lambda&_{q, \pm z}(p, \xi) \\
        &\!\!\!\!\!\!= \lbrack S_{\zeta_{\mp z}}(-\xi)\rbrack^{-1}\!
            \bra{\zeta_{\mp z}(-\xi)} Q_{q, \mp z}(0) \ket{q(p)}
        \!\lbrack S_{q}(p) \rbrack^{-1},
    \end{aligned}
    \end{equation}
    it is still expressed in its original form when solving the renormalization condition in \cref{eq:rixmom} numerically, and \cref{eq:app-amp-xmom} is only used to identify the corresponding renormalization factor as
    \begin{equation}
    \label{eq:app-z}
    \begin{aligned}
        Z&^{\xMOM}_{{\Lambda}_{q, n},\alpha\alpha^\prime}(p_{\mathrm{R}}, \xi_{\mathrm{R}}) \\
            &\equiv \frac{Z^{\xMOM}_{Q_{q,n}, \alpha\alpha^\prime}(p_{\mathrm{R}}, \xi_{\mathrm{R}})}
    		{\lbrack Z_{q}(p_{\mathrm{R}})\rbrack^{1/2}[Z^{\xMOM}_{\zeta_n}(\xi_{\mathrm{R}})]^{1/2}},
    \end{aligned}
    \end{equation}
    where $\alpha$, $\alpha^\prime$ are spin indices. 
    The remaining renormalization conditions in $\xMOM$ are approached similarly.
    Altogether, using \cref{eq:app-op-xmom}, the renormalization factor of the staple-shaped operator may then be computed via the renormalization factors in the auxiliary-field description. 
    \par Moreover, when computing the CS kernel via \cref{eq:kernel-wf-lattice}, renormalization factors with no spin structure cancel in the ratio\thinspace---\thinspace it is therefore sufficient to find any combination of them that fully encodes the mixing effects.
    $Z^{\xMOM}_{{\Lambda}_{q, n}}$ in \cref{eq:app-z}, as determined by solving \cref{eq:rixmom}, is one such combination. 
    \par For collinear configurations of $p^\mu_{\mathrm{R}}$ and $\xi^\mu_{\mathrm{R}}$ defined by $z \equiv p_\mathrm{R} \cdot \xi_{\mathrm{R}} / (|p_\mathrm{R}| |\xi_{\mathrm{R}}|) = \pm 1$, $Z^{\xMOM}_{{\Lambda}_{\pm z},\alpha\alpha^\prime}(p_{\mathrm{R}}, \xi_{\mathrm{R}})$ may be converted to $\MSbar$ via the conversion coefficient computed analytically in Landau gauge in continuum perturbation theory~\cite{Green:2020xco},
    \begin{widetext}
    \begin{equation}
        \label{eq:app-conversion-coeff}
        \begin{aligned}
        \big\lbrack C_{\pm z} &\big\rbrack^{ \MSbar}_{\xMOM}(\mu, p^\mu_\mathrm{R}, \xi^\mu_\mathrm{R}) \\
        &= \frac{1}{12} \mathrm{Tr}\Lambda_{Q_{\pm z}}^{\MSbar}(\mu, p^\mu_\mathrm{R}, \xi^\mu_\mathrm{R})   \\
    	&= 1 + \frac{\alpha_{\mathrm{s}}(p_\mathrm{R}) C_{\mathrm{F}}}{2\pi} 
    	\left(\!\!\left(
    		- 2\log 2 + \frac{3}{4} - \frac{1}{2} \frac{\sin y}{y}
    		- \frac{1}{4}\cos y  
    	-\left(
    			2\cos\frac{y}{2} 
    			+ \frac{y}{4} \sin\frac{y}{2}
    		\right) \mathrm{Ci}\left(\frac{y}{2}\right)
    		+ 2\mathrm{Ci}(y)
        \right)
    	\right.\\
    	&\left.\quad\!\!\!\!\!\!\qquad\qquad\qquad\qquad \mp i \left(
           			-\frac{1}{2 y} + \frac{1}{2} \frac{\cos y}{y} - \frac{1}{4} \sin y 
                    -\left(
           				2 \sin\frac{y}{2} - \frac{y}{4} \cos\frac{y}{2}
           			\right) \mathrm{Ci}\left(\frac{y}{2}\right)
           			+ 2 \mathrm{Si}(y)
           		\right)\!\!\right) 
             \\ &\quad\,\,\,\,\,+ \mathcal{O}(\alpha^2_{\mathrm{s}}(p_\mathrm{R})),
        \end{aligned}
        \end{equation}
        \end{widetext}
        where $y \equiv p_{\mathrm{R}} \cdot \xi_\mathrm{R}$, $C_{\mathrm{F}} = 4/3$ and $\mathrm{Si}(y) \equiv \int_{y}^{\infty} \frac{\sin(t)}{t}\mathrm{d}t$ and $\mathrm{Ci}(y) \equiv - \int_{y}^{\infty} \frac{\cos(t)}{t}\mathrm{d}t$ are the sine and cosine trigonometric integrals, respectively.
        The dependence on $\mu$ vanishes in Landau gauge at NLO.
        The conversion coefficient for $Z^{\xMOM}_{\Gamma\Gamma^\prime}(p_{\mathrm{R}}, \xi_{\mathrm{R}}, a)$ in \cref{eq:mod-quasi-wf-renorm} is given by 
        $C^{\MSbar}_{\xMOM} =
        \big\lbrack C_{- z} \big\rbrack^{ \MSbar *}_{\xMOM}\big\lbrack C_{+ z} \big\rbrack^{ \MSbar}_{\xMOM}$.
    \par As mentioned in \cref{sec:numerical-investigation-renorm}, the mixing effects induced by $Z^{\xMOM}_{\Gamma\Gamma^\prime}(p_{\mathrm{R}}, \xi_{\mathrm{R}})$ on $p^\mu_{\mathrm{R}}$ and $\xi^\mu_{\mathrm{R}}$ receive additional contributions not expected in lattice perturbation theory at one-loop order for noncollinear configurations of $p^\mu_{\mathrm{R}}$ and $\xi^\mu_{\mathrm{R}}$~\cite{Constantinou:2017sej}.
    These mixing effects are illustrated \cref{fig:app-mixing}.
    When $p_\mathrm{R} \cdot \xi_\mathrm{R} = 0$, additional mixing contributions appear at $10\%$-level.
    When $p_{\mathrm{R}}^\mu$ has components both collinear with and perpendicular to $\xi_{\mathrm{R}}^\mu$, the number of mixing contributions is larger, but the magnitude of each is reduced. 
    Since $\xMOM$ is an off-shell momentum scheme, contributions to mixing other than those induced by the staple-shaped operator renormalization itself are possible and may be relevant to explain the additional contributions~\cite{Politzer:1980me,Martinelli:1994ty,Ebert:2019tvc}.
    Notably, the additional contributions are significantly smaller than those observed in the $\MOM$ scheme in previous works~\cite{Shanahan:2019zcq}.

\section{Matching corrections}
\label{app:matching}
\par The quasi-TMD WF factorization formula from the discussion of power corrections in \cref{sec:theory} is given by~\cite{Ji:2019sxk,Ji:2021znw,Deng:2022gzi}
\begin{align}\label{eq:fact}
	&\frac{\tilde \phi^{\pm}(x, b_T,\mu, P^z)}{\sqrt{S_r(b_T,\mu)}} = H^{\pm}(x, P^z, \mu)
    \nonumber\\
	&\qquad\!\!\!\!\! \times \exp\!\left[{\frac{1}{4}}\!\left(\ln{\frac{(2xP^z)^2}{\zeta}} + \ln{\frac{(2\bar{x}P^z)^2}{  \zeta}}\right)\!\gamma_q(b_T,\mu)\right] \nonumber\\
	&\qquad\!\!\!\!\! \times \phi^{\pm}(x, b_T, \mu, \zeta) + \mathrm{p.c.},
\end{align}
where matching holds independently of the suppressed flavor indices, Dirac structure indices, and the renormalization scheme label up to power corrections, denoted $\mathrm{p.c.}$\footnote{Note that the CS evolution part in the matching formula differs from that in Refs.~\cite{Ji:2019sxk,Ji:2021znw,Deng:2022gzi} by a suppressed imaginary part in the exponential, which depends on the soft factor subtraction. 
    The imaginary part is suppressed here because it does not affect the extraction of the CS kernel.
    }
The reduced soft factor $S_r(b_T,\mu)$~\cite{Ji:2019sxk} ensures that the infrared physics is the same as that of the physical TMD WF. 
The $\pm$ label denotes the $\pm \hat{z}$ displacement of the transverse Wilson line relative to the quarks in the staple-shaped operator used to define the quasi-TMD WF. 
Only the $+\hat{z}$ displacement is shown in \cref{fig:staple-shaped-operator} and used in the determination of the CS kernel, and the $\pm$ label is omitted throughout the main text; the label is made explicit for completeness in the following discussion of the matching correction. 
The matching kernel $H^{\pm}(x, P^z, \mu)$ is given by~\cite{Vladimirov:2020ofp}
\begin{equation}
\label{eq:Hcc}
    H^{\pm}(x, P^z, \mu) = C^{\pm}_\phi(\mu, xP^z) C^{\pm}_\phi(\mu, \bar{x} P^z) 
\end{equation}
where the coefficients $C^\pm_\phi$ can be derived from the matching of a heavy-to-light current in the heavy-quark effective theory to soft-collinear effective theory~\cite{Vladimirov:2020ofp}.
\subsection{Fixed-order matching corrections}
\label{app:FO}
\par A fixed-order matching correction in \cref{eq:matching-correction} requires matching coefficients $C_\phi(\mu, p^z)$ computed in a perturbative expansion
\begin{equation}
    \label{eq:matching-pert}
    C^{\pm}_{\phi}(\mu, p^z) = 1 + \sum_{n=1} a_s^{n}(\mu) C^{\pm, (n)}_\phi(\mu, p^z),
\end{equation}
where $a_s(\mu) \equiv \alpha_s(\mu)/4 \pi$ and $\alpha_s(\mu)$ is determined by running from $\alpha_s(\mu_0 = \SI{2}{\GeV})$ as detailed in \cref{app:resum}.
At NNLO, the logarithmic ratio of these coefficients in the matching correction is expanded as
\begin{equation}
\label{eq:matching-nnlo-2}
\begin{aligned}
	\delta&\gamma^{\rm NNLO-II}(\mu,x, P_1^z, P_2^z) \\
    &=- {\frac{1}{\ln(P_1^z/P_2^z)}} \\
    &\quad \times 
    \bigg\lbrace a_s(\mu) \Big(\!C^{\pm, (1)}_\phi(\mu, xP_1^z) - C^{\pm, (1)}_\phi(\mu, xP_2^z)\!\Big) \\
    &\quad \quad -
    \frac{a^2_s(\mu)}{2}
        \Big(\!
            \big[C^{\pm, (1)}_\phi(\mu, xP_1^z)\big]^{\!2} \!-\!\big[C^{\pm, (1)}_\phi(\mu, xP_2^z)\big]^{\!2} \\
            &\qquad\qquad - 2 \big(C^{\pm, (2)}_\phi(\mu, xP_1^z) - 
            C^{\pm, (2)}_\phi(\mu, xP_2^z)\big) \\
    &\qquad+ (x \leftrightarrow \bar{x})\bigg\rbrace.
\end{aligned}
\end{equation}
While taking a naive logarithmic ratio of NNLO matching coefficients,
\begin{figure}[t]
    \centering    \includegraphics[width=0.46\textwidth]{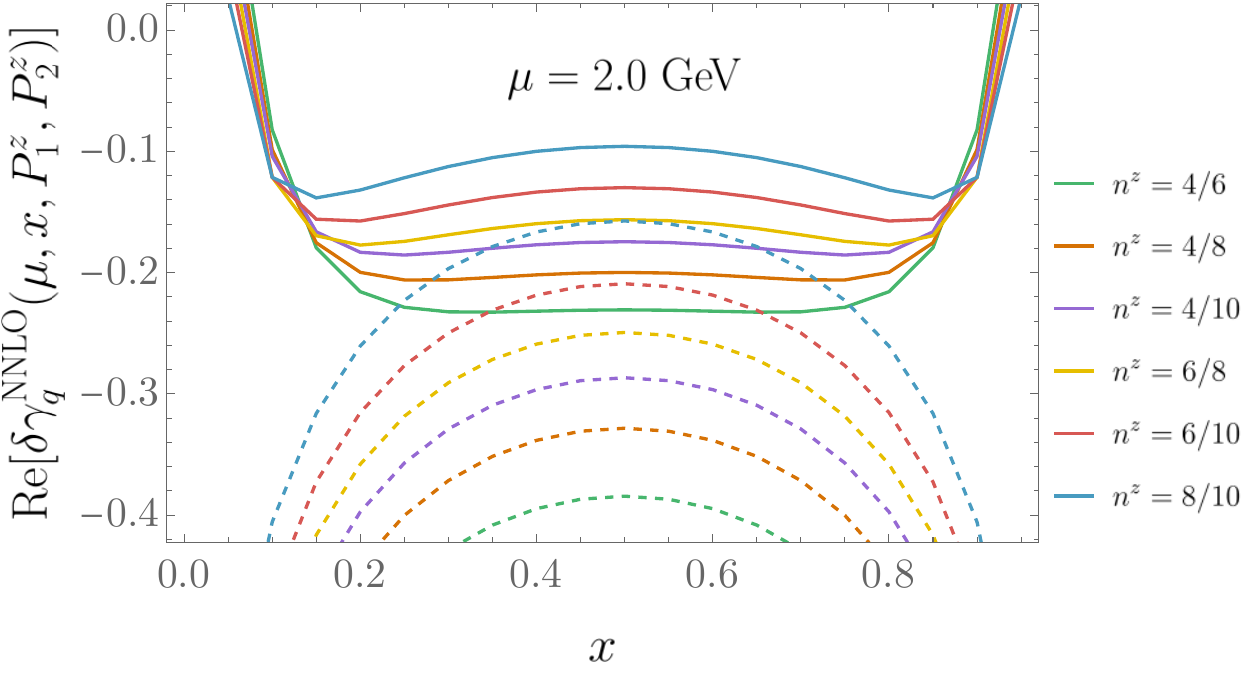}
    \includegraphics[width=0.46\textwidth]{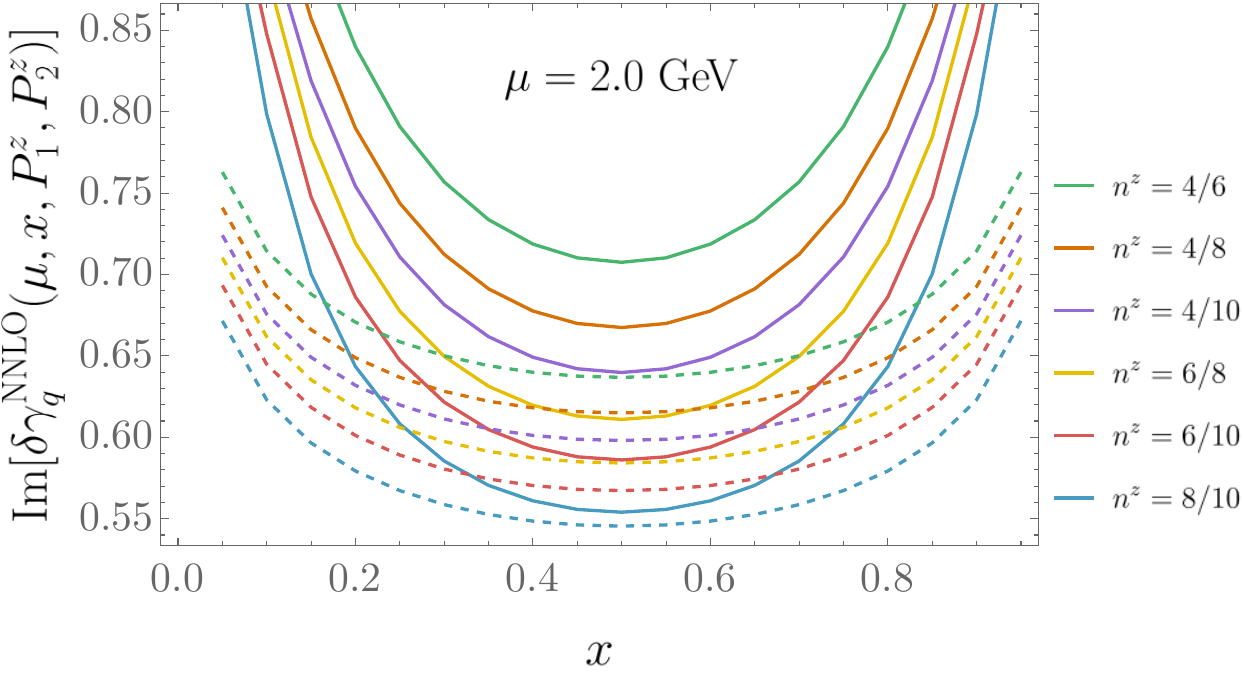}
         \caption{\label{fig:nnloIvsII} Matching correction to the CS kernel for different momentum pairs $n_1^z/n_2^z$ and $\mu=2$ GeV. The solid and dashed lines show results for NNLO-I and NNLO-II, respectively.}
\end{figure}
\begin{equation}
\label{eq:matching-nnlo-1}
\begin{aligned}
	\delta&\gamma^{\rm NNLO-I}(\mu,x, P_1^z, P_2^z) \\
    &= {\frac{1}{\ln(P_1^z/P_2^z)}}\\
	&\quad \times 
    \ln
        \frac
            {1 + a_s(\mu) C_\phi^{\pm, (1)}(\mu, xP_1^z) + a_s^2(\mu) C_\phi^{\pm, (2)}(\mu, xP_1^z)}
            {1 + a_s(\mu) C_\phi^{\pm, (1)}(\mu, xP_2^z) + a_s^2(\mu) C_\phi^{\pm, (2)}(\mu, xP_2^z)},
\end{aligned}
\end{equation}
differs from the correction in \cref{eq:matching-nnlo-2} only by higher-order terms, in the kinematic regime of this study the discrepancy is significant, as illustrated in \cref{fig:nnloIvsII}.
Consistent with the scaling of power corrections, $\delta\gamma^{\rm NNLO-I}$ and $\delta\gamma^{\rm NNLO-II}$ converge at larger momenta, but the rates of convergence and the sign and magnitude of $x$-dependent corrections differ between real and imaginary parts. 
The same conclusions apply to the NLO matching corrections, for which terms of order $a_s^2(\mu)$ are dropped in \cref{eq:matching-nnlo-2} and \cref{eq:matching-nnlo-1}.
\begin{figure*}[t]
    \centering    
    \includegraphics[width=0.46\textwidth]{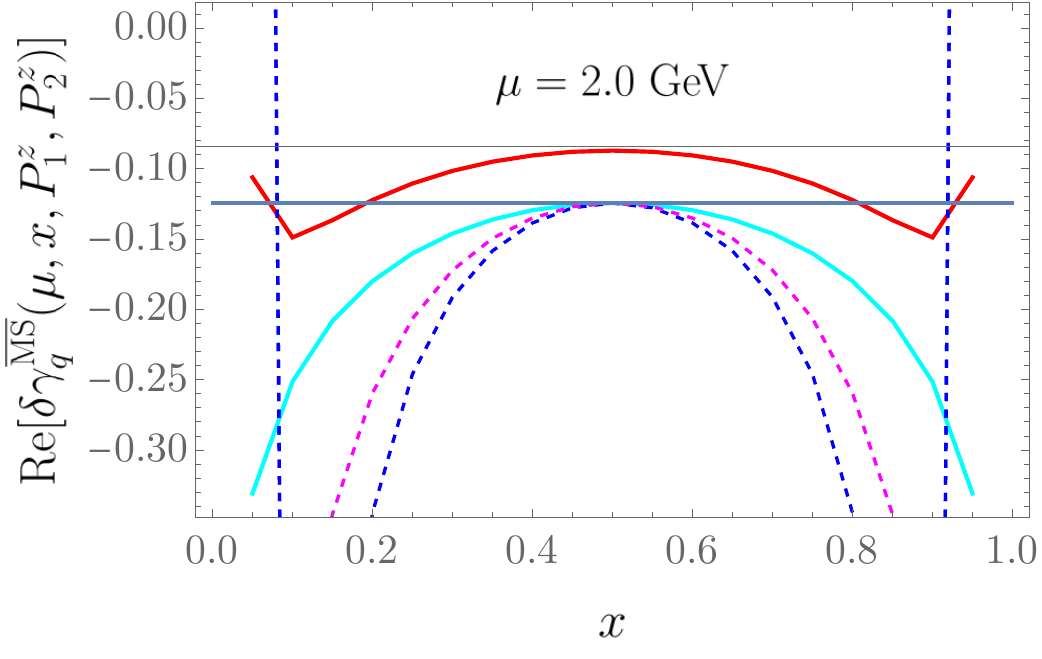}
    \hspace{20pt}
    \includegraphics[width=0.46\textwidth]{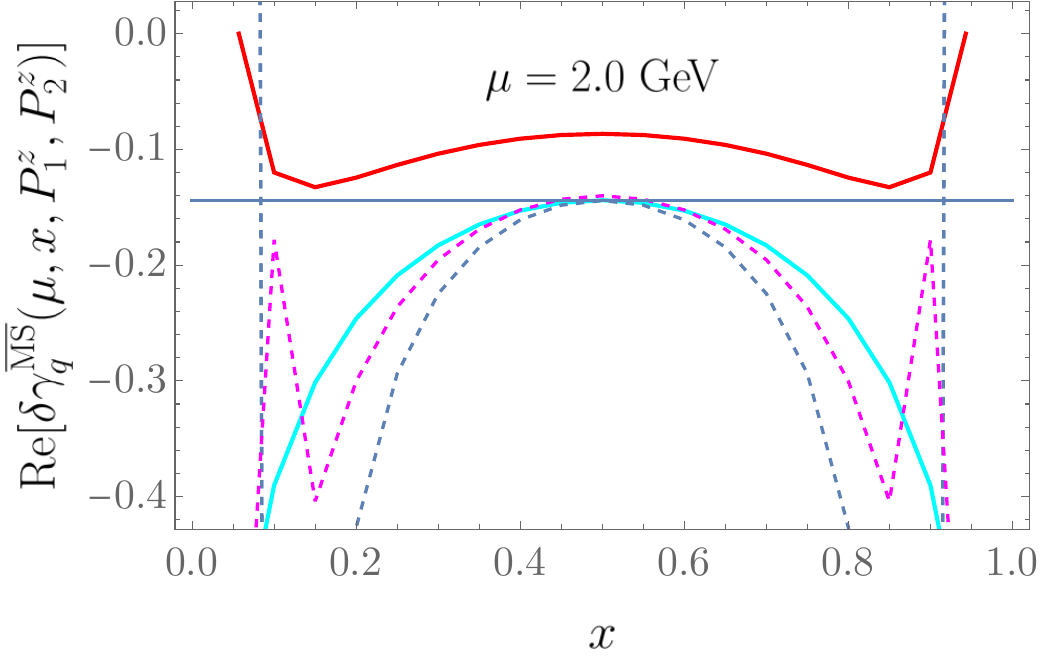}
         \caption{\label{fig:comp} Matching correction to the CS kernel at $(P_1^z,P_2^z)=(1.999,2.001)$ GeV and $\mu=2$ GeV. Left panel: NLO and NLL. Right panel: NNLO and NNLL. The fixed-order I and II, resummation I and II are represented by red, cyan, dashed blue and dashed magenta lines, respectively. 
         The noncusp anomalous dimension is represented by the blue solid line.
         }
    \end{figure*}
As discussed further in \cref{app:resum} and illustrated in \cref{fig:comp}, corrections $\delta\gamma^{\rm NLO-II}$ and $\delta\gamma^{\rm NNLO-II}$ are in a better agreement with results expected from the RG equations of $C^{\pm}_{\phi}(\mu, p^z)$.
For this reason, the fixed-order results with a naive logarithmic ratio are not used in the determination of the CS kernel.
\begin{figure*}[t]
    \centering    
    \includegraphics[width=0.46\textwidth]{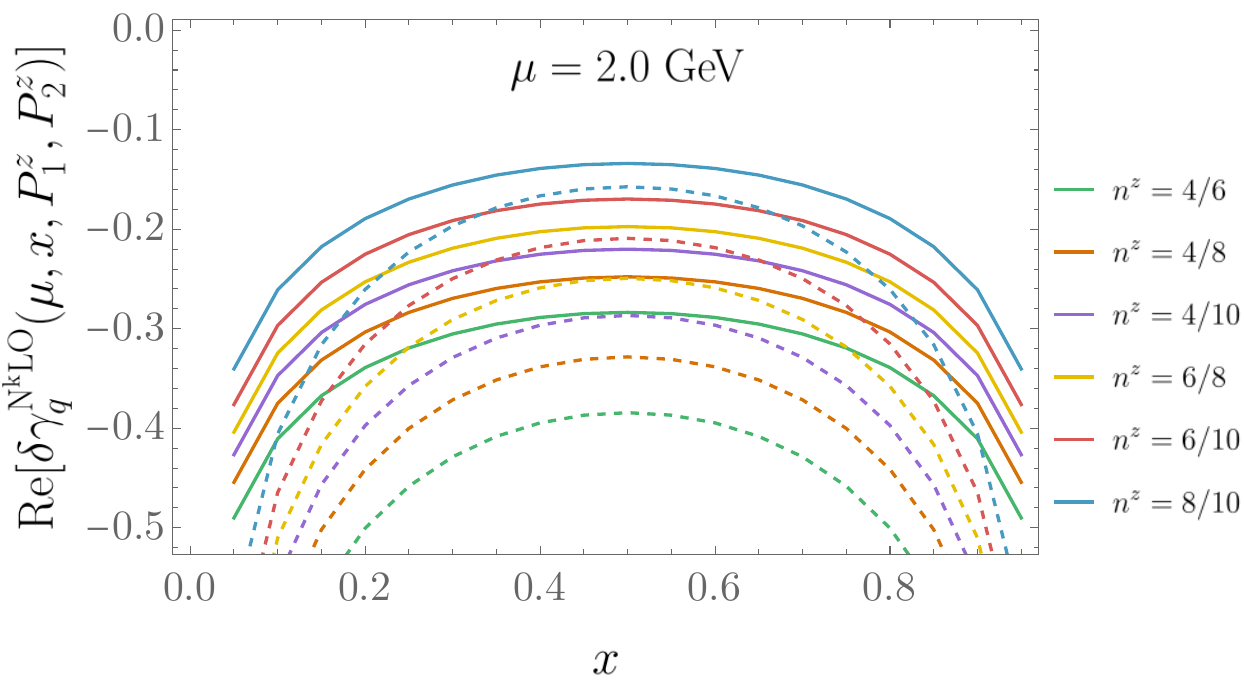}
    \hspace{20pt}
    \includegraphics[width=0.46\textwidth]{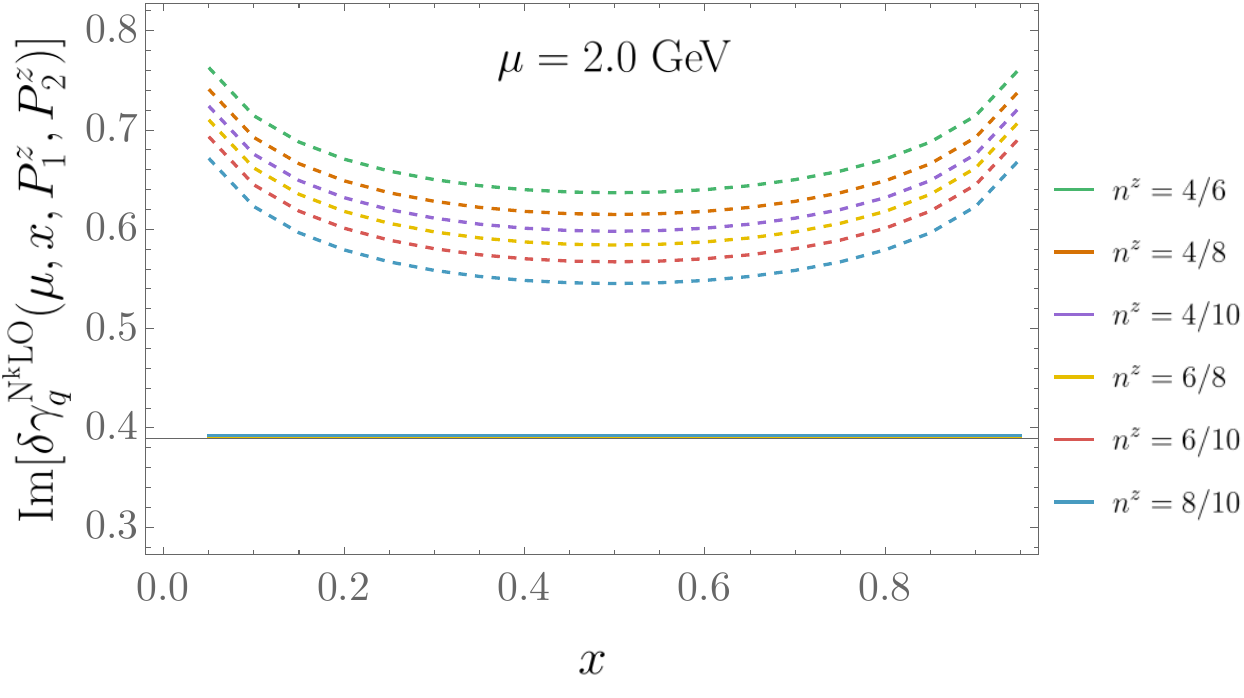}
         \caption{Matching corrections to the CS kernel at NLO (solid lines) and NNLO (dashed lines).
         \label{fig:nlovsnnlo}
         }
    \end{figure*}
\par The difference between $\delta\gamma^{\rm NLO-II}$ and $\delta\gamma^{\rm NNLO-II}$, illustrated in Fig.~\ref{fig:nlovsnnlo}, indicates expected convergence in the real component of the matching correction at moderate $x$. 
However, matching corrections converge poorly in the imaginary component.
This is in agreement with NLO results in Ref.~\cite{LPC:2022ibr} and may be explained by a larger sensitivity of $\mathrm{Im}(\delta\gamma_q(\mu,x, P_1^z, P_2^z))$ to power corrections at small $b_T$, as discussed further in \cref{app:finite}. 
\par The matching coefficients needed for the NNLO matching correction are given explicitly below, with $C_F = 4/3$, $C_A = 3$, $n_f =4$, and $\zeta(n)$ denoting the Riemann zeta function. 
At NLO, $C^{(1)}_\phi(\mu, p^z)$ has been calculated~\cite{Ji:2021znw,Deng:2022gzi} to be
\begin{align}
\label{eq:fo-1}
	C^{\pm,(1)}_\phi(\mu, p^z) 
	&= C_F\left[ - {\frac{1}{2}}(\mathbf{L}_z^{\pm})^2 + \mathbf{L}_z^{\pm} - 2 - {\frac{5\pi^2}{12}} \right],
\end{align}
where
\begin{align}
\label{eq:logs}
	\mathbf{L}_z^{\pm} & 
    = \ln {\frac{-(2p^z)^2 \pm i0}{\mu^2}} = \ln {\frac{(2p^z)^2}{\mu^2}} \pm i\pi.
\end{align}
The NNLO coefficients $C^{\pm, (2)}_\phi(\mu, p^z)$ for quasi-TMD WFs can be extracted from the corresponding results for quasi-TMD parton distribution functions (quasi-TMD PDFs), for which a factorization analogous to that in \cref{eq:fact} holds~\cite{Ebert:2018gzl,Ebert:2019okf,Ji:2019ewn,Ebert:2022fmh}.
The matching kernel for quasi-TMD PDFs has been calculated at NLO~\cite{Ji:2014hxa,Ebert:2018gzl,Ebert:2019okf,delRio:2023pse} and recently at NNLO~\cite{delRio:2023pse,Ji:2023pba}.
The real part of the coefficient $C^\pm_\phi(\mu, p^z)$ is equal to the square root of the matching kernel for the quasi-TMD PDF with the identification of $\zeta_z=(2p^z)^2$.
Obtained in this way, $C^{\pm,(1)}_\phi(\mu, p^z)$ is consistent with \cref{eq:fo-1}, and $C^{\pm,(2)}_\phi(\mu, p^z)$ is given by
\begin{widetext}
\begin{equation}
\label{eq:fo-2}
\begin{aligned}
	C^{\pm,(2)}_\phi(\mu, p^z) 
	   &= {C_F\over 2}\left\{{C_F\over 4}(\mathbf{L}_z^{\pm})^4 
        - \left(C_F-\frac{11}{9}C_A+\frac{2}{9} n_f\right)(\mathbf{L}_z^{\pm})^3\right.\\
        &\quad+ \left[\left(3+\frac{5\pi ^2}{12}\right) C_F + \left(\frac{\pi ^2}{3}-\frac{100}{9}\right) C_A +\frac{16}{9} n_f\right] (\mathbf{L}_z^{\pm})^2 \\
        &\quad-\left[\left(\frac{11 \pi ^2}{2}-24 \zeta (3)\right)C_F 
        + \left(22 \zeta (3)-\frac{44 \pi ^2}{9}-\frac{950}{27}\right) C_A + \left(\frac{152}{27}+\frac{8 \pi ^2}{9}\right)n_f\right] \mathbf{L}_z^{\pm} \\
	   &\quad+ \left(-30 \zeta (3)+\frac{65 \pi ^2}{3}-\frac{167 \pi ^4}{144}-16\right)C_F + \left(\frac{241 \zeta (3)}{9}+\frac{53 \pi ^4}{60}-\frac{1759 \pi ^2}{108}-\frac{3884}{81} \right)C_A \\
	   &\qquad\left. + \left(\frac{2 \zeta (3)}{9}+\frac{113 \pi ^2}{54}+\frac{656}{81} \right)n_f\right\}.
\end{aligned}
\end{equation}
\end{widetext}
\subsection{Resummation of momentum logarithms}
\label{app:resum}
\par The resummation of the matching coefficients discussed in \cref{sec:theory} is enabled by their RG evolution equations~\cite{Ji:2019ewn,Ebert:2022fmh},
    \begin{align}
    \label{eq:gamma-mu}
    &\begin{aligned}
	  & \gamma_\mu^\pm(\mu, p^z) 
          \equiv {d\ln C_\phi^\pm(\mu, p^z)\over d\ln \mu} \\
	       &\qquad\qquad = \Gamma_{\rm cusp}[a_s(\mu)] \mathbf{L}_z^\pm + \gamma_\mu[a_s(\mu)],
    \end{aligned} \\
    \label{eq:gamma-C}
    &\begin{aligned}
	&   \gamma^\pm_C(\mu, p^z) 
        \equiv {d\ln C^\pm_\phi(\mu, p^z)\over d\ln p_z} \\
	    &\qquad = 2\int_ {2p^z}^\mu {d\mu'\over \mu'} 
        \Big(\Gamma_{\rm cusp}[\alpha_s(\mu')] 
       + \gamma^\pm_C[\alpha_s(2p^z)]\Big),
    \end{aligned}
    \end{align}
    where $\gamma^{\pm}_\mu(\mu, p^z)$ and $\gamma^{\pm}_C(\mu, p^z)$ are the virtuality and momentum anomalous dimensions of $C^{\pm}_\phi(\mu, p^z)$, respectively,
    $\gamma_\mu(a_s(\mu))$ and $\gamma^{\pm}_C(a_s(2 p^z))$ denote initial values in the solutions to the RG equations, and
    \begin{equation}
    \label{eq:cusp}
        \Gamma_{\mathrm{cusp}}(\alpha_{s}(\mu)) = \dv{\gamma^{\pm}_\mu(\mu, p^z)}{\ln p^z} = \dv{\gamma^{\pm}_C(\mu, p^z)}{\ln \mu}.
    \end{equation}
  is the cusp anomalous dimension. 
  \par The anomalous dimension $\gamma^\pm_C(\mu, p^z)$ in \cref{eq:gamma-C} may be used to approximate the matching correction in \cref{eq:matching-correction} in the limit of $P_1^z \to P_2^z$.
  As illustrated in \cref{fig:comp}, this approximation is used to select a fixed-order expansion of the matching correction in \cref{eq:matching-nnlo-2} over that in  \cref{eq:matching-nnlo-1}.
  Finally, the relation
   \begin{align}
	   \gamma_C^\pm(\mu, p^z) = - \gamma^\pm_\mu(\mu, p^z) + \beta(a_s) {\partial \ln C_\phi^\pm(\mu, p^z)\over \partial a_s}
    \end{align}
   may be used to cross-check explicit perturbative results for $\gamma^\pm_C(\mu, p^z)$ and $\gamma^\pm_\mu(\mu, p^z)$ detailed further below.
\par In terms of the anomalous dimensions, the resummation kernel $K_\phi\left(\mu_0(\mu, p^z), \mu\right)$ in \cref{eq:matching-resummation} is given by
\begin{equation}
\label{eq:resummation-kernel-II}
 \begin{aligned}
 K&^{\pm\mathrm{II}}_\phi\big(2 p^z, \mu\big) \\
    &=  2 K_\Gamma(2p^z, \mu) 
                - K_{\gamma_\mu}(2p^z, \mu) 
                \mp \eta(2p^z, \mu)
\end{aligned}
\end{equation}
for $(p_0^z, \mu_0) = (p^z, 2p^z)$ and 
\begin{equation}
\label{eq:resummation-kernel-I}
 \begin{aligned}
 K&^{\pm\mathrm{I}}_\phi\big(2 p^z, \mu\big) = 2 K_\Gamma(2 p^z, \mu) 
                - K_{\gamma^\pm_C}(2 p^z, \mu)
\end{aligned}
\end{equation}
for $(p_0^z, \mu_0) = (\mu/2, \mu)$, where
\begin{align}
     \label{eq:matching-k-gamma}
     K_{\gamma_{\mu}}(\mu_0, \mu)
        &= \int_{\alpha_{s}(\mu_0)}^{\alpha_s(\mu)}
            \dfrac{\dd{\alpha_s}}{\beta(\alpha_s)}
                \gamma_{\mu}(\alpha_s), \\
     \label{eq:matching-k-gamma-C}
     K_{\gamma^\pm_{C}}(\mu_0, \mu)
        &= \int_{\alpha_{s}(\mu_0)}^{\alpha_s(\mu)}
            \dfrac{\dd{\alpha_s}}{\beta(\alpha_s)}
                \gamma^{\pm}_{C}(\alpha_s), \\
    \label{eq:matching-k-Gamma}
    K_\Gamma(\mu_0, \mu) 
        &=
            \int_{\alpha_{s}(\mu_0)}^{\alpha_s(\mu)}
            \dfrac{\dd{\alpha_s}}{\beta(\alpha_s)}
                \Gamma_{\mathrm{cusp}}(\alpha_s) \nonumber \\ 
            &\qquad\qquad\times \int_{\alpha_s(\mu_0)}^{\alpha_{s}}  \dfrac{\dd{\alpha^\prime_s}}{\beta(\alpha^\prime_s)},
        \\
    \label{eq:matching-eta}
    \eta_\Gamma(\mu_0, \mu) 
        &= i\pi\int_{\alpha_{s}(\mu_0)}^{\alpha_s(\mu)}
           \dfrac{\dd{\alpha_s}}{\beta(\alpha_s)}
                \Gamma_{\mathrm{cusp}}(\alpha_s),
    \end{align}
    and $\beta[\alpha_s(\mu)]\equiv \dv{\alpha_s(\mu)}{\ln\mu}$ is the QCD $\beta$-function.
\begin{figure}[t]
    \centering    \includegraphics[width=0.46\textwidth]{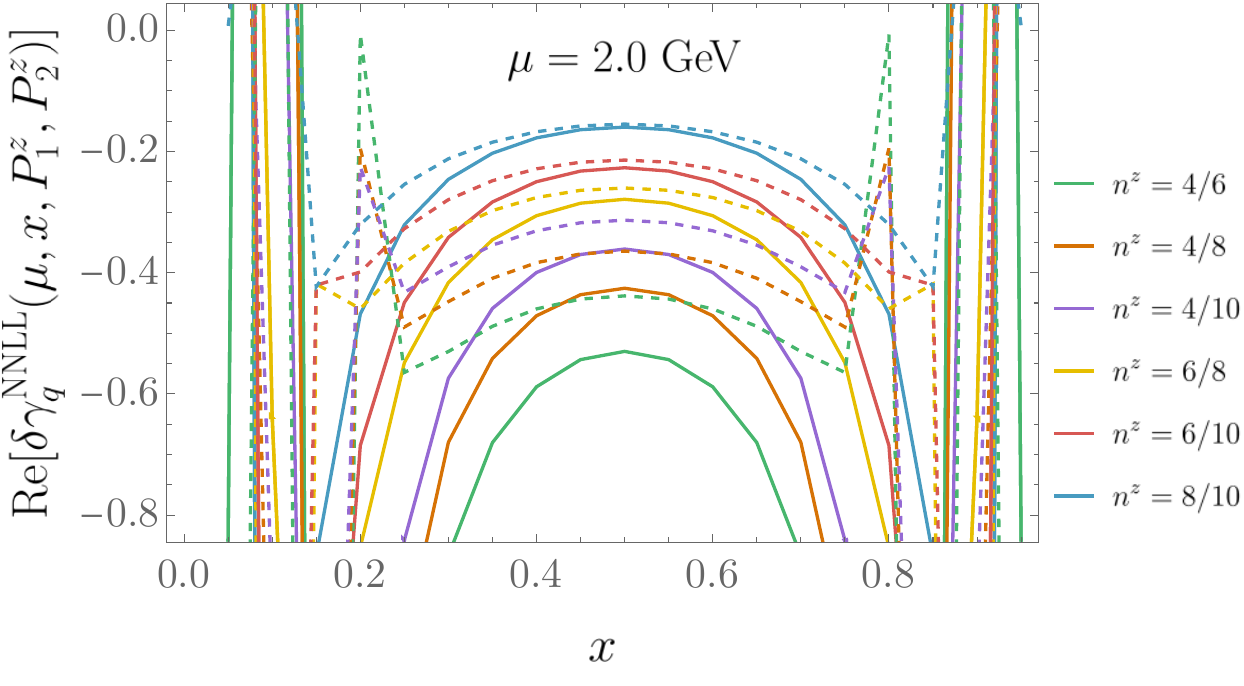}
    \includegraphics[width=0.46\textwidth]{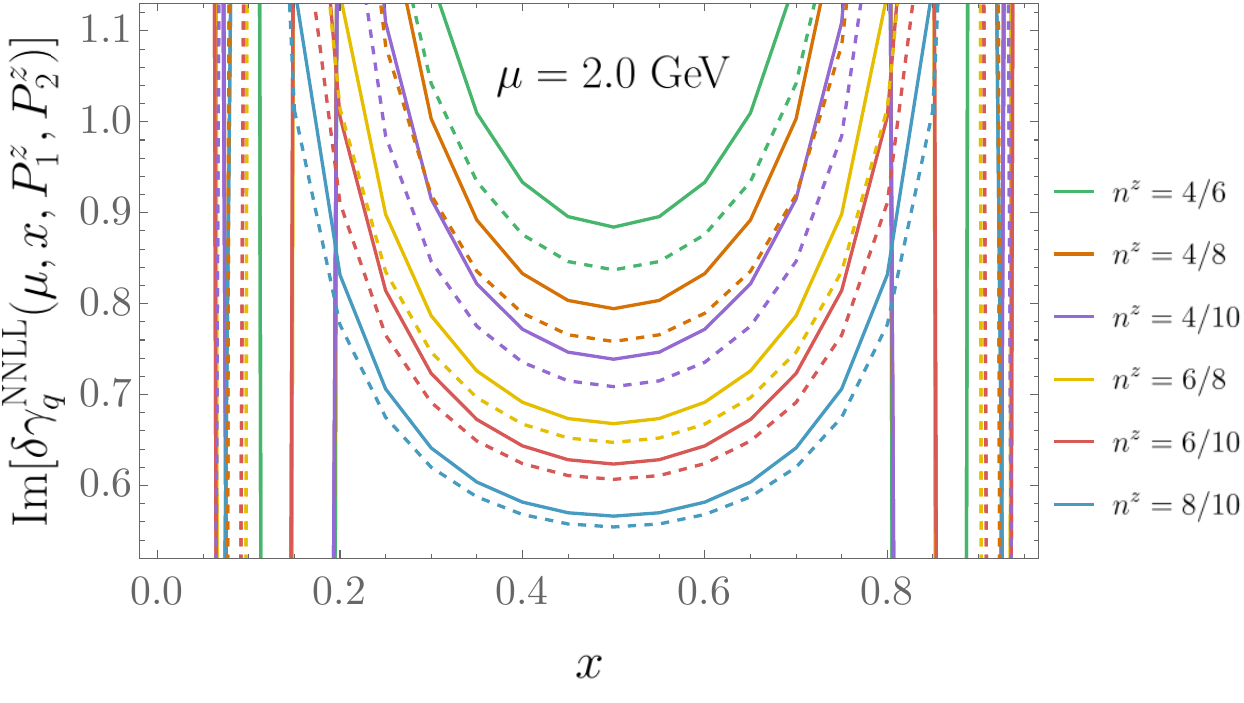}
         \caption{Real and imaginary parts of resummed matching corrections to the CS kernel $\delta\gamma^{\MSbar,\,\mathrm{N}^k\mathrm{LL}}_q(\mu,x, P_1^z, P_2^z)$ at NNLL in the two resummation schemes: NNLL-I (solid) and NNLL-II (dashed), as defined by a choice of initial scale in \cref{eq:matching-resum-I,eq:matching-resum-II}.
         The corrections are shown in different colors at different momentum pairs $n_1^z/ n_2^z$ and $\mu=2$ GeV.
         \label{fig:nnllIvsII}
         }
\end{figure}
\par The resummed matching coefficients corresponding to $K^\mathrm{I}_\phi$ and $K^\mathrm{II}_\phi$ are given according to \cref{eq:matching-correction-NNLO} by
\begin{equation}
    \label{eq:matching-resum-II}
    \begin{aligned}
	   \delta&\gamma_q^{\mathrm{II}}(\mu,x, P_1^z, P_2^z) \\
	   &= - \dfrac{1}{\ln(P_{1}^{z}/P_{2}^{z})}  \bigg(\ln\frac{C_\phi(2x P_1^z, x P^z_1)}{C_\phi(2x P_2^z, x P^z_2)}
         \\
        &\quad- \left(\!K^{\mathrm{II}}_\phi\!\left(2 x P^z_1, \mu\right) - K^{\mathrm{II}}_\phi\!\left(2 x P^z_2, \mu\right)\!\right) \\
        &\quad+ (x \leftrightarrow \bar{x})\!
        \bigg),
    \end{aligned}
    \end{equation}
where the logarithmic ratio of initial-scale matching coefficients is expanded in $a_s(\mu)$ as in \cref{eq:matching-nnlo-2} for the fixed-order case, and
\begin{equation}
    \label{eq:matching-resum-I}
    \begin{aligned}
	   \delta&\gamma_q^{\mathrm{I}}(\mu,x, P_1^z, P_2^z) \\
	   &= \dfrac{1}{\ln(P_{1}^{z}/P_{2}^{z})}\bigg(\!K^{\mathrm{I}}_\phi\!\left( 2 x P^z_1, \mu \right) - K^{\mathrm{I}}_\phi\!\left(2 x P^z_2, \mu \right) \\
        &\quad+ (x \leftrightarrow \bar{x})\!
        \bigg).
\end{aligned}
\end{equation}
\fig{nnllIvsII} compares matching corrections in the two schemes at NNLL in the kinematic regime used in this work to determine the CS kernel. 
The differences between the resummations decrease at larger momenta, consistent with the decreasing $\alpha_s$.
Since the ratios calculated from the lattice are renormalization group invariant and independent of the $\MSbar$ scale $\mu$, the natural choice of the initial scales should be proportional to the hard parton momentum in the quasi-TMD WFs, e.g., $(p_0^z, \mu_0)=(p^z,2p^z)$.
Therefore, in this work the resummed matching corrections are determined in scheme II.
\par To obtain the resummed matching corrections, all functions comprising $K_\phi$ are computed perturbatively in $a_s(\mu)$,
    \begin{align}
        \beta[\alpha_s(\mu)] &= - 2\alpha_s(\mu) \sum_{n=0}^\infty a_s^{n+1}(\mu)\beta_n, \\
	    \Gamma_{\rm cusp}[a_s(\mu)] &= \sum_{n=0}^\infty a_s^{n+1}(\mu) \Gamma_n, \\
        \gamma_\mu[a_s(\mu)] &= \sum_{n=0}^\infty a_s^{n+1}(\mu)\gamma^{\mu}_{n}, \\
        \gamma_C^\pm[a_s(\mu)] &= \sum_{n=0}^\infty a_s^{n+1}(\mu)\gamma^{C\pm}_{n}.
    \end{align}
    \begin{ruledtabular}
    \begin{table}[t]
	\centering
	\begin{tabular}{cccccc}
	Accuracy & $K_\Gamma$ & $K_{\gamma_C}$ & $K_{\gamma_\mu}$ & $\eta$ & $C_\phi$ \\
        \hline
	NLL  & 2    & 1 & 1 & 1 & 0 \\
	NNLL & 3	& 2 & 2 & 2 & 1 \\
	\end{tabular}
    \caption{Loop orders of each term comprising the resummed matching coefficient defined in \cref{eq:matching-resummation} at a given accuracy.
    The loop orders of the beta function $\beta[\alpha_s(\mu)]$ and the coupling $a_s(\mu)$ are equal to the loop order of the term they are used in. 
    All the functions are defined in \cref{app:resum}.
    }
    \label{tab:res}
    \end{table}
    \end{ruledtabular}
    A resummation of $C^\pm_\phi(\mu, p^z)$ from $C^\pm_\phi(\mu_0, p^z_0)$ of a given accuracy corresponds to a consistent set of loop orders chosen for $C^\pm_\phi(\mu_0, p^z_0)$ and the functions above, with $a_{s}(\mu)$ run from $a_{s}(\mu_0 = \SI{2}{\GeV})$ as detailed further below.
    Examples for NLL and NNLL resummations are provided in \cref{tab:res}.
    Explicitly, the following perturbative results are used for the NLL and NNLL resummations.
    The $\beta$-function is given by 
    \begin{align}
    \label{eq:beta}
	&\beta_0 = {11\over 3}C_A - {4\over 3}T_Fn_f, \\
	&\beta_1 = {34\over3}C_A^2 - \left({20\over3}C_A + 4C_F\right) T_Fn_f, \\
	&\begin{aligned}
        \beta_2 &= {2857\over 54}C_A^3  \\ 
        &\quad+ \left(2C_F^2 - {205\over 9}C_FC_A - {1415\over 27}C_A^2 \right)T_Fn_f \\
	    &\quad+ \left({44\over 9} C_F+ {158\over 27}C_A \right)T_F^2 n_f^2\,,
    \end{aligned}
    \end{align}
    where $T_F=1/2$.
The cusp anomalous dimension $\Gamma_{\rm cusp}$,  computed to four-loop order~\cite{Korchemsky:1987wg,Moch:2004pa,Henn:2019swt,vonManteuffel:2020vjv,Moult:2022xzt,Duhr:2022yyp}, is given by
\begin{align}\label{eq:cusp-pert}
	&\Gamma_0 = 2C_F,\qquad \\
    &\Gamma_1 = 2C_F\left[\left({67\over 9} - {\pi^2\over3}\right) C_A - {20\over9}T_F n_f \right], \\
    &\begin{aligned}
	\Gamma_2 &= 2C_F \left[C_A^2\left(\!{245\over 6} - {134\pi^2\over 27} + {11\pi^4\over 45} + {22\over3}\zeta(3)\!\right) \right.\\
	&\quad + C_AT_Fn_f \left(-{418\over 27} + {40\pi^2\over 27} - {56\over 3}\zeta(3)\right) \\
	& \quad \left. + C_F T_F n_f \left(-{55\over3} + 16\zeta(3)\right) - {16\over 27}T_F^2n_f^2\right].
    \end{aligned}
\end{align}
The noncusp anomalous dimensions are given in terms of $\gamma_n^\mu$ and $\gamma_n^{C\pm} \equiv \gamma_n^C \mp i\pi \gamma_n^{C*}$.
Like the matching coefficients discussed in \cref{app:FO}, they can be extracted from the recently calculated NNLO matching kernel of the quasi TMD PDFs~\cite{delRio:2023pse,Ji:2023pba} and are given by
\begin{align}
	&\gamma^\mu_0= -2C_F,\\
	&\begin{aligned}
    \gamma^\mu_1 
        &= C_F\left[  C_F\left(-4+{14\over 3}\pi^2-24\zeta(3)\right) \right.\\
	       &\quad\left.+ C_A\left(-\frac{554}{27}-\frac{11 \pi ^2}{6}+22 \zeta (3)\right) \right. \\
        &\quad\left.+ n_f\left(\frac{80}{27}+\frac{\pi ^2}{3}\right)\right]\,,
    \end{aligned}
\end{align}
and
\begin{align}
\label{eq:noncusp}
	&\gamma^{C\pm}_0= 2C_F \mp i\pi \Gamma_0\,,\\
	&\begin{aligned}
	    \gamma^{C\pm}_1 &= C_F\left[  C_F\left(4-{14\over 3}\pi^2+24\zeta(3)\right) \right. \\
	   &\quad\left. + C_A\left(\frac{950}{27}+\frac{11 \pi ^2}{9}-22 \zeta (3)\right)\right. \\
    &\quad\left. + n_f\left(-\frac{152}{27}-\frac{2 \pi ^2}{9}\right)\right]\\
	   &\quad  \mp i\pi \left[\Gamma_1 + \beta_0(2 \mathrm{Re}[\gamma_0^C] - \Gamma_0)\right],
    \end{aligned}
\end{align}
respectively, where the imaginary part is inferred from the logarithm $\mathbf{L}_z^\pm$ in \cref{eq:logs} of the fixed-order result. 
\par The corresponding perturbative expressions of resummation kernels for the NNLL resummation are~\cite{Stewart:2010qs}
\begin{equation}
\begin{aligned}
K&^{\rm NNLL}_{\gamma_\mu}(\mu_0, \mu) \\
	&= - {\gamma^\mu_0\over 2\beta_0}\left[\ln r + a_s(\mu_0) \left({\gamma_1^\mu\over \gamma_0^\mu} - {\beta_1\over \beta_0} \right) (r-1)  \right],
\end{aligned}
\end{equation}
\begin{equation}
\begin{aligned}
K&^{\rm NNLL}_{\gamma_C}(\mu_0,\mu) \\
	&= - {\gamma^C_0\over 2\beta_0}\left[\ln r + a_s(\mu_0) \left({\gamma_1^C\over \gamma_0^C} - {\beta_1\over \beta_0} \right) (r-1)  \right],
\end{aligned}
\end{equation}
\begin{equation}
\begin{aligned}
K&^{\rm NNLL}_\Gamma(\mu_0, \mu) \\
	&= - {\Gamma_0\over 4\beta_0^2}\left\{{1 \over a_s(\mu_0)} \left(1-{1\over r} - \ln r\right) \right.\\
	&\quad + \left({\Gamma_1\over \Gamma_0} - {\beta_1\over \beta_0} \right)(1-r+\ln r) + {\beta_1\over 2\beta_0}\ln^2r\\
	&\quad  + a_s(\mu_0)\left[ \left({\beta_1^2\over \beta_0^2} - {\beta_2\over \beta_0}\right)\left({1-r^2\over 2} + \ln r\right) \right.\\\
	&\qquad \left. + \left( {\beta_1 \Gamma_1\over \beta_0\Gamma_0} - {\beta_1^2\over \beta_0^2}\right)(1-r+r \ln r)\right.\\
	&\qquad \left.\left. - \left({\Gamma_2\over \Gamma_0} - {\beta_1 \Gamma_1 \over \beta_0\Gamma_0}\right) {(1-r)^2\over 2}\right] \right\},
\end{aligned}
\end{equation}
and
\begin{equation}
\begin{aligned}
\eta&^{\rm NNLL}(\mu_0, \mu) \\
	&= - i \pi {\Gamma_0\over 2\beta_0}\left[\ln r + a_s(\mu_0) \left({\Gamma_1\over \Gamma_0} - {\beta_1\over \beta_0} \right) (r-1)  \right],
\end{aligned}
\end{equation}
where $r \equiv a_s(\mu)/a_s(\mu_0)$ and the running coupling at $\mu$ is given at NNLO order by
\begin{equation}
\label{eq:as-pert}
\begin{aligned}
	{1\over a_s(\mu)} &= {X\over a_s(\mu_{0})} + {\beta_1\over \beta_0} \ln X \\ 
    &\quad+ a_s(\mu_{0}) \left[{\beta_2\over \beta_0}\left(1- {1\over X} \right) \right.\\
	&\qquad\qquad\quad \left. + {\beta_1^2\over \beta_0^2}\left({\ln X\over X} + {1\over X} -1 \right) \right],
\end{aligned}
\end{equation}
where $X\equiv 1+ \beta_0  a_s(\mu)\ln(\mu_0^2/\mu^2)$, and $\alpha_{s}(\mu_0 = \SI{2}{\GeV}) \approx 0.293$ is determined as prescribed in Ref.~\cite{Burkert:2008rj}.
N$^{3}$LO terms require $a_s(\mu)$ at NNLO, and NNLO terms at NLO.
Finally, for the NNLL resummation, the logarithmic ratio of initial-scale coefficients in \cref{eq:matching-resum-II} is expanded as in \cref{eq:matching-nnlo-2} to NLO.
\subsection{Estimate of \texorpdfstring{$b_T$}{bT}-dependent corrections
\label{app:finite}
}
\par The validity of the factorization formula in \Eq{fact} requires that $xP^z b_T \gg 1$ and $(1-x)P^zb_T \gg 1$.
Within the kinematic range of $P^z$ and $b_T$ used in this work, such conditions are not sufficiently satisfied, especially at small $b_T$, and considerable power corrections are expected.
\par Nonetheless, a factorization should exist for some range $x\in[x_{\rm min}, x_{\rm max}]$ for all values of $b_T$, as long as $xP^z, \bar{x}P^z\gg \Lambda_{\rm QCD}$.
If $P^z b_T \gg 1$, a factorization into TMDs applies; 
if $P^z b_T \ll 1$, then it is reduced to a collinear factorization. 
One may conjecture a factorization formula that interpolates between collinear and TMD factorizations, written schematically at finite $P^z$ as
\begin{equation}
\label{eq:fact2}
\begin{aligned}
	\tilde \phi&^\pm(x, b_T,\mu, P^z) \\ 
 &= \int_0^1 dy\ H^{\overline{\rm MS}\pm}_\phi(x, y;\bar{x},\bar{y}; xP^z,\bar{x}P^z,b_T,\mu)\\
	&\quad \times \exp\left[{1\over4}\left({(2xP^z)^2\over \zeta} + {(2\bar{x}P^z)^2\over \zeta}\right)\gamma_\zeta(b_T,\mu)\right]\\
	&\quad \times \phi(y, b_T, \mu, \zeta),
 \end{aligned}
\end{equation}
where the matching kernel has a large $P^z b_T$ expansion for  $P^zb_T\gtrsim1$,
\begin{equation}
\begin{aligned}
	& H^{\overline{\rm MS}\pm}_\phi(x, y;\bar{x},\bar{y}; xP^z,\bar{x}P^z,b_T,\mu) \\
	&\xrightarrow{P^zb_T\gg 1} C_\phi^\pm(\mu, xP^z) C_\phi^{\pm}(\mu, \bar{x}P^z) \delta(x-y) \\
	&\qquad\qquad\quad+ 
    \left[\delta C_\phi^\pm(b_T, xP^z) C_\phi^{\pm}(b_T, \bar{x}P^z) \right. \\ 
    &\qquad\qquad\qquad\quad\left.+
        (x\to \bar{x})\right]\delta(x-y) \\
	&\qquad\qquad\quad+  \delta H^{\overline{\rm MS}\pm}_\phi(x, y;\bar{x},\bar{y}; xP^z,\bar{x}P^z,b_T)\,,
\end{aligned}
\end{equation}
where $\delta C_\phi^\pm(b_T,xP^z)$ and $\delta H^{\overline{\rm MS}\pm}_\phi(x, y;\bar{x},\bar{y}; xP^z,\bar{x}P^z,b_T)$ denote power- or exponentially-suppressed terms such as $1/(x P^z b_T)$ and $1/(\bar{x} P^z b_T)$ or $\exp(-x P^z b_T)$ and $\exp(-\bar{x} P^z b_T)$.
\par For the purposes of estimating the significance of the finite-$b_\tran$ correction, the contribution $\delta H^{\overline{\rm MS}\pm}_\phi$ to the above matching kernel is neglected and its study is left to future work. 
The matching kernel then reduces to
\begin{equation}
\begin{aligned}
	& H^{\overline{\rm MS}\pm}_\phi(x, y;\bar{x},\bar{y}; xP^z,\bar{x}P^z,b_T,\mu) \\
	&\xrightarrow{P^z\gg b_T^{-1}}  C_\phi^{\pm\unexp}(b_T,\mu, xP^z) C_\phi^{\pm\unexp}(b_T,\mu, \bar{x}P^z) \\
    &\qquad\qquad\quad\times \delta(x-y),
\end{aligned}
\end{equation}
where the $(p^zb_T)$-unexpanded coefficient $C_\phi^{\pm\unexp}(b_T,\mu, p^z) = C_\phi^\pm(\mu, p^z) + \delta C_\phi^\pm(b_T,p^z)$
has a perturbative expansion analogous to that of $C_\phi^\pm(\mu, p^z)$ in \cref{eq:matching-pert}. and has been calculated at NLO in Refs.~\cite{Ebert:2019okf,Deng:2022gzi}.
\par Explicitly, the NLO contribution $C_\phi^{\pm\unexp(1)}(b_T,\mu,p^z)$ is given by
\begin{equation}
\begin{aligned}
	C&_\phi^{\pm\unexp(1)}(b_T,\mu,p^z) \\
	&= C_F\left\{ {1\over\sqrt{\pi}}\left[{1\over p^z b_T}G^{2,2}_{2,4}\left({(p^z b_T)^2\over 4}\Bigg| 
	\begin{array}{c}
	 {3\over 2},{3\over2}\\
	 {3\over 2},{3\over 2},0,{1\over 2}
	\end{array}
	 \right)\right.\right.\\
	 &\quad  \left.-{p^zb_T\over2} G^{2,3}_{3,5}\left({(p^z b_T)^2\over 4}\Bigg| 
	\begin{array}{c}
	 {1\over 2},{1\over2},{1\over2}\\
	 {1\over 2},{1\over 2},-{1\over2},-{1\over2},0
	\end{array}
	 \right)\right]\\
	&\quad - \left(-{1\over2}\ln^2{b_T^2\mu^2\over b_0^2} + \ln{b_T^2\mu^2\over b_0^2} \left(1- \ln{\zeta_z\over \mu^2}\right) - {\pi^2\over 12}\right) \\
	& \quad \left. \pm i\pi\left[ 2\mbox{Ei}(-p^zb_T) - {1-e^{-p^zb_T}\over p^zb_T}  - \ln{\zeta_z\over \mu^2} +1\right]  \right\},
\end{aligned}
\end{equation}
where $b_0=2e^{-\gamma_E}$, $\mbox{Ei}(z) \equiv - \int_{-z}^\infty dt\ {e^{-t}\over t}$ is the exponential integral function, and $G^{m,n}_{p,q}\big(z \big\vert \substack{a_1, \ldots, a_p \\ b_1, \ldots, b_q} \big)$ is the Meijer $G$-function. 
    \begin{figure}[t]
    \centering    
    \includegraphics[width=0.46\textwidth]{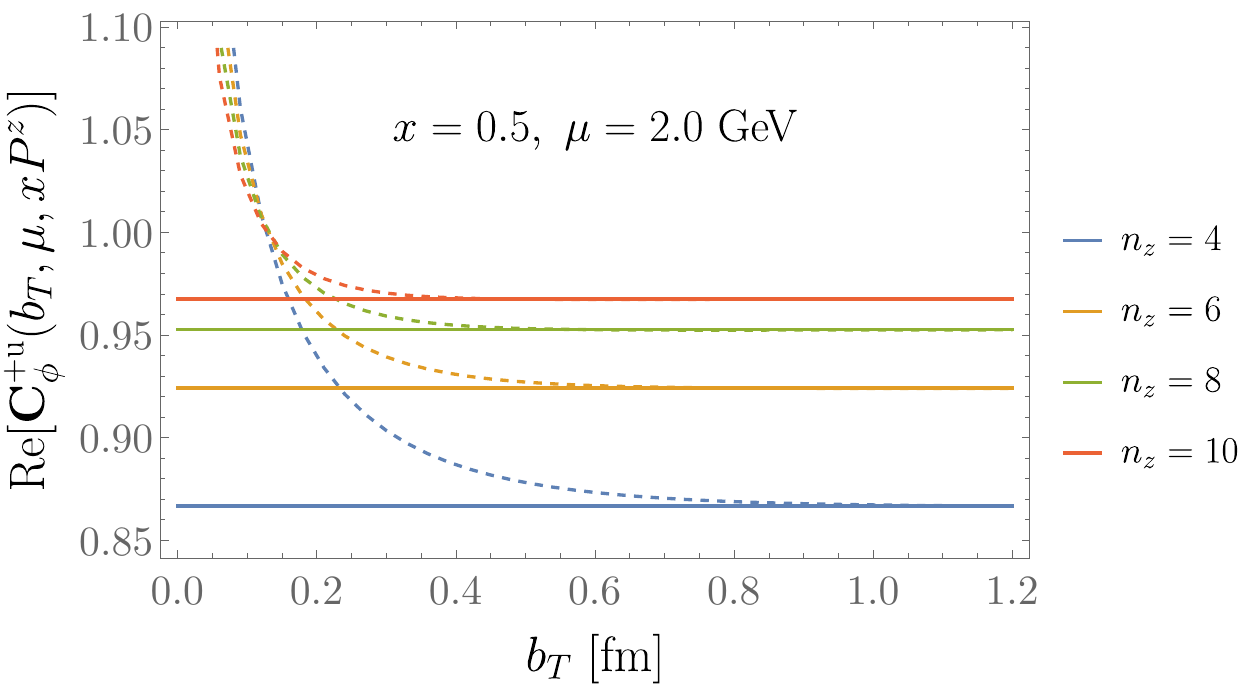}
    \includegraphics[width=0.46\textwidth]{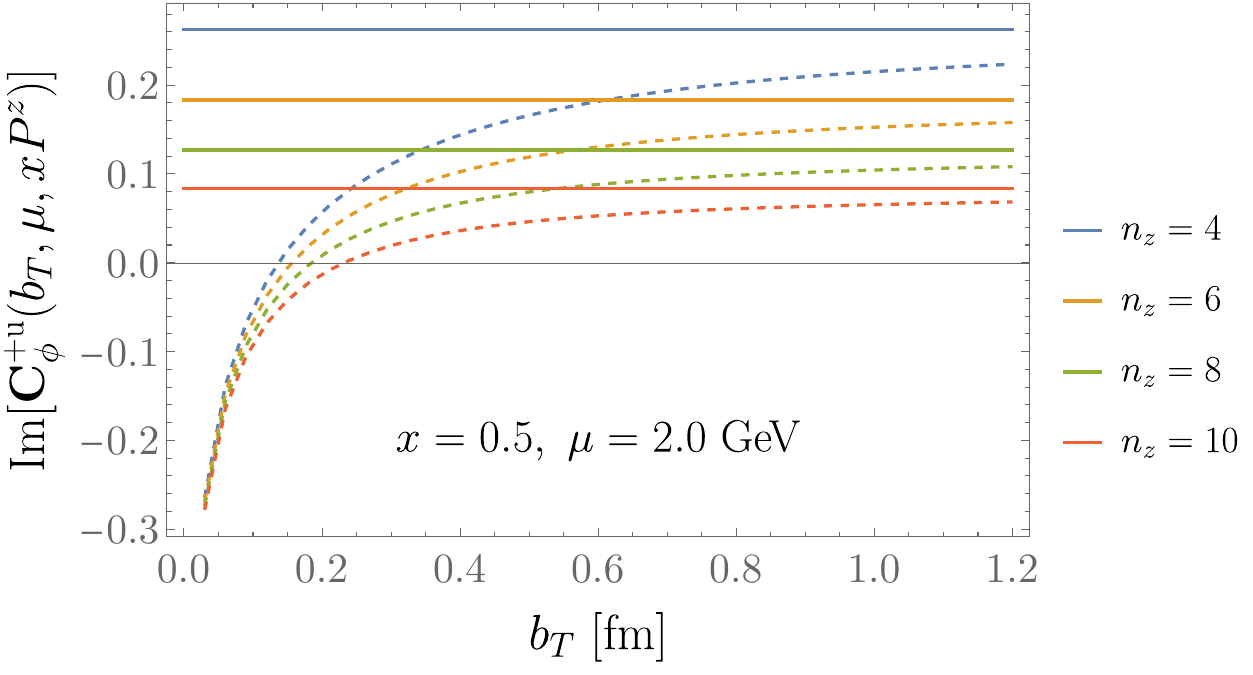}
         \caption{\label{fig:unexp} NLO matching coefficient with (solid) and without (dashed) expansion at large $P^zb_T$ and $x=0.5$ and $\mu=2$ GeV.
         }
    \end{figure}
The unexpanded coefficient $C_\phi^{\pm{\unexp}}(b_\tran,\mu, xP^z)$ and the corresponding perturbative correction to the CS kernel $\delta\gamma^{\rm uNLO}_q(b_\tran, x,  P_1^z, P_2^z, \mu)$
\begin{figure}[t]
    \centering    \includegraphics[width=0.46\textwidth]{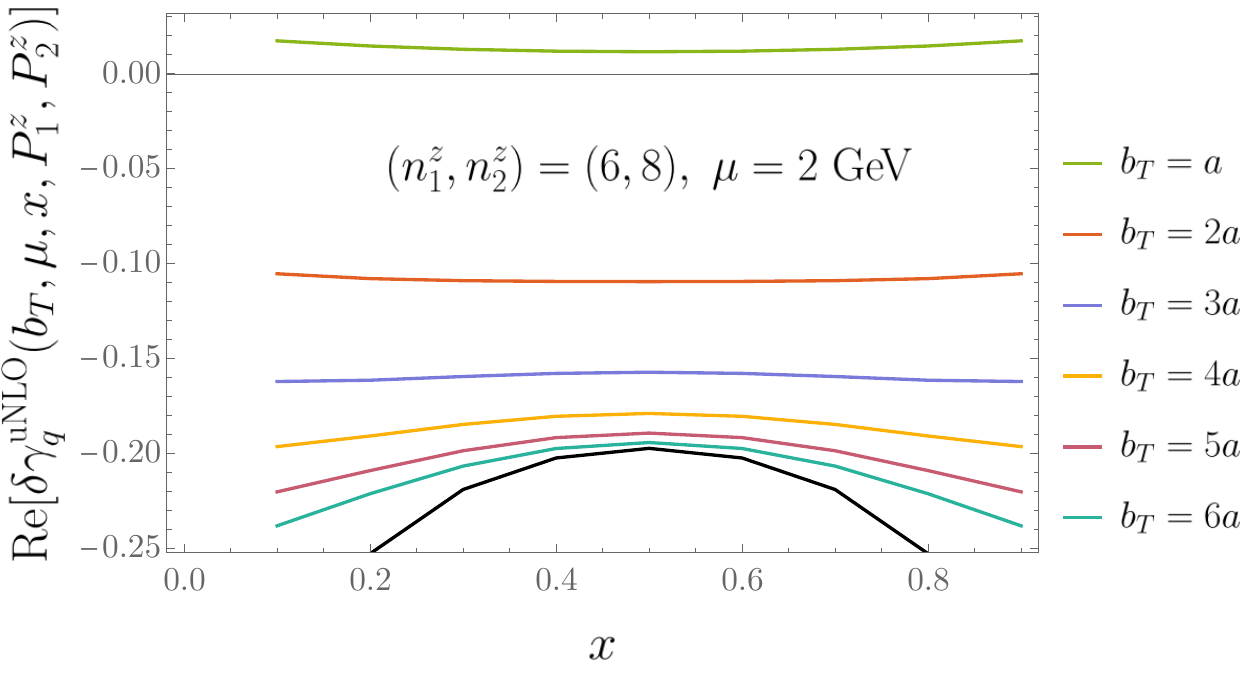}
    \includegraphics[width=0.46\textwidth]{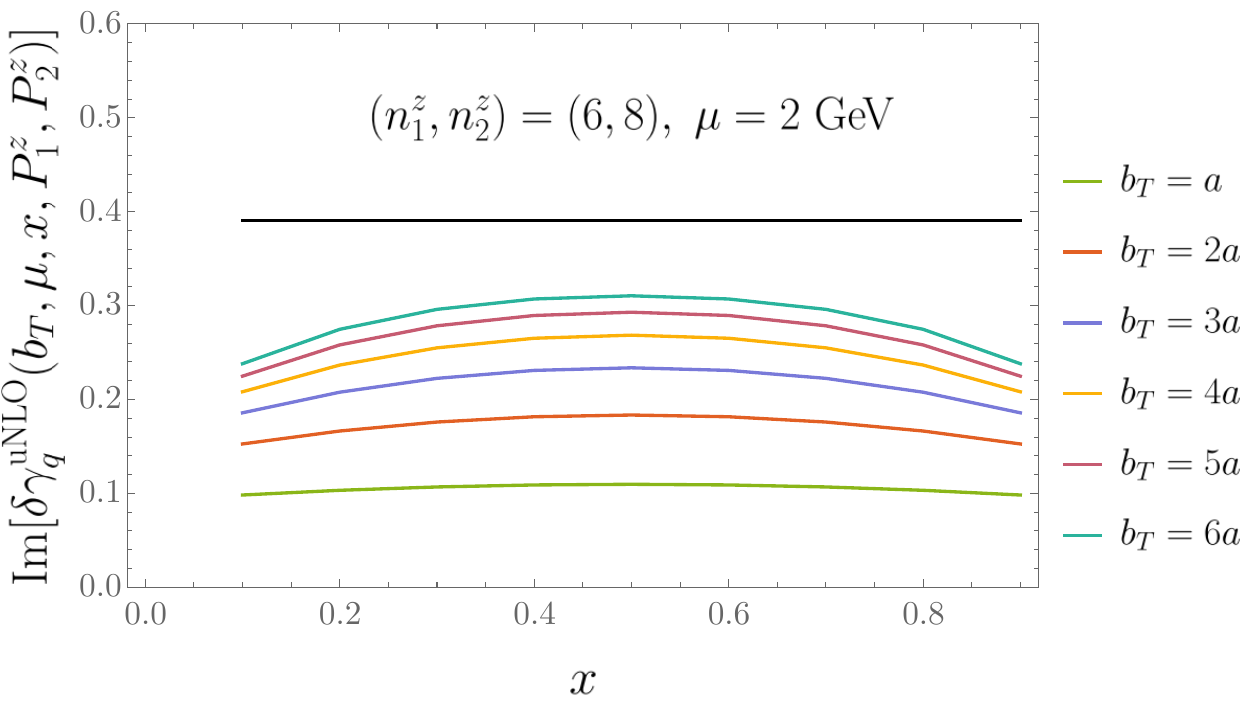}
    \caption{\label{fig:CSunexp} uNLO matching correction to the CS kernel without expansion at large $P^zb_T$ at momentum pair $(n_1^z, n_2^z)=(6,8)$ and $\mu=2$ GeV. The black line represents the NLO correction.}
    \end{figure}
are shown as a function of $x$ in \cref{fig:unexp,fig:CSunexp}, respectively.
The estimated corrections are consistent with the different rates of convergence observed in real and imaginary parts for fixed-order and resummed corrections in \cref{fig:nlovsnnlo,fig:nllvsnnll}, respectively.
In the real part, the corrections become negligible for $b_\tran \gtrsim \SI{0.4}{\femto\meter}$, except for the pair of smallest momenta used in this work.
In the imaginary part, the corrections are large for the entire kinematic range of this study. 

\section{Additional examples for \texorpdfstring{\cref{sec:numerical-investigation}}{Section III}}
\label{sec:app:plots}
\par This section collates examples of intermediate analysis steps in the numerical calculation of the CS kernel, supplementing \cref{sec:numerical-investigation}.
\par Supplementing \cref{fig:analysis-1}, additional examples of the $\MSbar$-renormalized quasi-TMD WFs $W^{\MSbar}_\Gamma(b_\tran, \mu, x, P^{z}, \ell)$ are illustrated in \cref{fig:wf_ms_pz4_gamma7_a,fig:wf_ms_pz4_gamma7_b,fig:wf_ms_pz4_gamma11_a,fig:wf_ms_pz4_gamma11_b,fig:wf_ms_pz6_gamma7_a,fig:wf_ms_pz6_gamma7_b,fig:wf_ms_pz6_gamma11_a,fig:wf_ms_pz6_gamma11_b,fig:wf_ms_pz8_gamma7_a,fig:wf_ms_pz8_gamma7_b,fig:wf_ms_pz8_gamma11_a,fig:wf_ms_pz8_gamma11_b,fig:wf_ms_pz10_gamma7_a,fig:wf_ms_pz10_gamma7_b,fig:wf_ms_pz10_gamma11_a,fig:wf_ms_pz10_gamma11_b}.
\par
\begin{figure*}[t]
    \centering
        \includegraphics[width=0.46\textwidth]{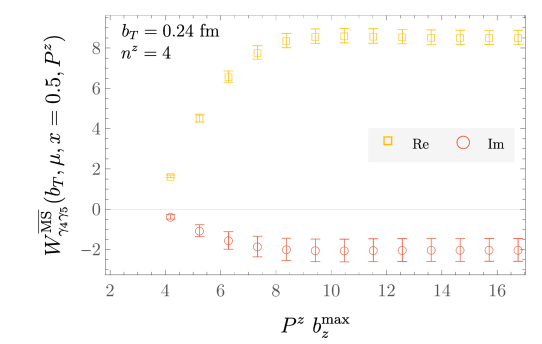}   
        \hspace{20pt}
        \includegraphics[width=0.46\textwidth]{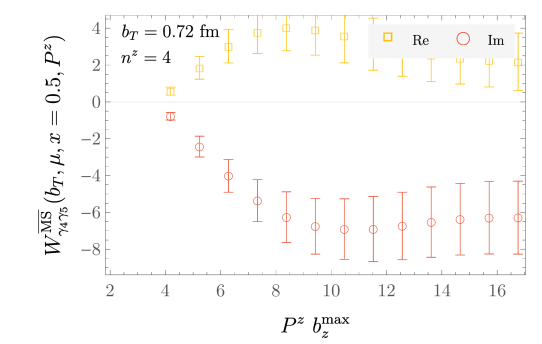}
        \includegraphics[width=0.46\textwidth]{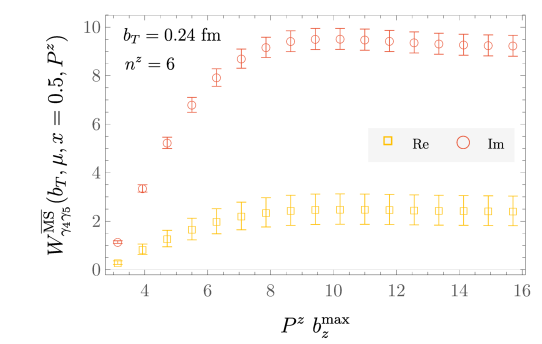}   
        \hspace{20pt}
        \includegraphics[width=0.46\textwidth]{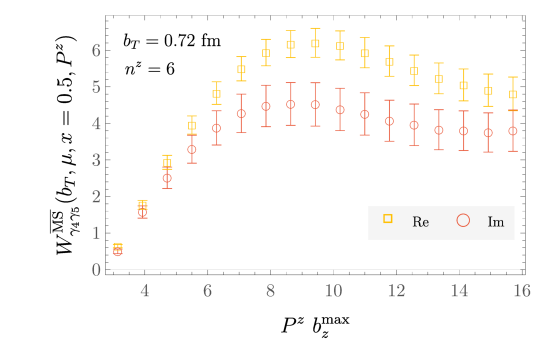}
        \includegraphics[width=0.46\textwidth]{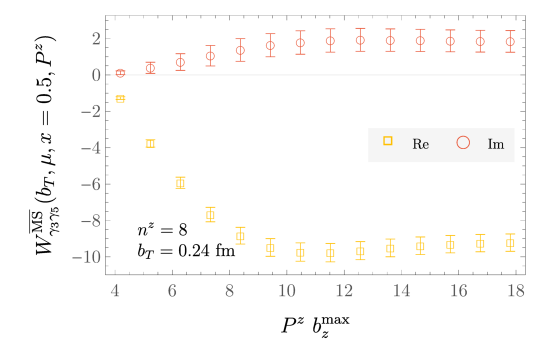}   
        \hspace{20pt}
        \includegraphics[width=0.46\textwidth]{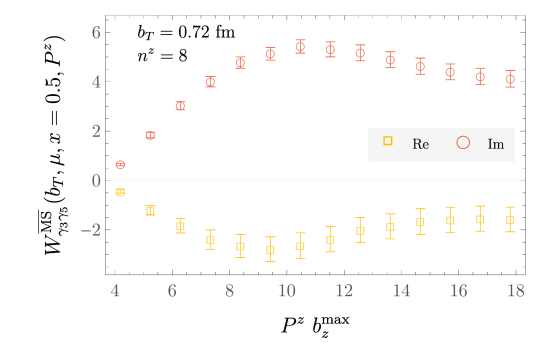}
        \includegraphics[width=0.46\textwidth]{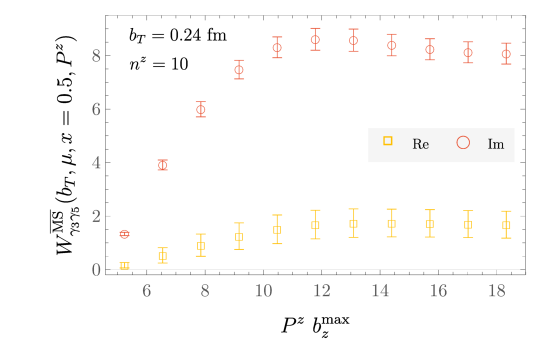} 
        \hspace{20pt}
        \includegraphics[width=0.46\textwidth]{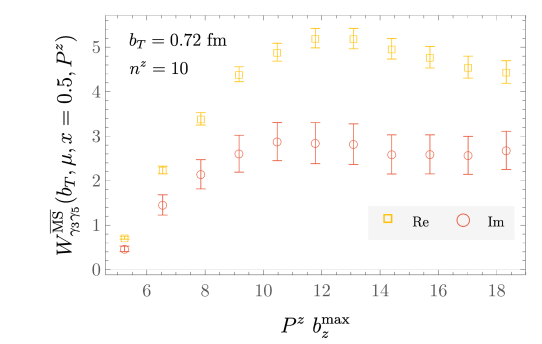}  
        \caption{As in \cref{fig:ft-truncation-robust}, for different combinations of $\Gamma$, $b_\tran$, and $P^z = \frac{2\pi}{L} n^z$.
        \label{fig:wf_dft_trunc}
        }
\end{figure*}

\begin{figure*}[t]
    \centering
        \includegraphics[width=0.46\textwidth]{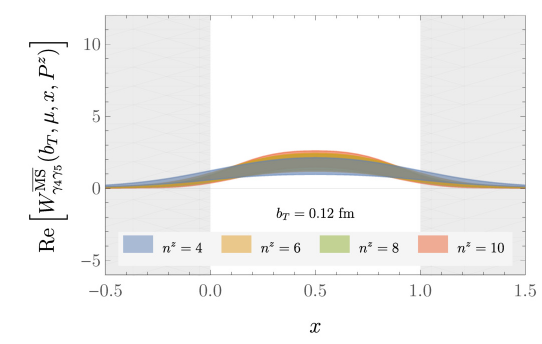}   
        \hspace{20pt}
        \includegraphics[width=0.46\textwidth]{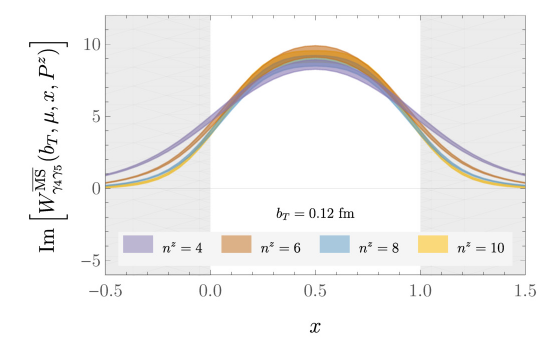}   
        \includegraphics[width=0.46\textwidth]{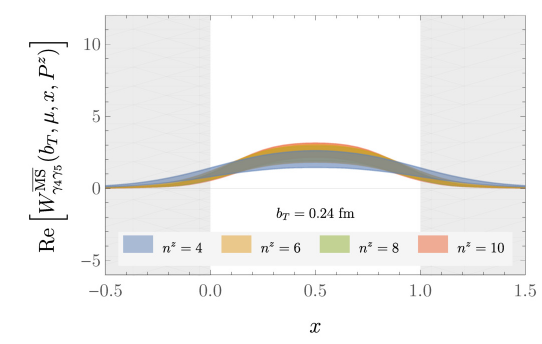}   
        \hspace{20pt}
         \includegraphics[width=0.46\textwidth]{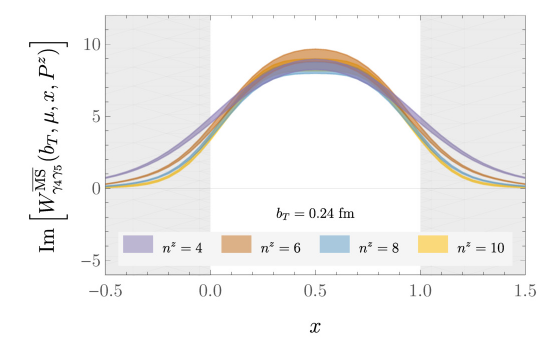}   
        \includegraphics[width=0.46\textwidth]{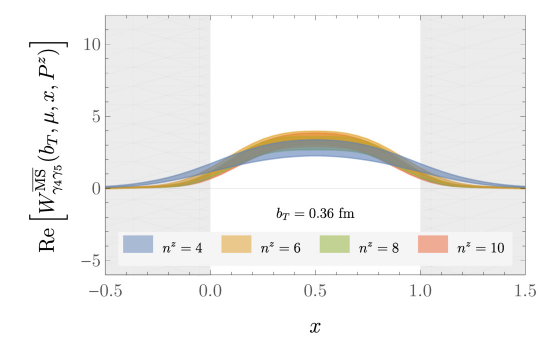}   
        \hspace{20pt}
        \includegraphics[width=0.46\textwidth]{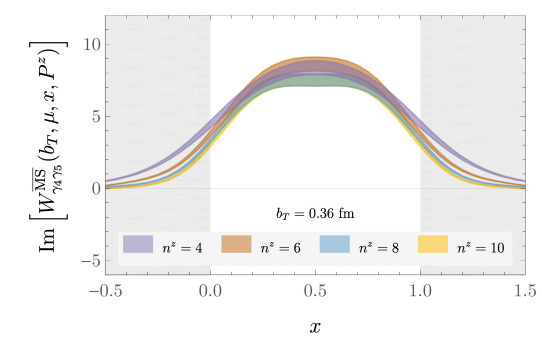}   
        \includegraphics[width=0.46\textwidth]{WF_DFT_MS_gamma7_bT4_vs_x_re}   
        \hspace{20pt}
        \includegraphics[width=0.46\textwidth]{WF_DFT_MS_gamma7_bT4_vs_x_im}   
        \caption{Examples of real and imaginary parts of the Fourier-transformed $\MSbar$-renormalized quasi-TMD WF ratios $W^{\MSbar}_{\Gamma}(b_\tran, \mu, x, P^z)$ defined in \cref{eq:wf-ratio-fourier}, for $\Gamma = \gamma_4 \gamma_5$ and $\SI{0.12}{\femto\meter} \leq b_\tran \leq \SI{0.36}{\femto\meter}$, where 
        $n^z$ labels the momentum $P^z = \frac{2\pi}{L} n^z$
        ($W^{\MSbar}_{\gamma_4\gamma_5}$ for $b_\tran = \SI{0.48}{\femto\meter}$ is illustrated in \cref{fig:analysis-2} in the main text).
        \label{fig:wf_ms_x_gamma7_a}}
\end{figure*}

\begin{figure*}[t]
    \centering
        \includegraphics[width=0.46\textwidth]{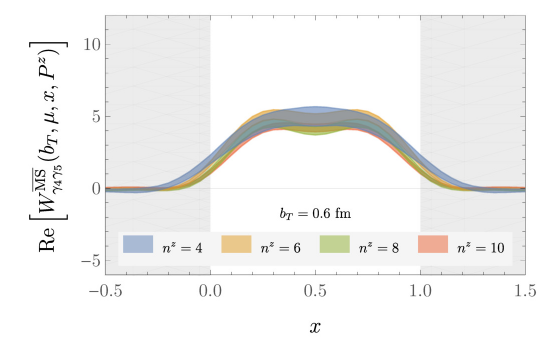}   
        \hspace{20pt}
        \includegraphics[width=0.46\textwidth]{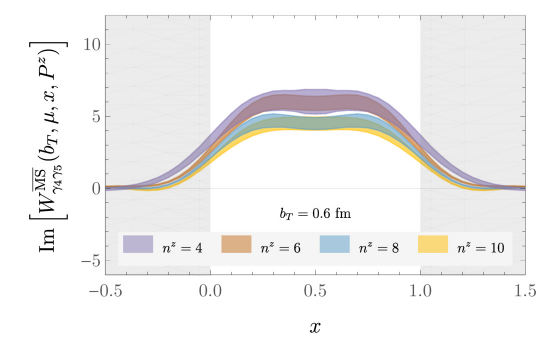}   
        \includegraphics[width=0.46\textwidth]{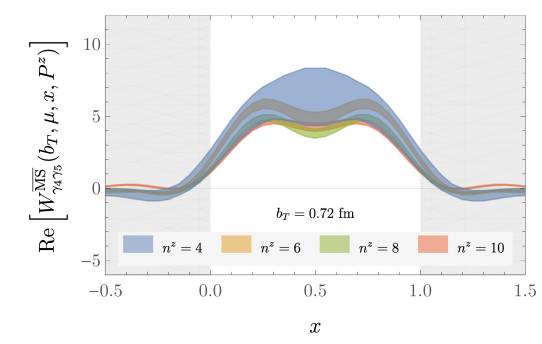}   
        \hspace{20pt}
        \includegraphics[width=0.46\textwidth]{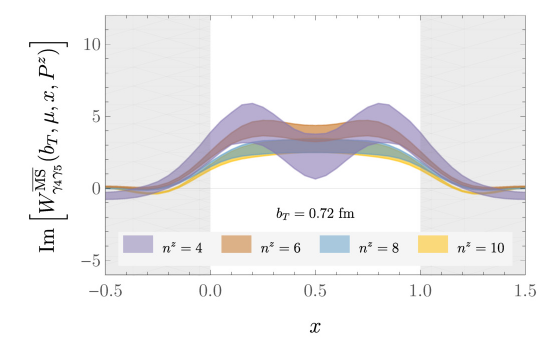}   
        \includegraphics[width=0.46\textwidth]{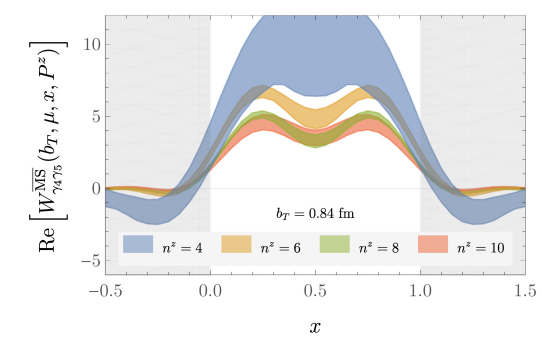}   
        \hspace{20pt}
        \includegraphics[width=0.46\textwidth]{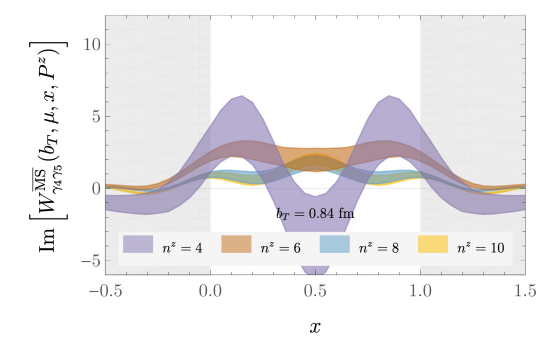}   
        \caption{As in \cref{fig:wf_ms_x_gamma7_a}, for $\SI{0.60}{\femto\meter} \leq b_\tran \leq \SI{0.84}{\femto\meter}$.
        \label{fig:wf_ms_x_gamma7_b}
        }
\end{figure*}

\begin{figure*}[t]
    \centering
        \includegraphics[width=0.46\textwidth]{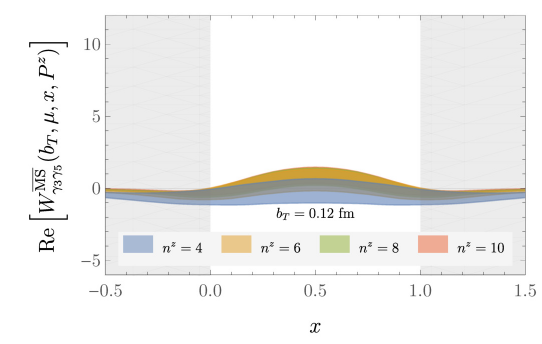}   
        \hspace{20pt}
        \includegraphics[width=0.46\textwidth]{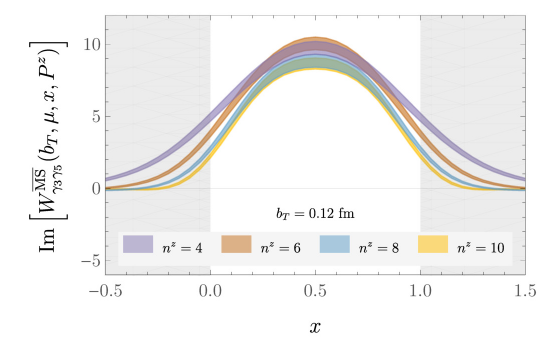}   
        \includegraphics[width=0.46\textwidth]{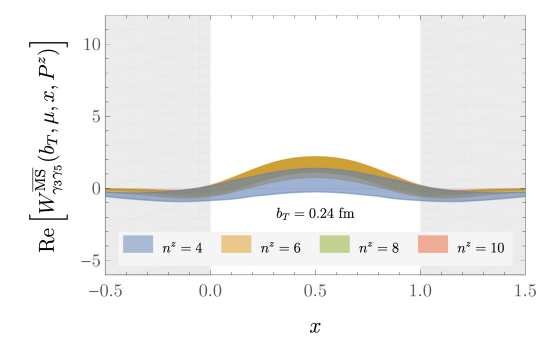}   
        \hspace{20pt}
        \includegraphics[width=0.46\textwidth]{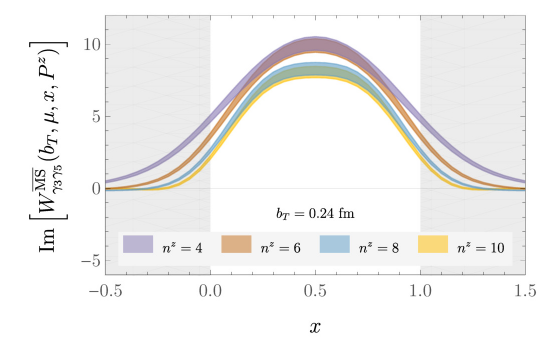}   
        \includegraphics[width=0.46\textwidth]{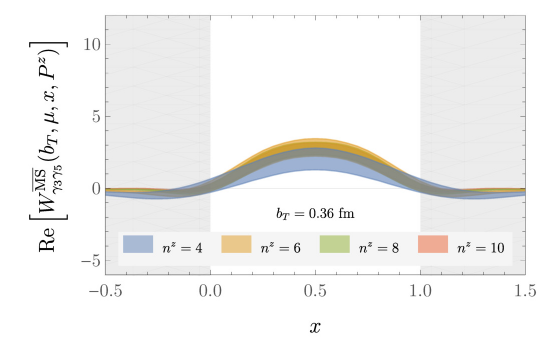}   
        \hspace{20pt}
        \includegraphics[width=0.46\textwidth]{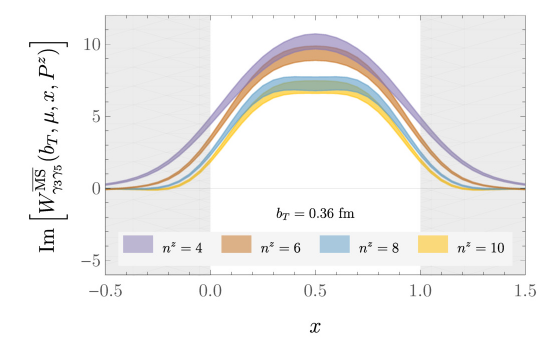}   
        \includegraphics[width=0.46\textwidth]{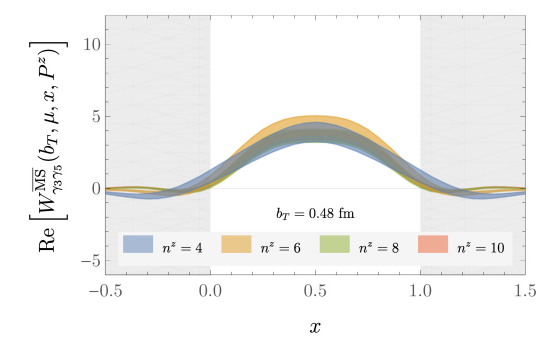}   
        \hspace{20pt}
        \includegraphics[width=0.46\textwidth]{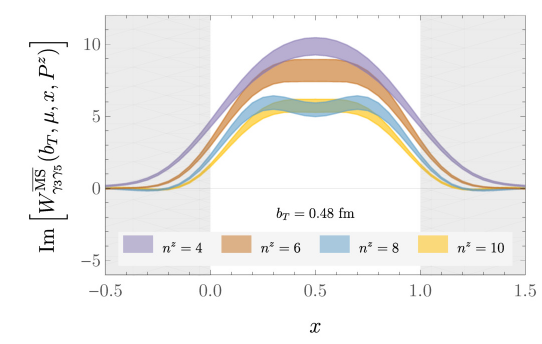}   
        \caption{As in \cref{fig:wf_ms_x_gamma7_a}, for $\Gamma = \gamma_3\gamma_5$ and $\SI{0.12}{\femto\meter} \leq b_\tran \leq \SI{0.48}{\femto\meter}$.
        \label{fig:wf_ms_x_gamma11_a} 
        }
\end{figure*}

\begin{figure*}[t]
    \centering
        \includegraphics[width=0.46\textwidth]{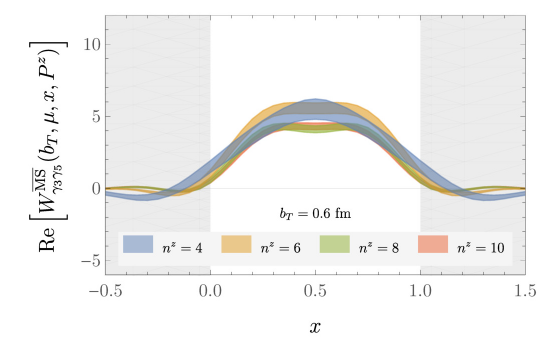}   
        \hspace{20pt}
        \includegraphics[width=0.46\textwidth]{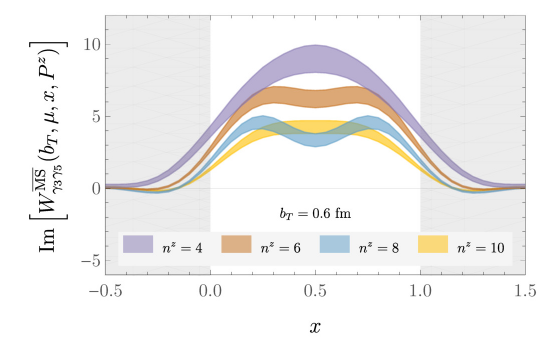}   
        \includegraphics[width=0.46\textwidth]{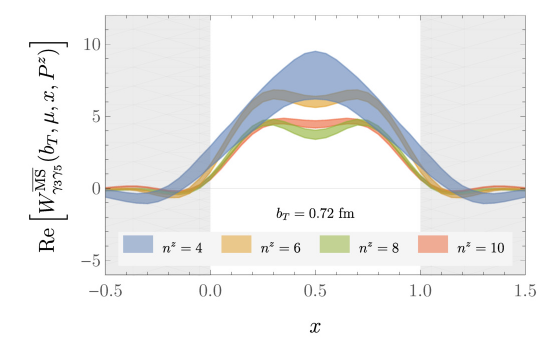}   
        \hspace{20pt}
        \includegraphics[width=0.46\textwidth]{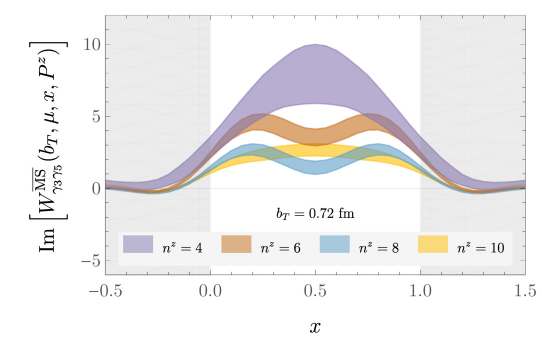}   
        \includegraphics[width=0.46\textwidth]{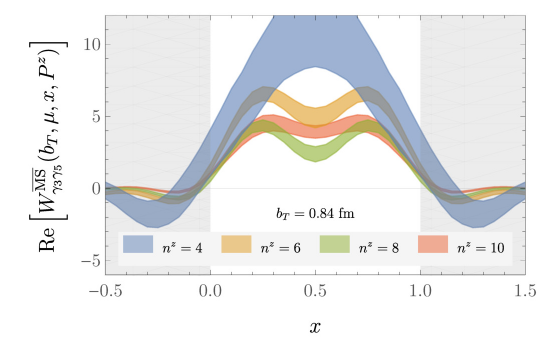}   
        \hspace{20pt}
        \includegraphics[width=0.46\textwidth]{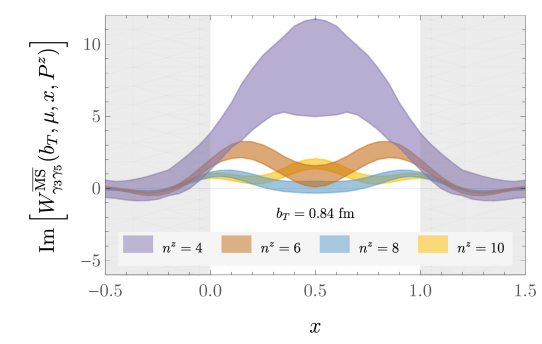}   
        \caption{As in \cref{fig:wf_ms_x_gamma7_a}, for $\Gamma = \gamma_3\gamma_5$. 
        \label{fig:wf_ms_x_gamma11_b}}
\end{figure*}

 Supplementing \cref{fig:ft-truncation-robust,fig:analysis-2}, additional examples of the Fourier-transformed $\MSbar$-renormalized quasi-TMD WF ratios $W^{\MSbar}_\Gamma(b_\tran, \mu, x, P^{z})$ are provided in \cref{fig:wf_dft_trunc} and \cref{fig:wf_ms_x_gamma7_a,fig:wf_ms_x_gamma7_b,fig:wf_ms_x_gamma11_a,fig:wf_ms_x_gamma11_b}, respectively.
\par
\begin{figure*}[t]
    \centering
        \includegraphics[width=0.46\textwidth]{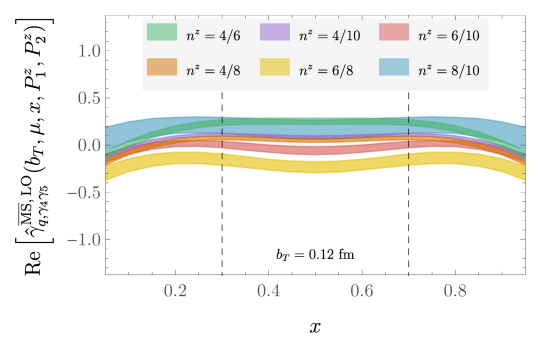}   
        \includegraphics[width=0.46\textwidth]{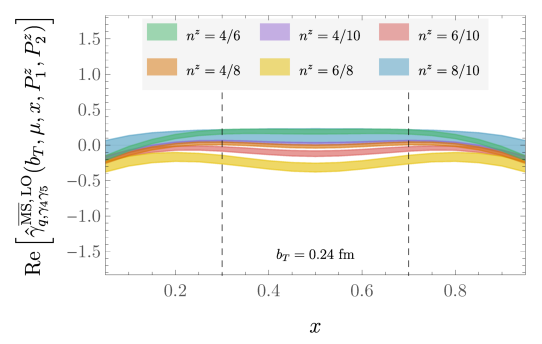}   
        \includegraphics[width=0.46\textwidth]{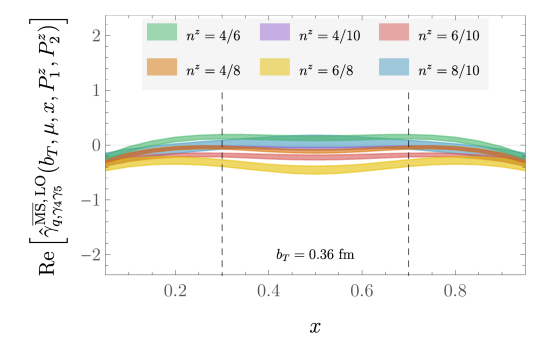}   
        \includegraphics[width=0.46\textwidth]{CS_MS_LO_gamma7_bT4_vs_x_re}   
        \includegraphics[width=0.46\textwidth]{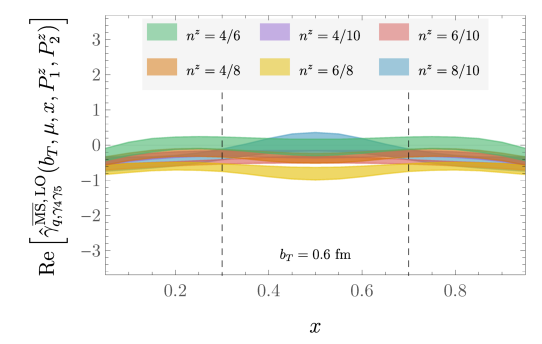}   
        \includegraphics[width=0.46\textwidth]{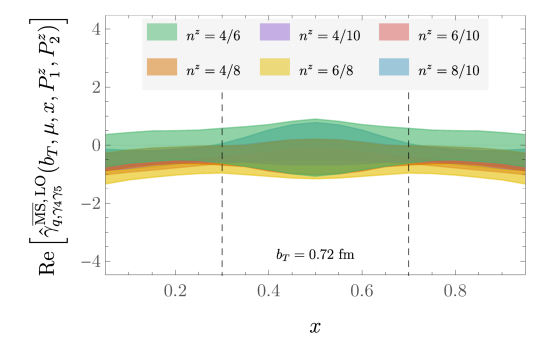}   
        \hspace{20pt}
        \includegraphics[width=0.46\textwidth]{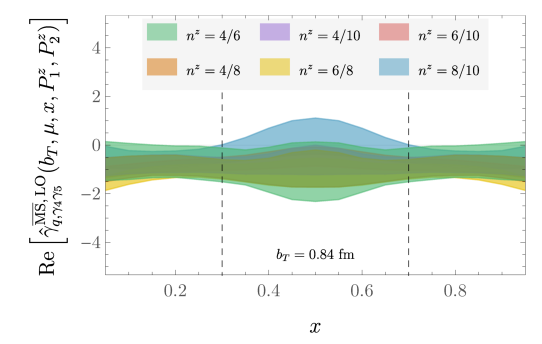}   
        \hspace{20pt}
        \caption{Examples of real parts of CS kernel estimators $\hat{\gamma}_{\Gamma}^{\MSbar}(b_{\tran}, x, P_1^z, P_2^z, \mu)$, computed as described in Section~\ref{sec:numerical-investigation-cs} with matching corrections at LO, for $\Gamma = \gamma_4\gamma_5$ and $\SI{0.12}{\femto\meter} \leq b_\tran \leq \SI{0.84}{\femto\meter}$ ($\mathrm{Re}[\hat{\gamma}_{\gamma_4\gamma_5}^{\MSbar}]$ for $b_\tran = \SI{0.48}{\femto\meter}$ is illustrated in \cref{fig:analysis-3} in the main text). 
        The black dashed lines enclose the region in $x$ used to determine the CS kernel.
        The notation $n^z = P_1^z/P_2^z$ displays momenta in lattice units.
        \label{fig:wf_cs_x_b1}}
\end{figure*}

\begin{figure*}[t]
    \centering
        \includegraphics[width=0.46\textwidth]{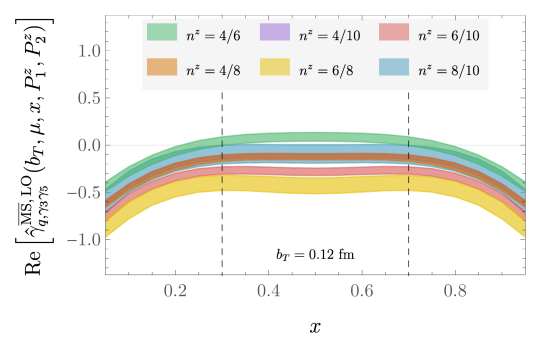}   
        \hspace{20pt}
        \includegraphics[width=0.46\textwidth]{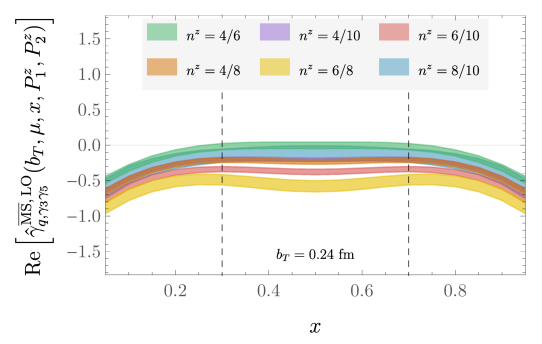}   
        \includegraphics[width=0.46\textwidth]{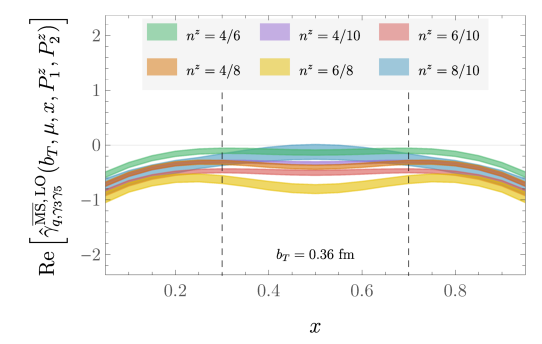}   
        \hspace{20pt}
        \includegraphics[width=0.46\textwidth]{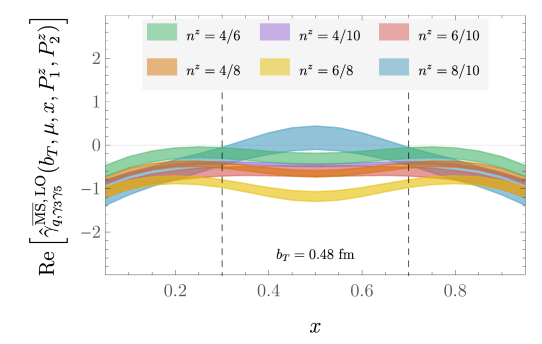}   
        \includegraphics[width=0.46\textwidth]{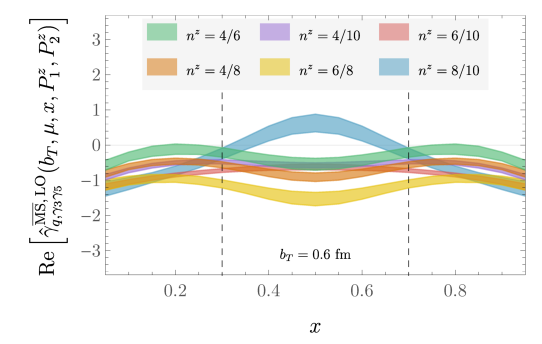}   
        \hspace{20pt}
        \includegraphics[width=0.46\textwidth]{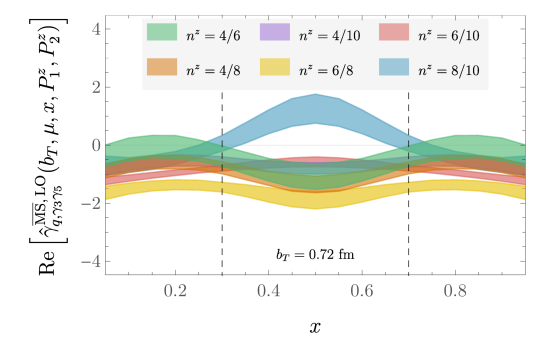}   
        \includegraphics[width=0.46\textwidth]{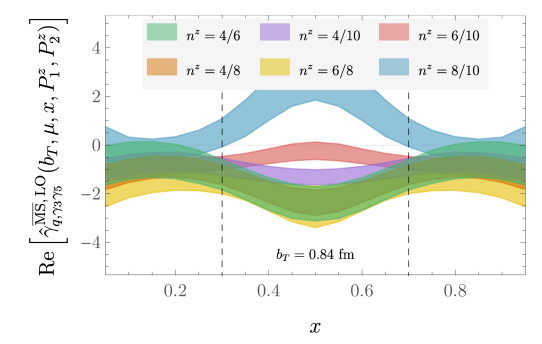}   
        \caption{As in \cref{fig:wf_cs_x_b1}, for $\Gamma = \gamma_3\gamma_5$.
        \label{fig:wf_cs_x_b2}}
\end{figure*}

\begin{figure*}[t]
    \centering
        \includegraphics[width=0.46\textwidth]{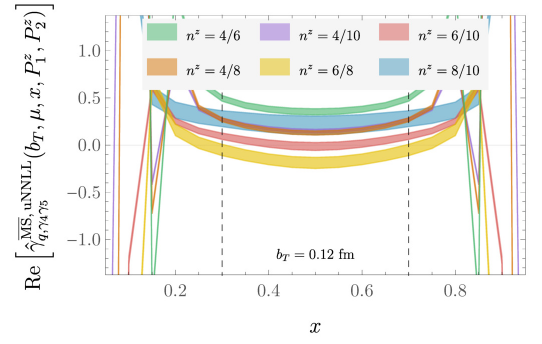}   
        \hspace{20pt}
        \includegraphics[width=0.46\textwidth]{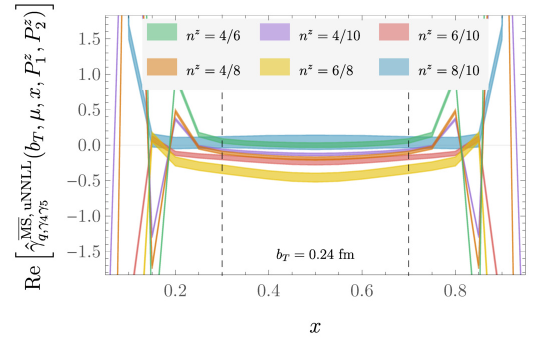}   
        \includegraphics[width=0.46\textwidth]{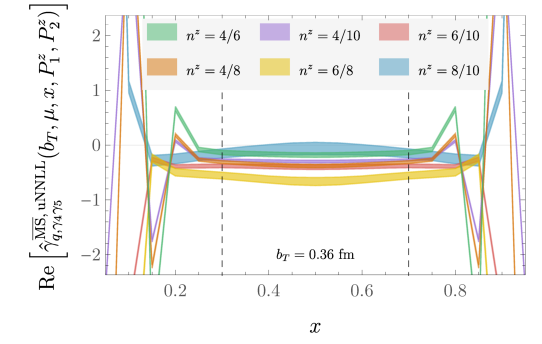}   
        \hspace{20pt}
        \includegraphics[width=0.46\textwidth]{CS_MS_bestNNLO_gamma7_bT4_vs_x_re}   
        \includegraphics[width=0.46\textwidth]{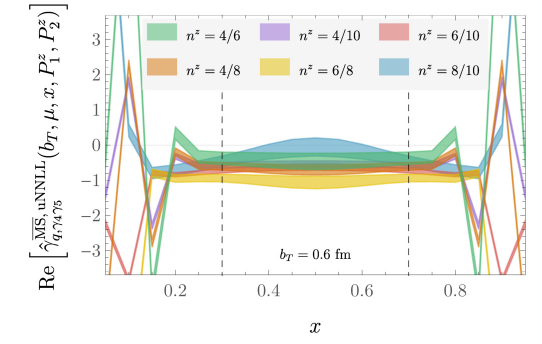}   
        %\hspace{20pt}
        \includegraphics[width=0.46\textwidth]{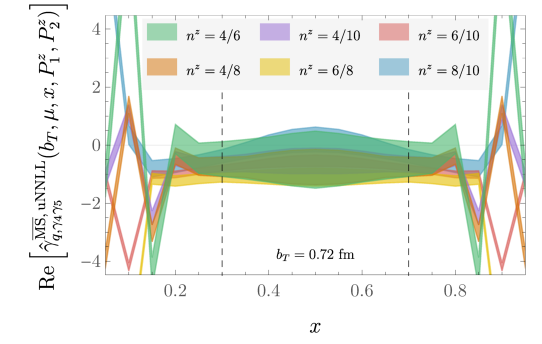}   
        \includegraphics[width=0.46\textwidth]{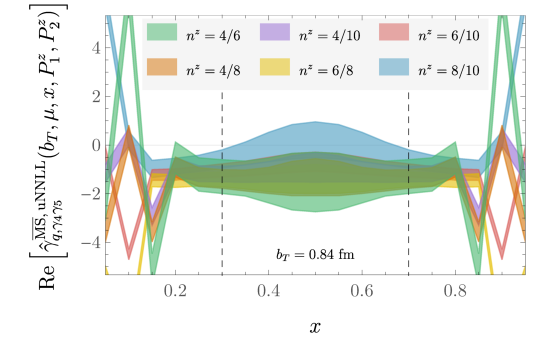}   
        \hspace{20pt}
        \caption{As in \cref{fig:wf_cs_x_b1}, with uNNLL matching ($\mathrm{Re}[\hat{\gamma}_{\gamma_4\gamma_5}^{\MSbar}]$ for $b_\tran = \SI{0.48}{\femto\meter}$ is illustrated in \cref{fig:analysis-3} in the main text).
        \label{fig:wf_cs_x_a1}}
\end{figure*}

\begin{figure*}[t]
    \centering
        \includegraphics[width=0.46\textwidth]{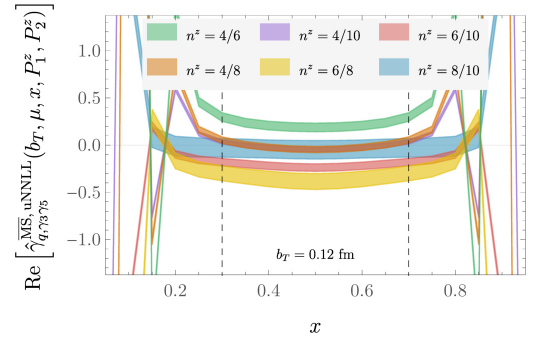}   
        \hspace{20pt}
        \includegraphics[width=0.46\textwidth]{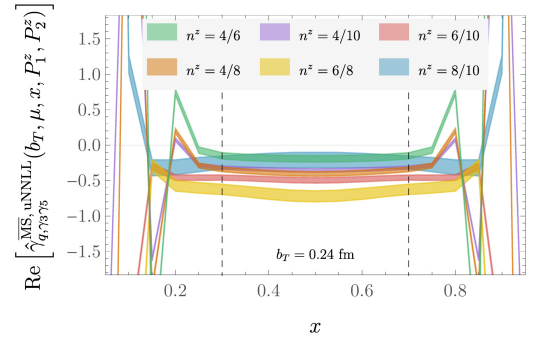}   
        \includegraphics[width=0.46\textwidth]{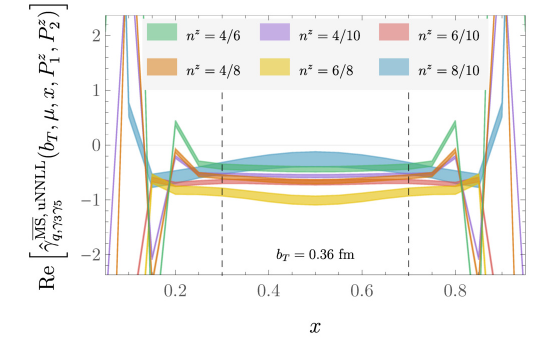}   
        \hspace{20pt}
        \includegraphics[width=0.46\textwidth]{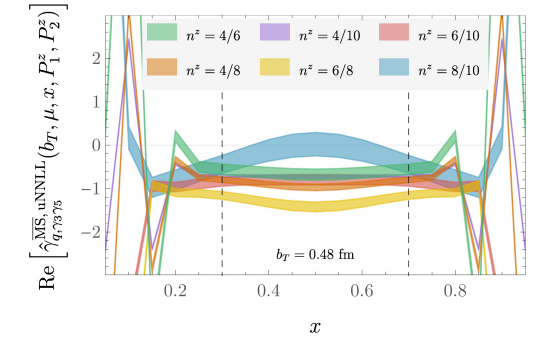}   
        \includegraphics[width=0.46\textwidth]{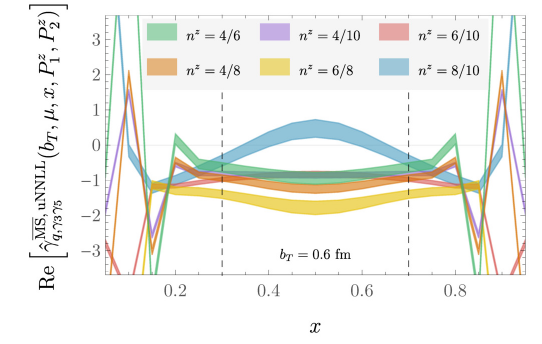}   
        \hspace{20pt}
        \includegraphics[width=0.46\textwidth]{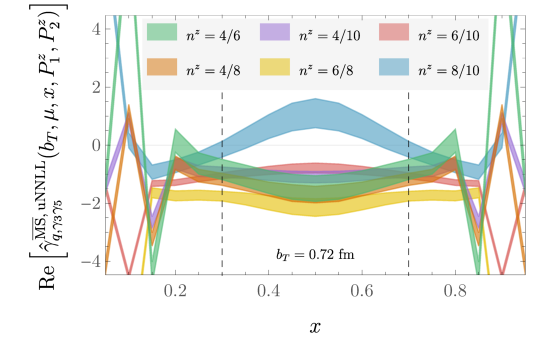}   
        \includegraphics[width=0.46\textwidth]{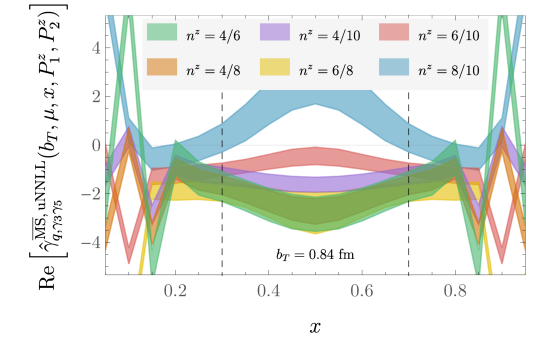}   
        \caption{As in \cref{fig:wf_cs_x_b1}, with uNNLL matching, for $\Gamma = \gamma_3\gamma_5$.
        \label{fig:wf_cs_x_a2}}
\end{figure*}

 Supplementing \cref{fig:analysis-3}, additional examples of real parts of CS kernel estimators $\mathrm{Re}\big\lbrack\hat{\gamma}_{\Gamma}^{\MSbar}(b_{\tran}, x, P_1^z, P_2^z, \mu)\big\rbrack$ are provided in \cref{fig:wf_cs_x_a1,fig:wf_cs_x_a2} with LO matching, and in \cref{fig:wf_cs_x_b1,fig:wf_cs_x_b2} with uNNLL matching.
\begin{figure*}[t]
    \centering
        \includegraphics[width=0.46\textwidth]{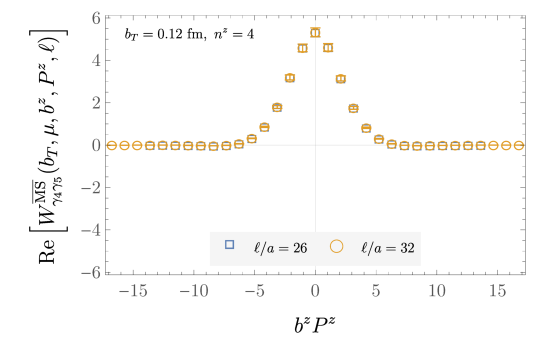}   
        \hspace{20pt}
        \includegraphics[width=0.46\textwidth]{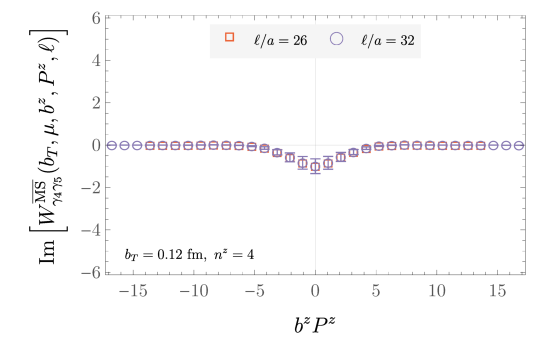}
        \includegraphics[width=0.46\textwidth]{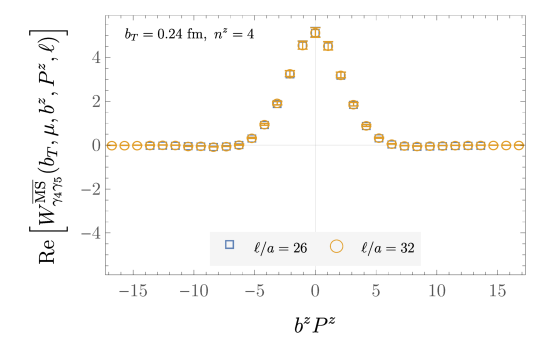} 
        \hspace{20pt}
        \includegraphics[width=0.46\textwidth]{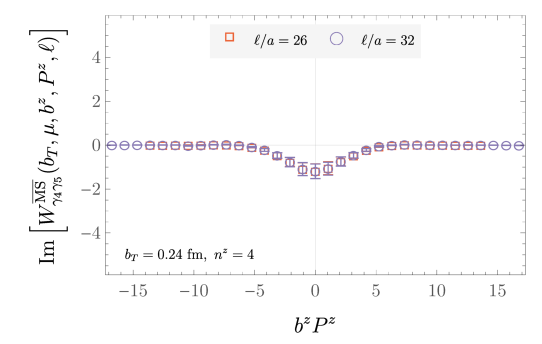} 
        \includegraphics[width=0.46\textwidth]{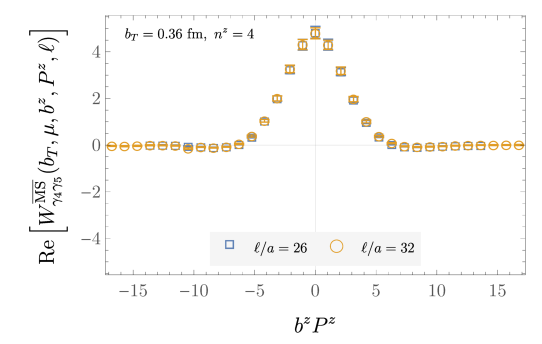} 
        \hspace{20pt}
         \includegraphics[width=0.46\textwidth]{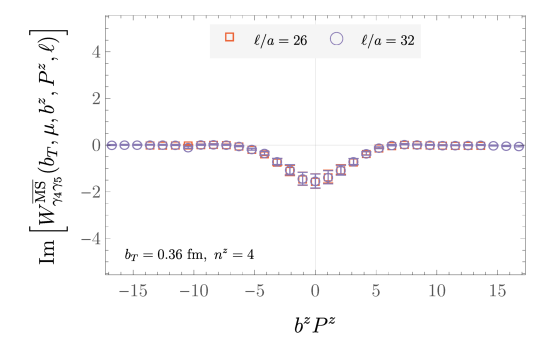} 
        \includegraphics[width=0.46\textwidth]{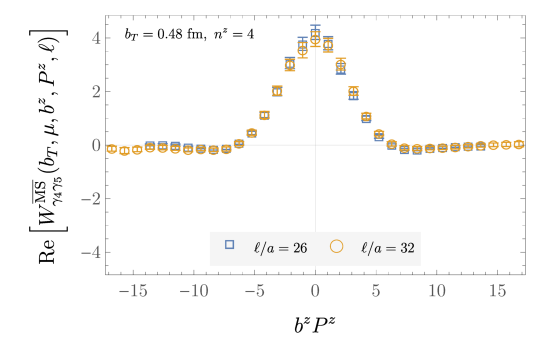} 
        \hspace{20pt}
        \includegraphics[width=0.46\textwidth]{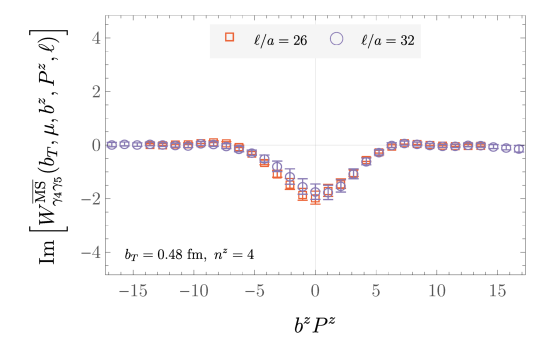} 
        \caption{Examples of real and imaginary parts the $\MSbar$-renormalized quasi-TMD WF ratios $W^{\MSbar}_{\Gamma}(b_\tran, \mu, b^z, P^z, \ell)$ defined in \cref{eq:quasi-wf-pos-ren}, for $\Gamma = \gamma_4 \gamma_5$, $P^z = \frac{2\pi}{L} n^z = \SI{0.86}{GeV}$ and $\SI{0.12}{\femto\meter} \leq b_\tran \leq \SI{0.48}{\femto\meter}$.
        Both $\ell/a$ at $P^z = \SI{0.86}{GeV}$ are chosen to be even to compare the data at matching points in $b^z P^z$ space.
        \label{fig:wf_ms_pz4_gamma7_a}
        }
\end{figure*}
\begin{figure*}[t]
    \centering
        \includegraphics[width=0.46\textwidth]{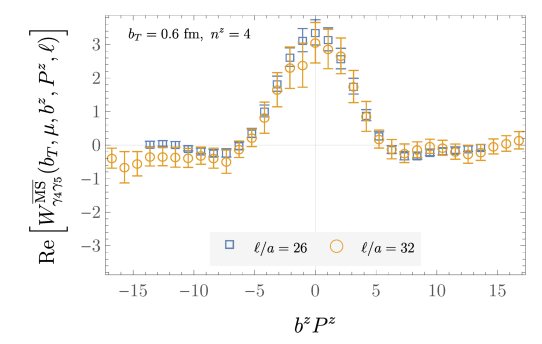} 
        \hspace{20pt}
        \includegraphics[width=0.46\textwidth]{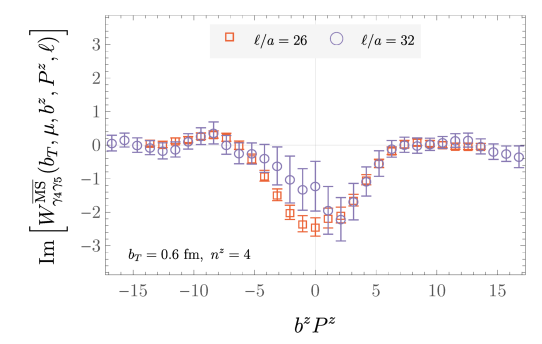} 
        \includegraphics[width=0.46\textwidth]{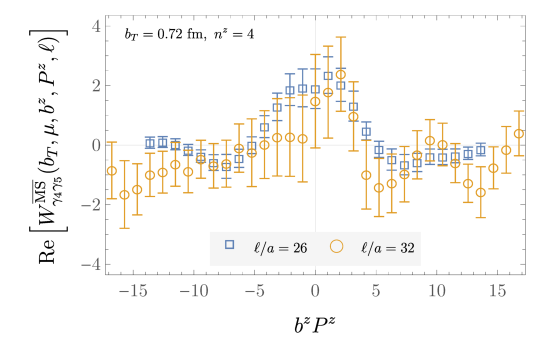} 
        \hspace{20pt}
        \includegraphics[width=0.46\textwidth]{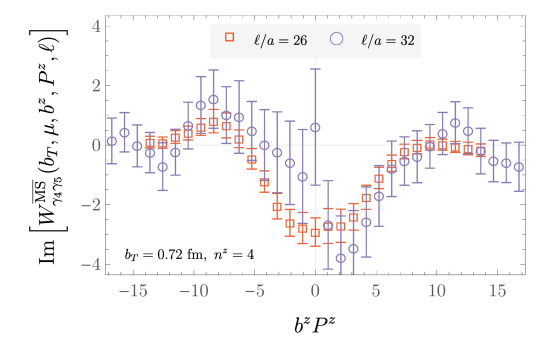} 
        \includegraphics[width=0.46\textwidth]{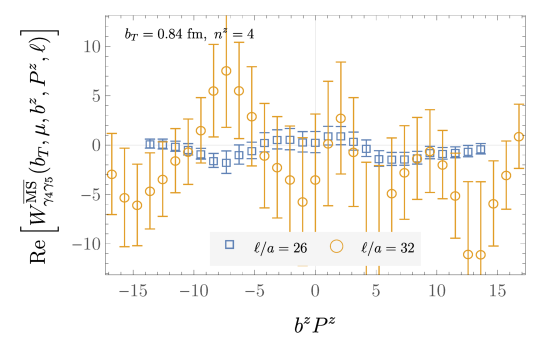}  
        \hspace{20pt}
        \includegraphics[width=0.46\textwidth]{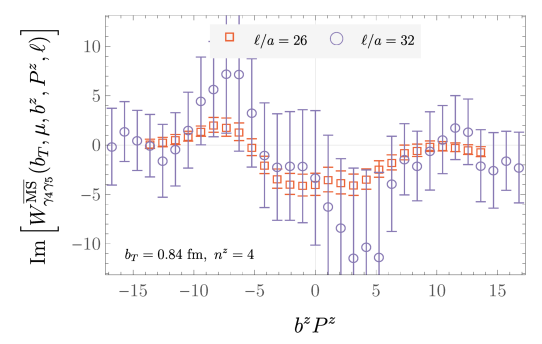}  
        %\includegraphics[width=0.46\textwidth]{WF_MS_pz4_gamma7_bT8_vs_bz_re} 
        %\hspace{20pt}
        %\includegraphics[width=0.46\textwidth]{WF_MS_pz4_gamma7_bT8_vs_bz_im} 
    \caption{As in \cref{fig:wf_ms_pz4_gamma7_a}, for $\SI{0.60}{\femto\meter} \leq b_\tran \leq \SI{0.84}{\femto\meter}$.
    \label{fig:wf_ms_pz4_gamma7_b}
    }
\end{figure*}

\begin{figure*}[t]
    \centering
        \includegraphics[width=0.46\textwidth]{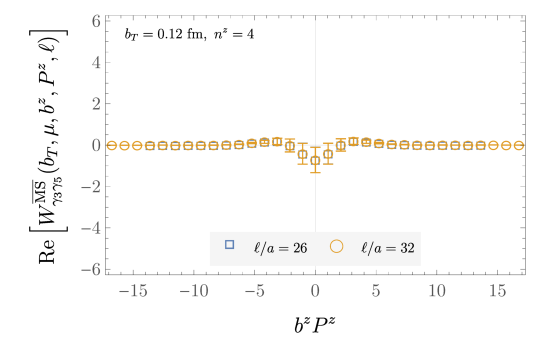}   
        \hspace{20pt}
        \includegraphics[width=0.46\textwidth]{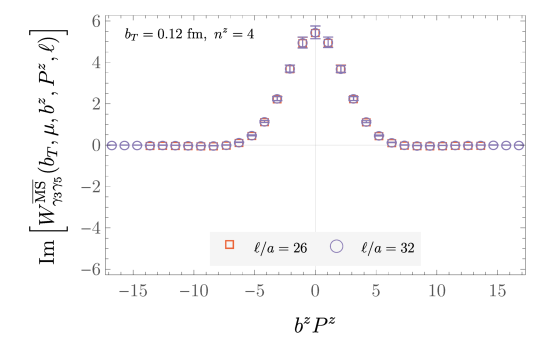}
        \includegraphics[width=0.46\textwidth]{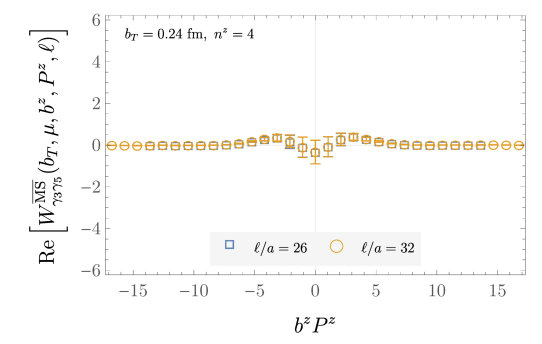} 
        \hspace{20pt}
        \includegraphics[width=0.46\textwidth]{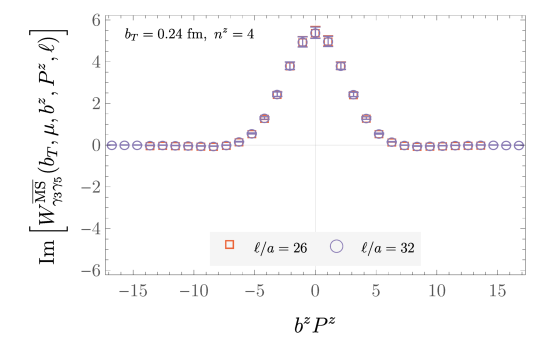} 
        \includegraphics[width=0.46\textwidth]{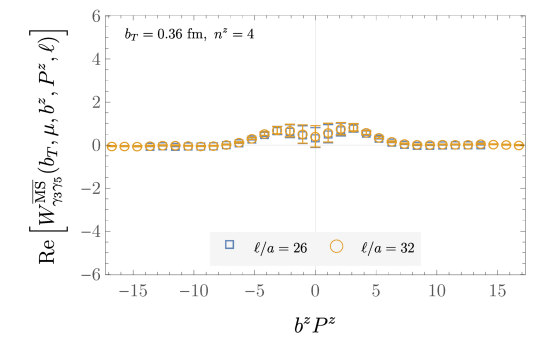} 
        \hspace{20pt}
         \includegraphics[width=0.46\textwidth]{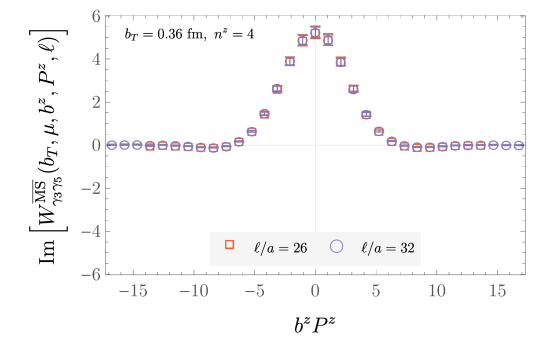} 
        \includegraphics[width=0.46\textwidth]{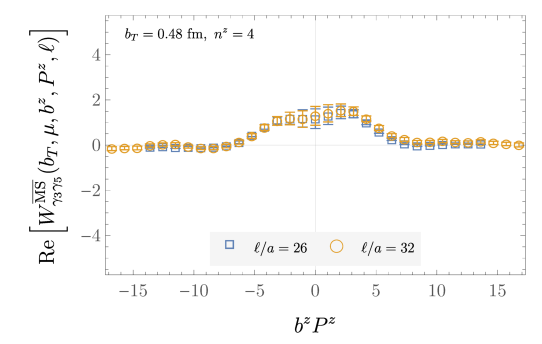} 
        \hspace{20pt}
        \includegraphics[width=0.46\textwidth]{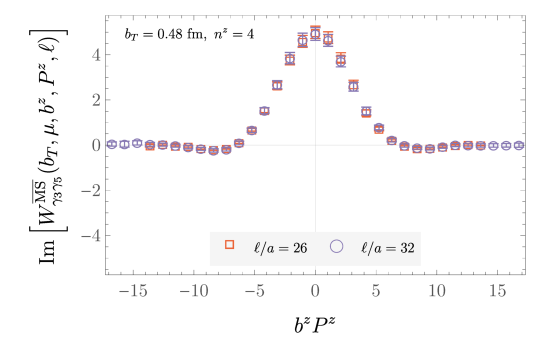} 
        \caption{As in \cref{fig:wf_ms_pz4_gamma7_a}, for $\Gamma = \gamma_3\gamma_5$. \label{fig:wf_ms_pz4_gamma11_a} 
        }
\end{figure*}
\begin{figure*}[t]
    \centering
        \includegraphics[width=0.46\textwidth]{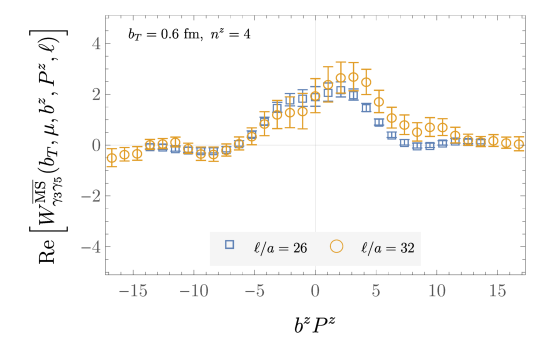} 
        \hspace{20pt}
        \includegraphics[width=0.46\textwidth]{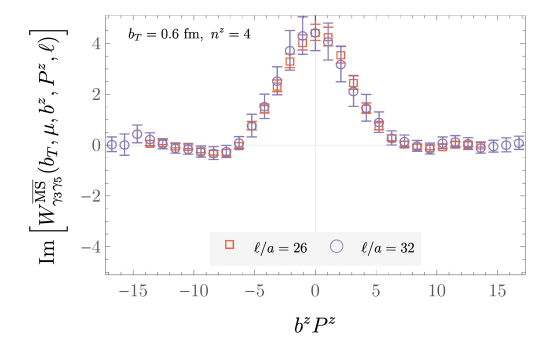} 
        \includegraphics[width=0.46\textwidth]{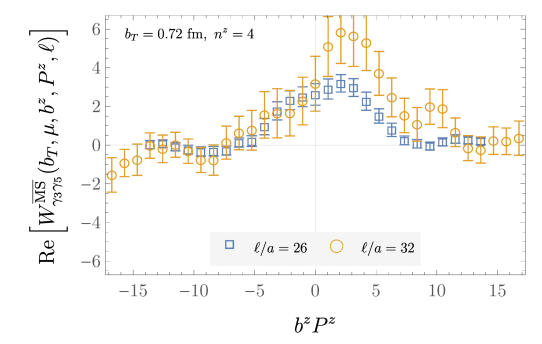} 
        \hspace{20pt}
        \includegraphics[width=0.46\textwidth]{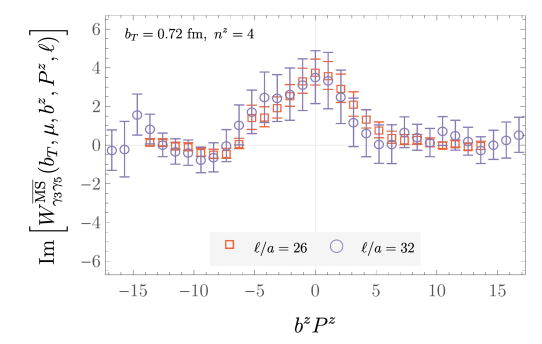} 
        \includegraphics[width=0.46\textwidth]{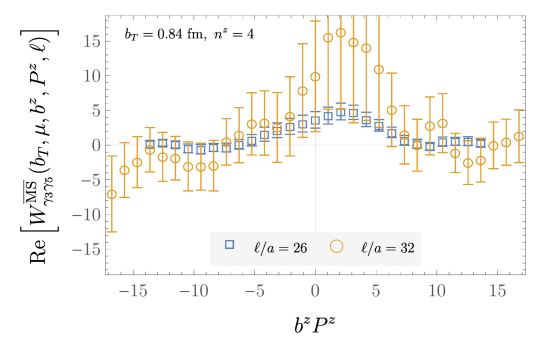}  
        \hspace{20pt}
        \includegraphics[width=0.46\textwidth]{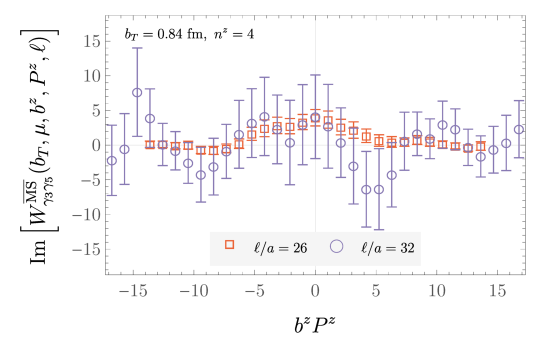}  
    \caption{As in \cref{fig:wf_ms_pz4_gamma7_a}, for $\Gamma = \gamma_3\gamma_5$ and $\SI{0.60}{\femto\meter} \leq b_\tran \leq \SI{0.84}{\femto\meter}$.
    \label{fig:wf_ms_pz4_gamma11_b}}
\end{figure*}
        
\begin{figure*}[t]
    \centering
        \includegraphics[width=0.46\textwidth]{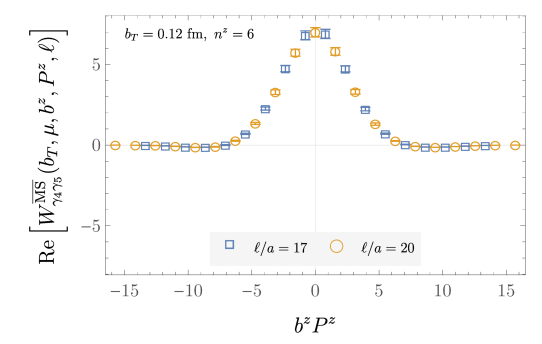}   
        \hspace{20pt}
        \includegraphics[width=0.46\textwidth]{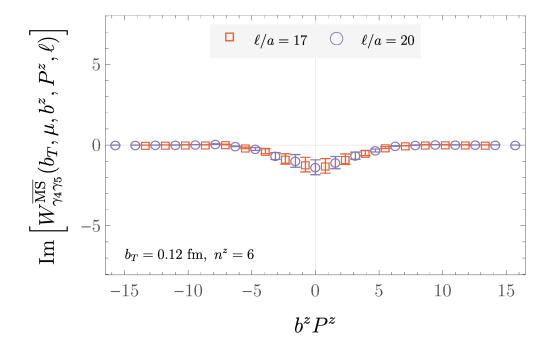}
        \includegraphics[width=0.46\textwidth]{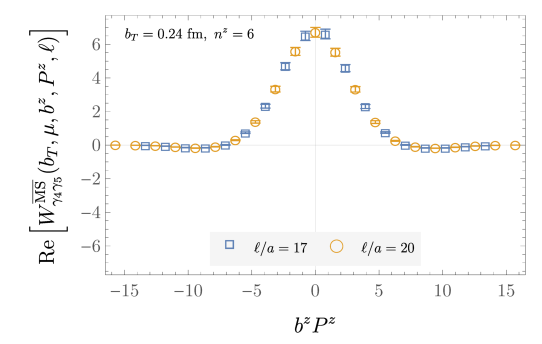} 
        \hspace{20pt}
        \includegraphics[width=0.46\textwidth]{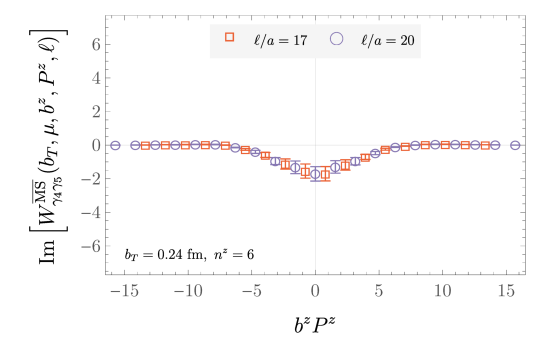} 
        \includegraphics[width=0.46\textwidth]{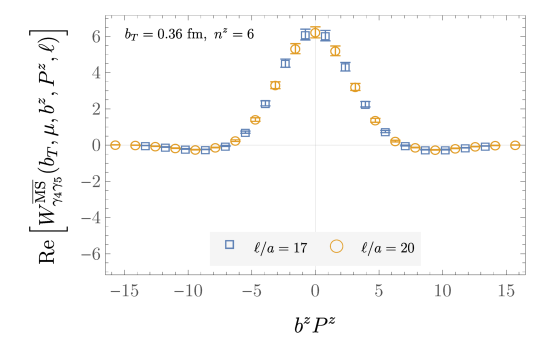} 
        \hspace{20pt}
        \includegraphics[width=0.46\textwidth]{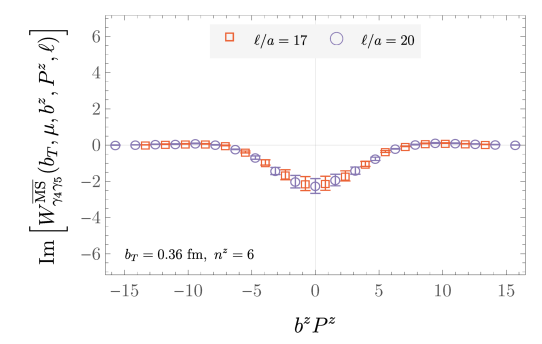} 
        \includegraphics[width=0.46\textwidth]{WF_MS_pz6_gamma7_bT4_vs_bz_re} 
        \hspace{20pt}
        \includegraphics[width=0.46\textwidth]{WF_MS_pz6_gamma7_bT4_vs_bz_im} 
        \caption{As in \cref{fig:wf_ms_pz4_gamma7_a}, for $P^z = \SI{1.29}{\GeV}$ ($W^{\MSbar}_{\gamma_4\gamma_5}$ for $b_\tran = \SI{0.48}{\femto\meter}$ is illustrated in \cref{fig:analysis-1} in the main text).
        \label{fig:wf_ms_pz6_gamma7_a}}
\end{figure*}
\begin{figure*}[t]
    \centering
        \includegraphics[width=0.46\textwidth]{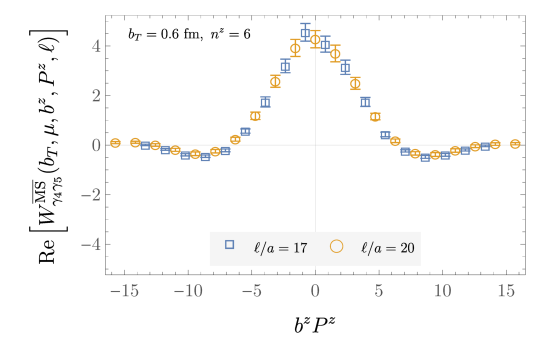} 
        \hspace{20pt}
        \includegraphics[width=0.46\textwidth]{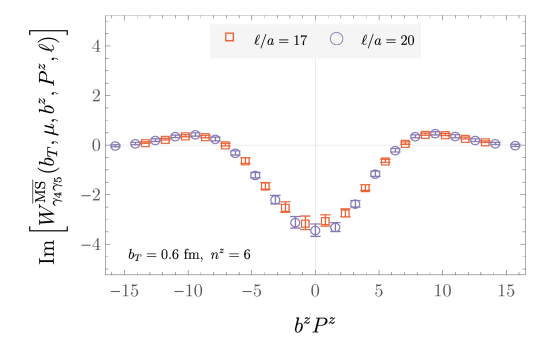} 
        \includegraphics[width=0.46\textwidth]{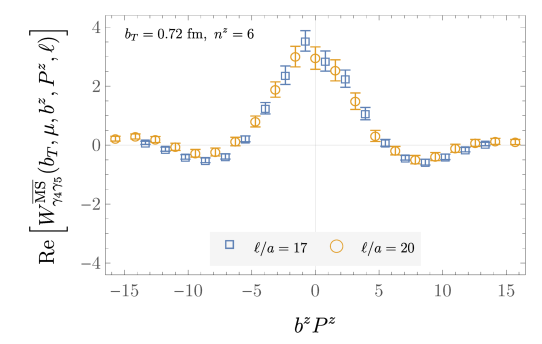} 
        \hspace{20pt}
        \includegraphics[width=0.46\textwidth]{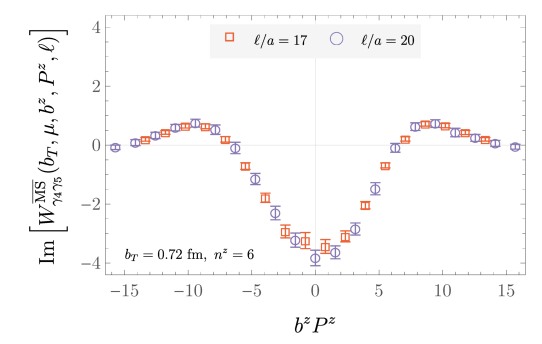} 
        \includegraphics[width=0.46\textwidth]{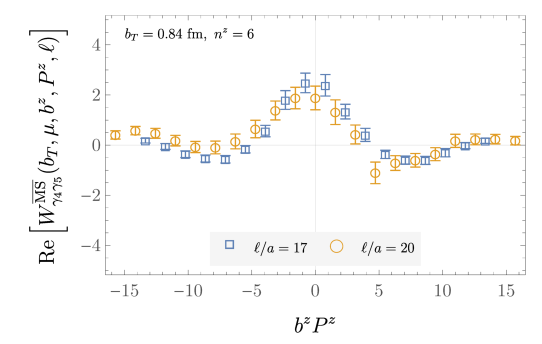}  
        \hspace{20pt}
        \includegraphics[width=0.46\textwidth]{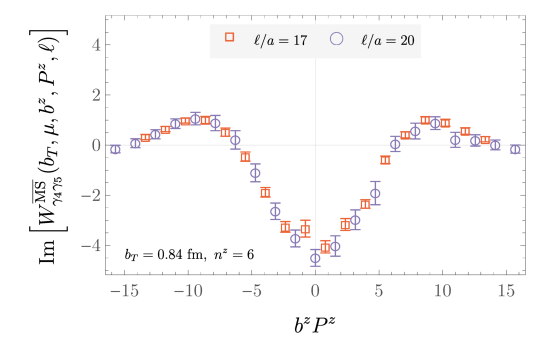}  
    \caption{As in \cref{fig:wf_ms_pz4_gamma7_a}, for $P^z = \SI{1.29}{\GeV}$ and $\SI{0.60}{\femto\meter} \leq b_\tran \leq \SI{0.84}{\femto\meter}$.
    \label{fig:wf_ms_pz6_gamma7_b}}
\end{figure*}

\begin{figure*}[t]
    \centering
        \includegraphics[width=0.46\textwidth]{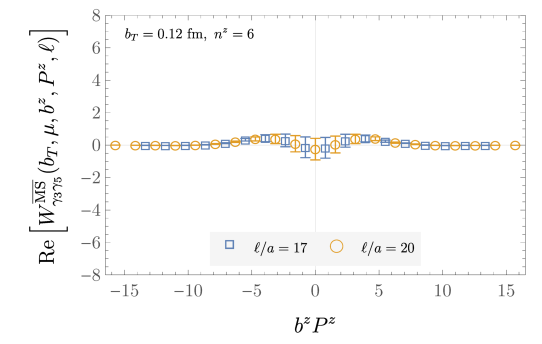}   
        \hspace{20pt}
        \includegraphics[width=0.46\textwidth]{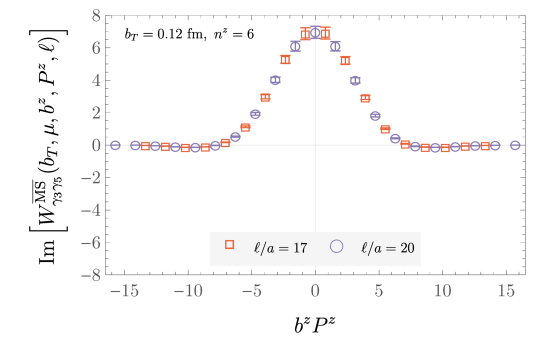}
        \includegraphics[width=0.46\textwidth]{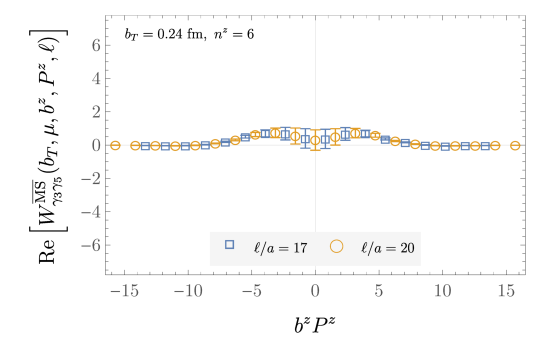} 
        \hspace{20pt}
        \includegraphics[width=0.46\textwidth]{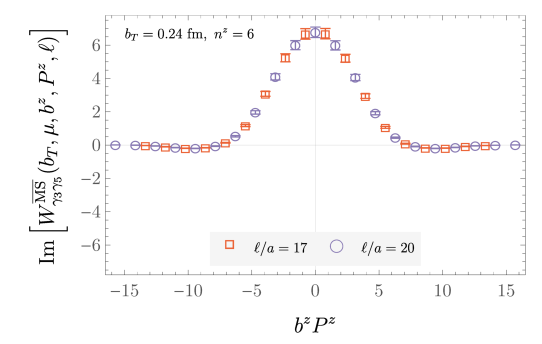} 
        \includegraphics[width=0.46\textwidth]{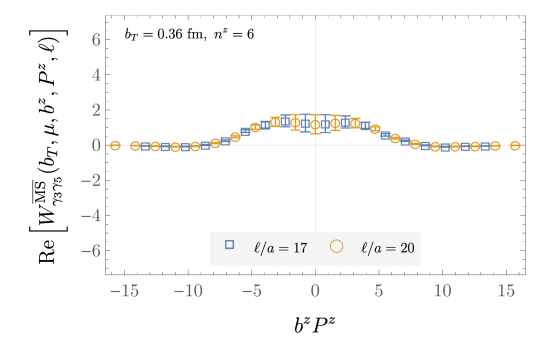} 
        \hspace{20pt}
         \includegraphics[width=0.46\textwidth]{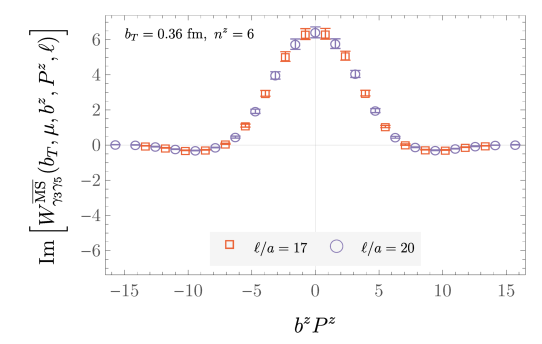} 
        \includegraphics[width=0.46\textwidth]{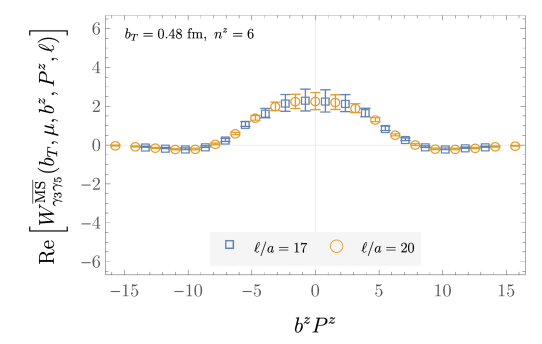} 
        \hspace{20pt}
        \includegraphics[width=0.46\textwidth]{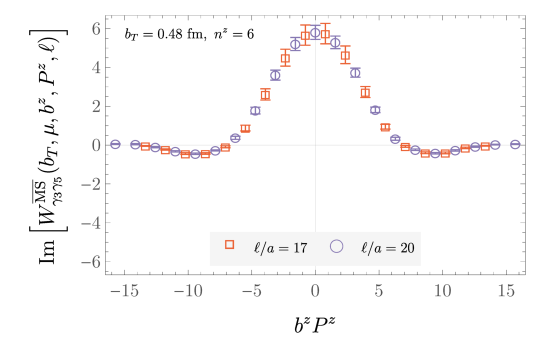} 
        \caption{As in \cref{fig:wf_ms_pz4_gamma7_a}, for $\Gamma = \gamma_3\gamma_5$ and $P^z = \SI{1.29}{\GeV}$.
        \label{fig:wf_ms_pz6_gamma11_a}}
\end{figure*}
\begin{figure*}[t]
    \centering
        \includegraphics[width=0.46\textwidth]{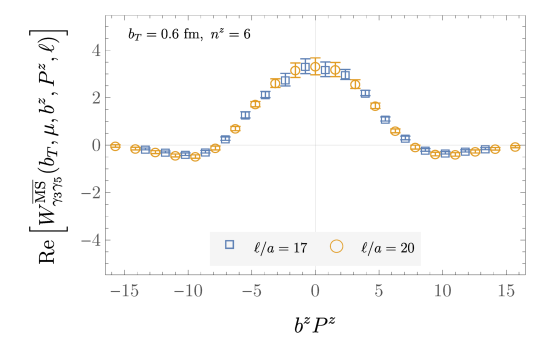} 
        \hspace{20pt}
        \includegraphics[width=0.46\textwidth]{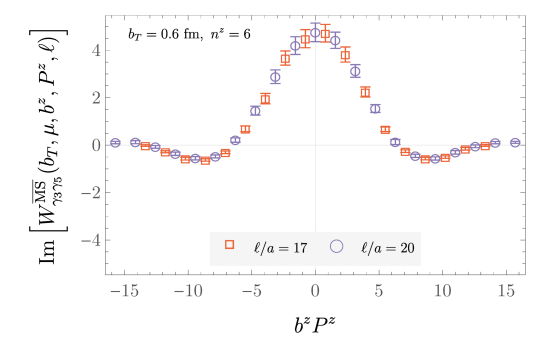} 
        \includegraphics[width=0.46\textwidth]{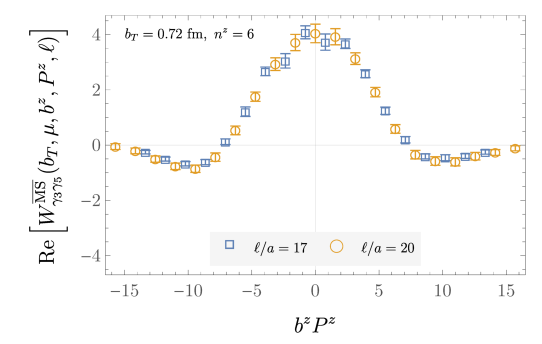} 
        \hspace{20pt}
        \includegraphics[width=0.46\textwidth]{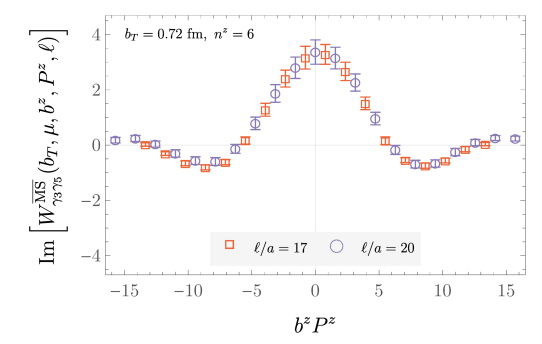} 
        \includegraphics[width=0.46\textwidth]{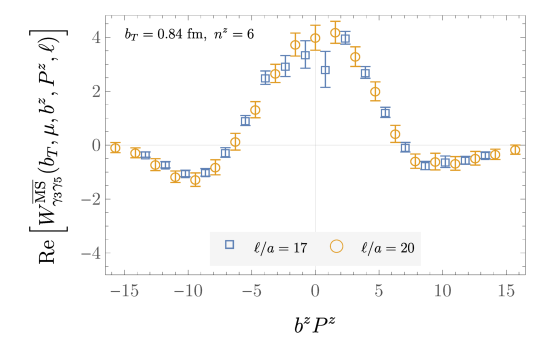}  
        \hspace{20pt}
        \includegraphics[width=0.46\textwidth]{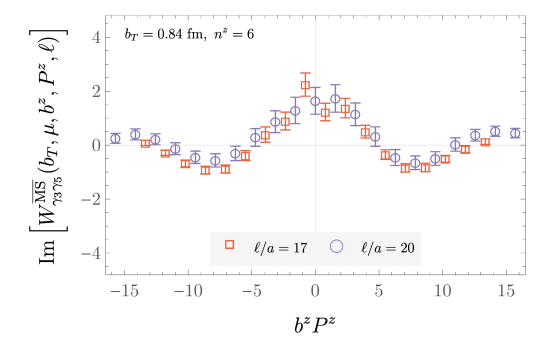}  
    \caption{As in \cref{fig:wf_ms_pz4_gamma7_a}, for $\Gamma = \gamma_3\gamma_5$, $P^z = \SI{1.29}{\GeV}$, and $\SI{0.60}{\femto\meter} \leq b_\tran \leq \SI{0.84}{\femto\meter}$.
    \label{fig:wf_ms_pz6_gamma11_b}
    }
\end{figure*}

\begin{figure*}[t]
    \centering
        \includegraphics[width=0.46\textwidth]{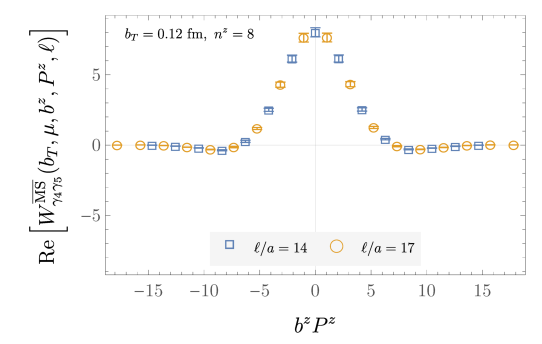}   
        \hspace{20pt}
        \includegraphics[width=0.46\textwidth]{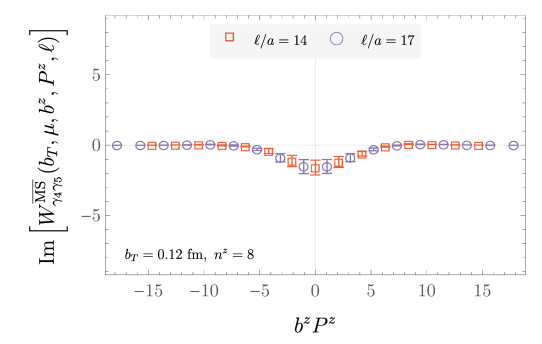}
        \includegraphics[width=0.46\textwidth]{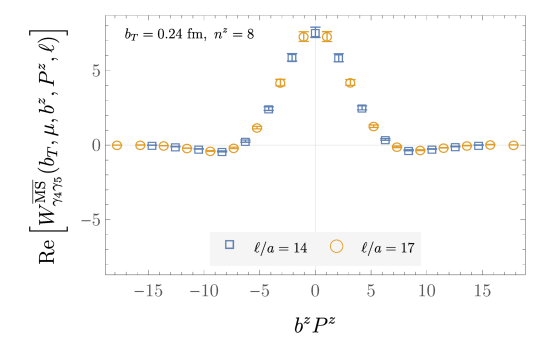} 
        \hspace{20pt}
        \includegraphics[width=0.46\textwidth]{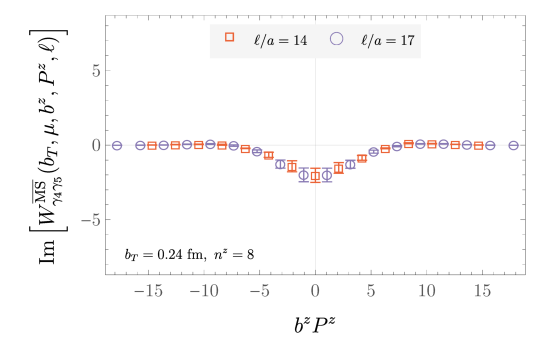} 
        \includegraphics[width=0.46\textwidth]{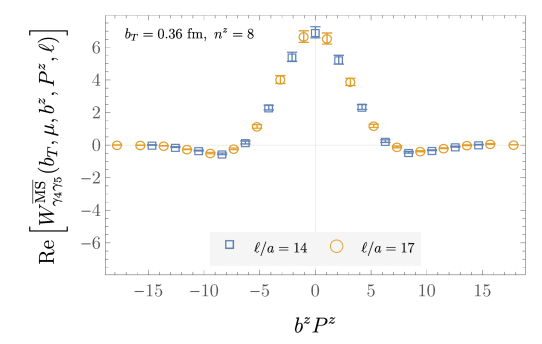} 
        \hspace{20pt}
         \includegraphics[width=0.46\textwidth]{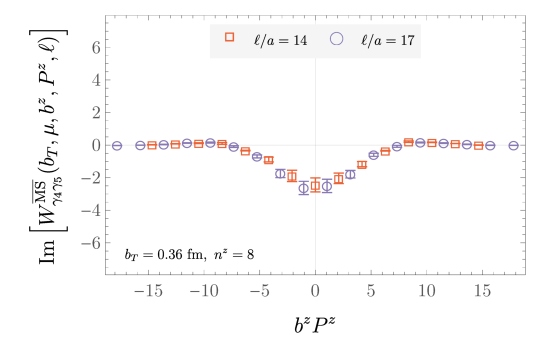} 
        \includegraphics[width=0.46\textwidth]{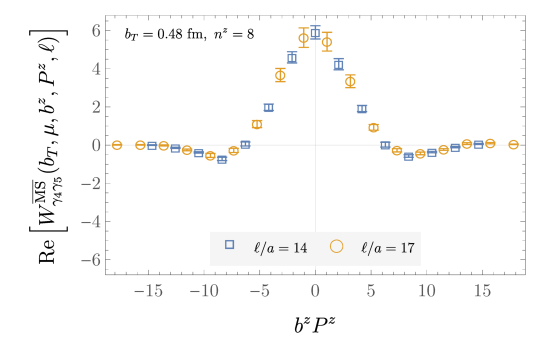} 
        \hspace{20pt}
        \includegraphics[width=0.46\textwidth]{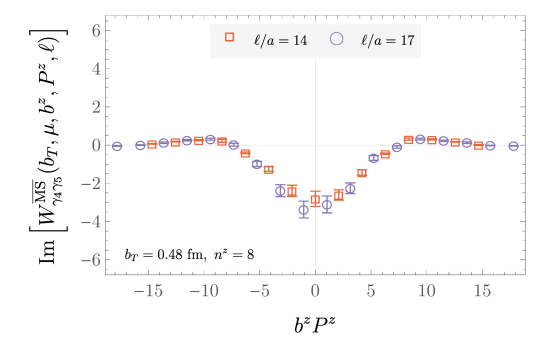} 
        \caption{As in \cref{fig:wf_ms_pz4_gamma7_a}, for $P^z = \SI{1.72}{\GeV}$.
        \label{fig:wf_ms_pz8_gamma7_a}}
\end{figure*}
\begin{figure*}[t]
    \centering
        \includegraphics[width=0.46\textwidth]{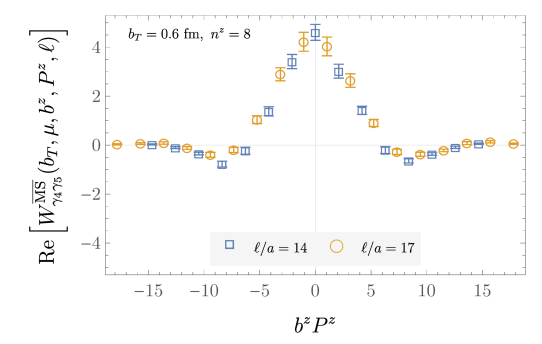} 
        \hspace{20pt}
        \includegraphics[width=0.46\textwidth]{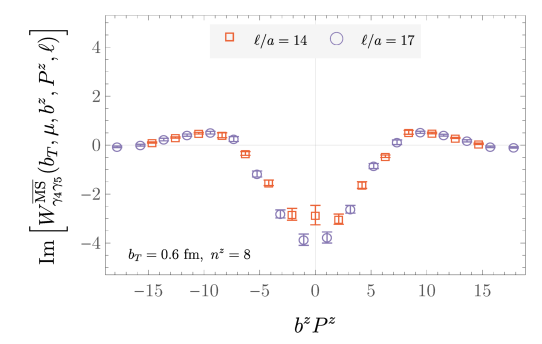} 
        \includegraphics[width=0.46\textwidth]{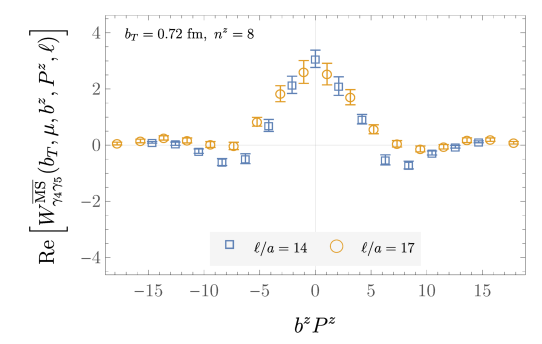} 
        \hspace{20pt}
        \includegraphics[width=0.46\textwidth]{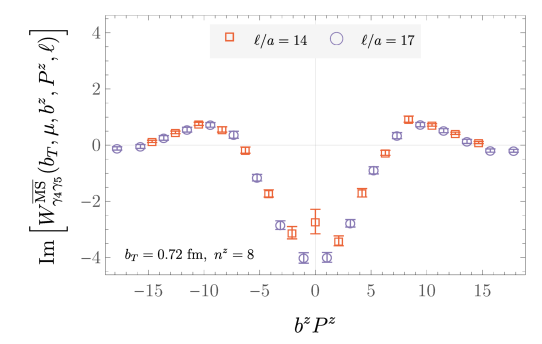} 
        \includegraphics[width=0.46\textwidth]{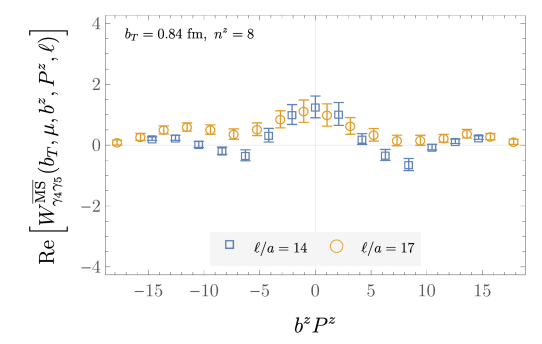}  
        \hspace{20pt}
        \includegraphics[width=0.46\textwidth]{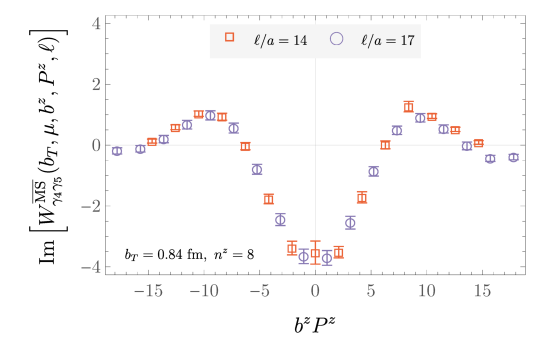}  
    \caption{As in \cref{fig:wf_ms_pz4_gamma7_a}, for $P^z = \SI{1.72}{\GeV}$, and $\SI{0.60}{\femto\meter} \leq b_\tran \leq \SI{0.84}{\femto\meter}$.
    \label{fig:wf_ms_pz8_gamma7_b}
    }
\end{figure*}

\begin{figure*}[t]
    \centering
        \includegraphics[width=0.46\textwidth]{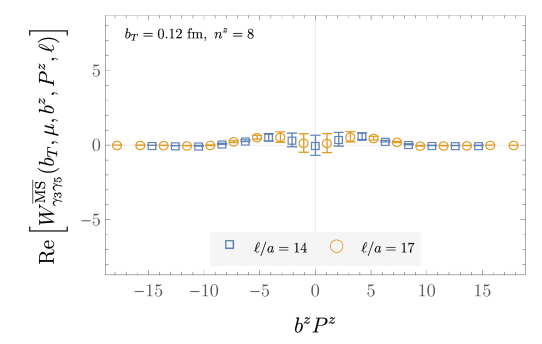}   
        \hspace{20pt}
        \includegraphics[width=0.46\textwidth]{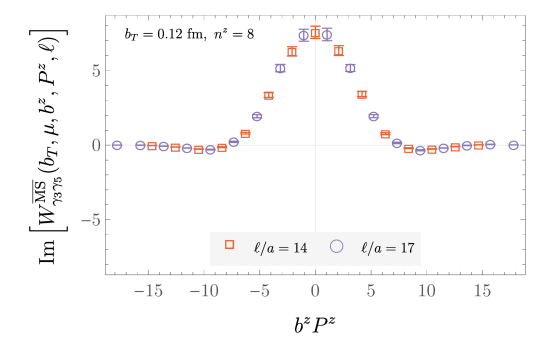}
        \includegraphics[width=0.46\textwidth]{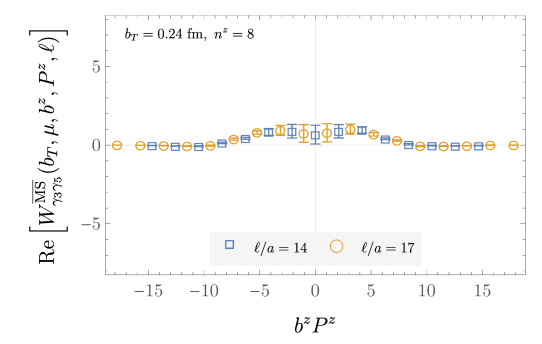} 
        \hspace{20pt}
        \includegraphics[width=0.46\textwidth]{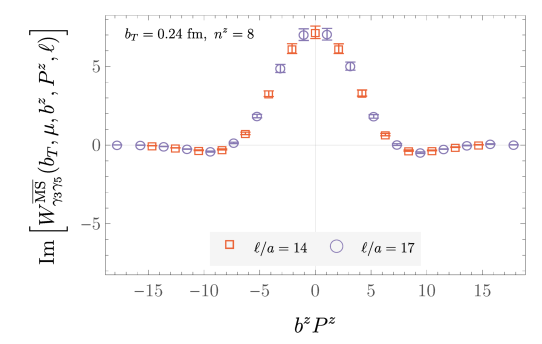} 
        \includegraphics[width=0.46\textwidth]{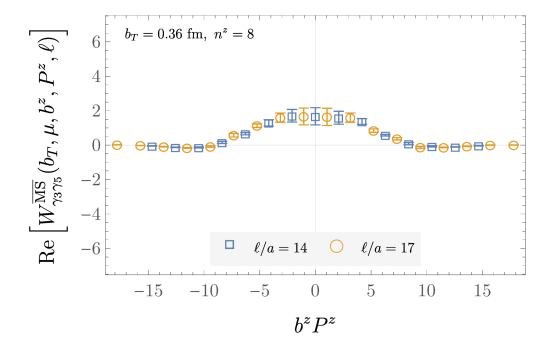} 
        \hspace{20pt}
         \includegraphics[width=0.46\textwidth]{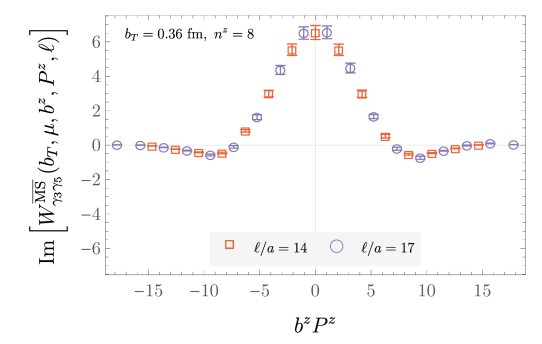} 
        \includegraphics[width=0.46\textwidth]{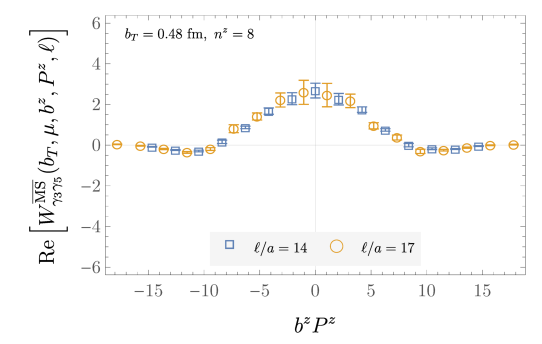} 
        \hspace{20pt}
        \includegraphics[width=0.46\textwidth]{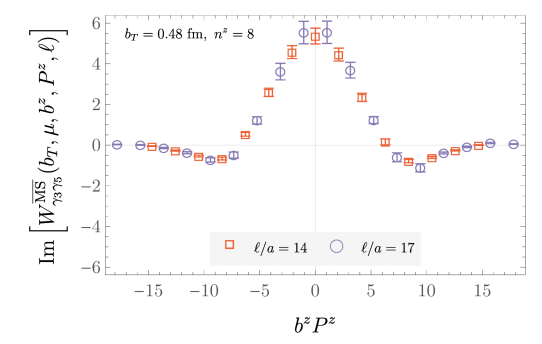} 
        \caption{As in \cref{fig:wf_ms_pz4_gamma7_a}, for $\Gamma = \gamma_3\gamma_5$ and $P^z = \SI{1.72}{\GeV}$.
        \label{fig:wf_ms_pz8_gamma11_a}}
\end{figure*}
\begin{figure*}[t]
    \centering
        \includegraphics[width=0.46\textwidth]{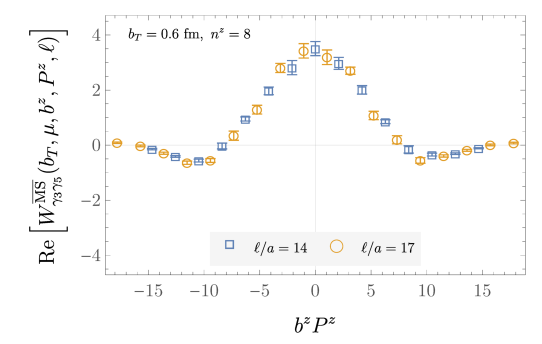} 
        \hspace{20pt}
        \includegraphics[width=0.46\textwidth]{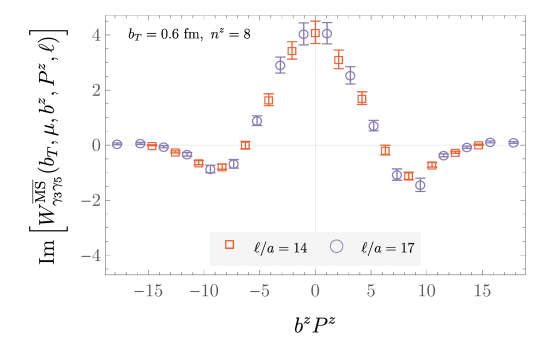} 
        \includegraphics[width=0.46\textwidth]{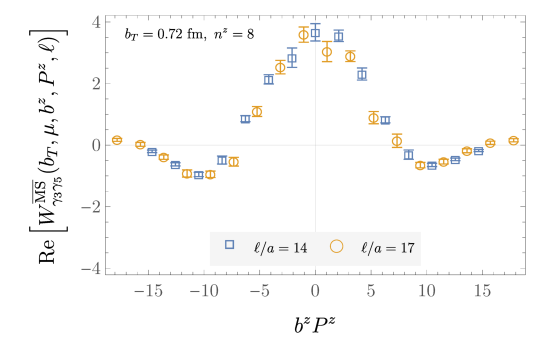} 
        \hspace{20pt}
        \includegraphics[width=0.46\textwidth]{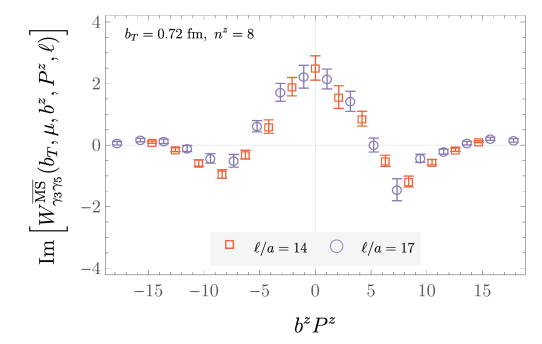} 
        \includegraphics[width=0.46\textwidth]{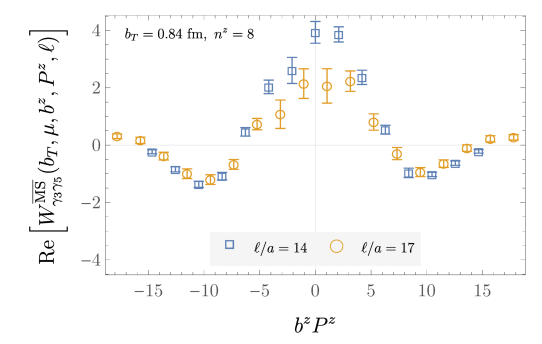}  
        \hspace{20pt}
        \includegraphics[width=0.46\textwidth]{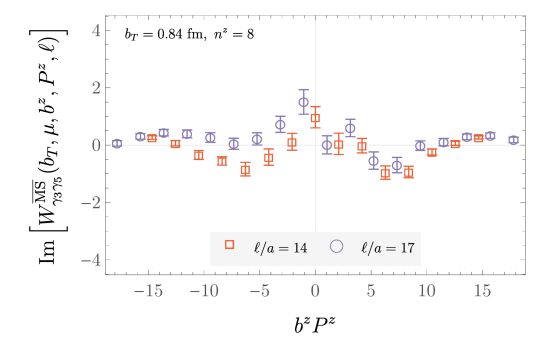}  
        %\includegraphics[width=0.46\textwidth]{WF_MS_pz8_gamma11_bT8_vs_bz_re} 
        %\hspace{20pt}
        %\includegraphics[width=0.46\textwidth]{WF_MS_pz8_gamma11_bT8_vs_bz_im} 
    \caption{As in \cref{fig:wf_ms_pz4_gamma7_a}, for $\Gamma = \gamma_3\gamma_5$, $P^z = \SI{1.72}{\GeV}$, and $\SI{0.60}{\femto\meter} \leq b_\tran \leq \SI{0.84}{\femto\meter}$.
    \label{fig:wf_ms_pz8_gamma11_b}}
\end{figure*}

\begin{figure*}[t]
    \centering
        \includegraphics[width=0.46\textwidth]{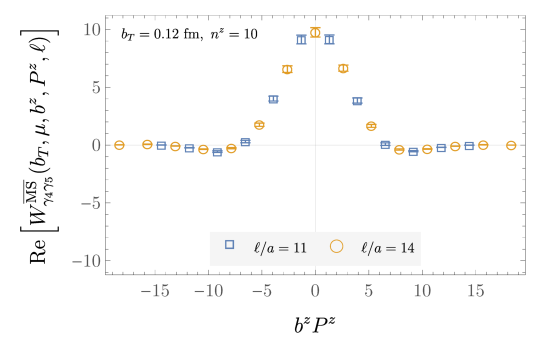}   
        \hspace{20pt}
        \includegraphics[width=0.46\textwidth]{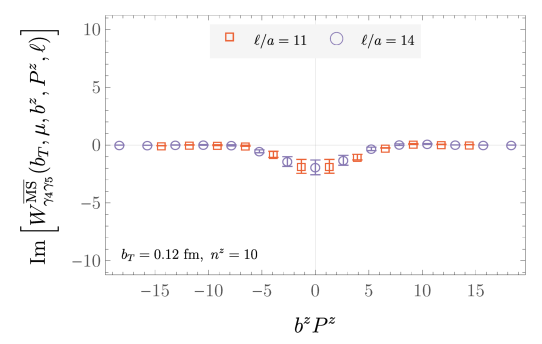}
        \includegraphics[width=0.46\textwidth]{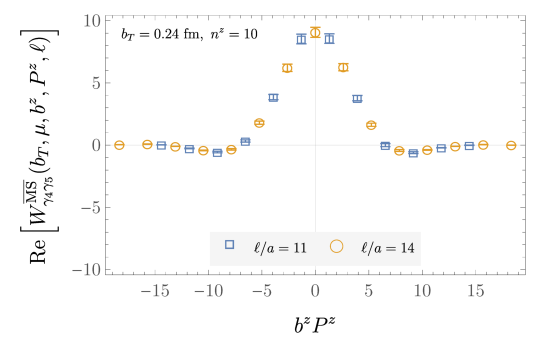} 
        \hspace{20pt}
        \includegraphics[width=0.46\textwidth]{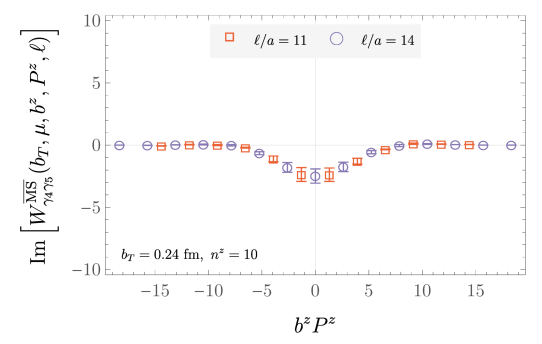} 
        \includegraphics[width=0.46\textwidth]{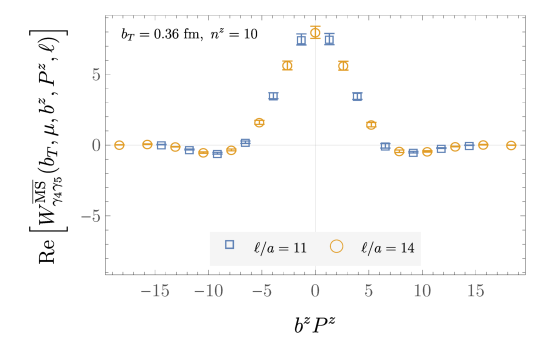} 
        \hspace{20pt}
         \includegraphics[width=0.46\textwidth]{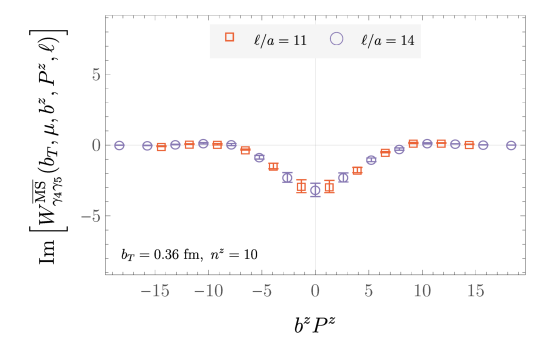} 
        \includegraphics[width=0.46\textwidth]{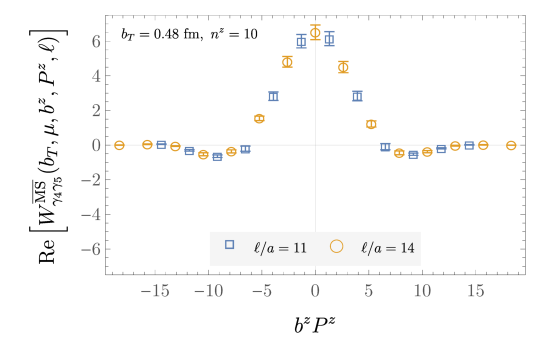} 
        \hspace{20pt}
        \includegraphics[width=0.46\textwidth]{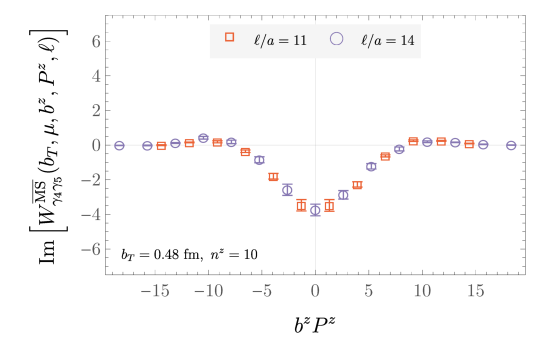} 
        \caption{As in \cref{fig:wf_ms_pz4_gamma7_a}, for  $P^z = \SI{2.15}{\GeV}$.
        \label{fig:wf_ms_pz10_gamma7_a}}
\end{figure*}
\begin{figure*}[t]
    \centering
        \includegraphics[width=0.46\textwidth]{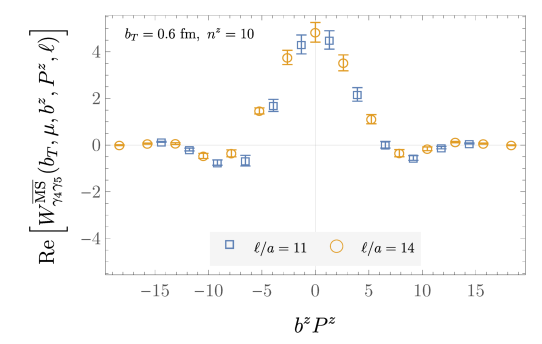} 
        \hspace{20pt}
        \includegraphics[width=0.46\textwidth]{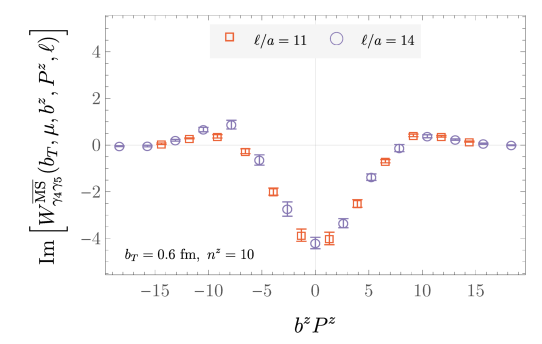} 
        \includegraphics[width=0.46\textwidth]{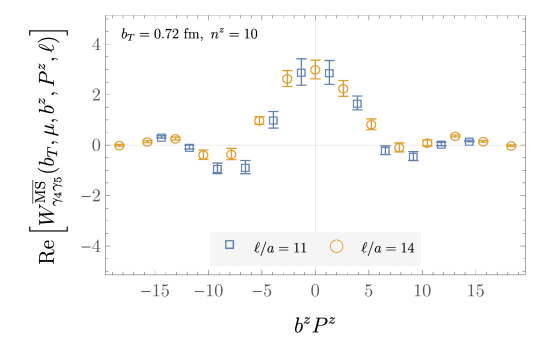} 
        \hspace{20pt}
        \includegraphics[width=0.46\textwidth]{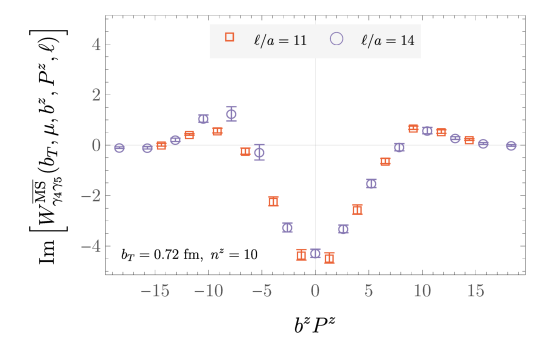} 
        \includegraphics[width=0.46\textwidth]{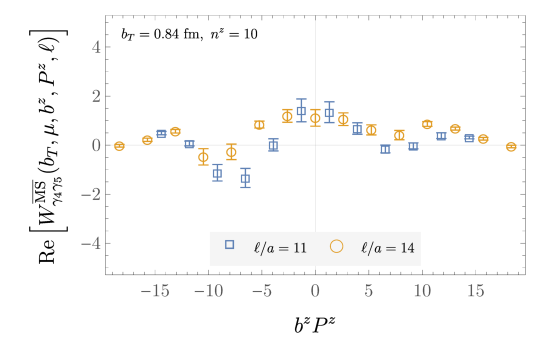}  
        \hspace{20pt}
        \includegraphics[width=0.46\textwidth]{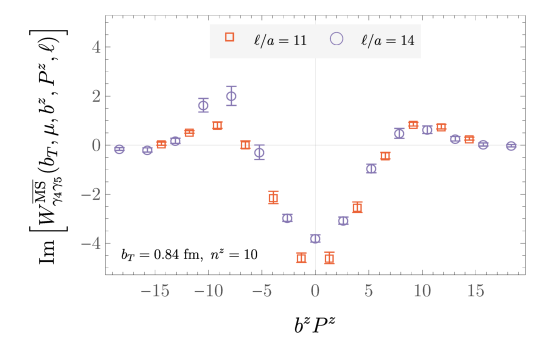}  
        %\includegraphics[width=0.46\textwidth]{WF_MS_pz10_gamma7_bT8_vs_bz_re} 
        %\hspace{20pt}
        %\includegraphics[width=0.46\textwidth]{WF_MS_pz10_gamma7_bT8_vs_bz_im} 
    \caption{As in \cref{fig:wf_ms_pz4_gamma7_a}, for $P^z = \SI{2.15}{\GeV}$ and $\SI{0.60}{\femto\meter} \leq b_\tran \leq \SI{0.84}{\femto\meter}$.
    \label{fig:wf_ms_pz10_gamma7_b}}
\end{figure*}

\begin{figure*}[t]
    \centering
        \includegraphics[width=0.46\textwidth]{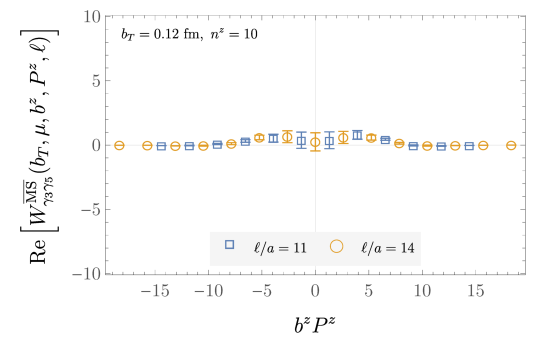}   
        \hspace{20pt}
        \includegraphics[width=0.46\textwidth]{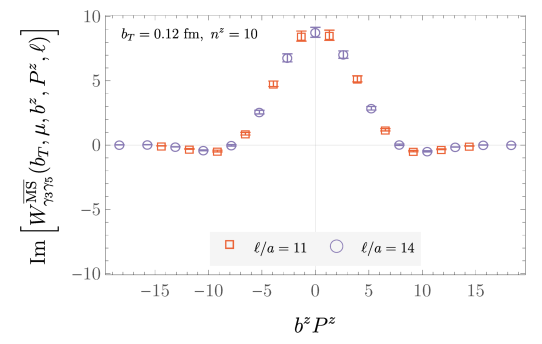}
        \includegraphics[width=0.46\textwidth]{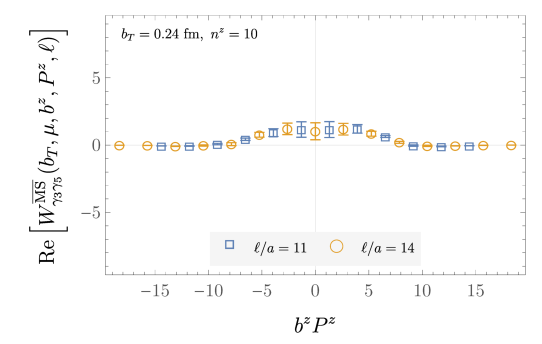} 
        \hspace{20pt}
        \includegraphics[width=0.46\textwidth]{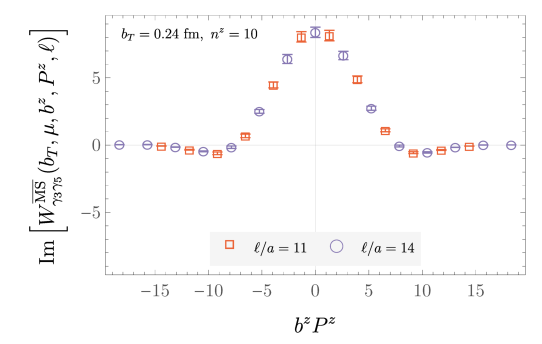} 
        \includegraphics[width=0.46\textwidth]{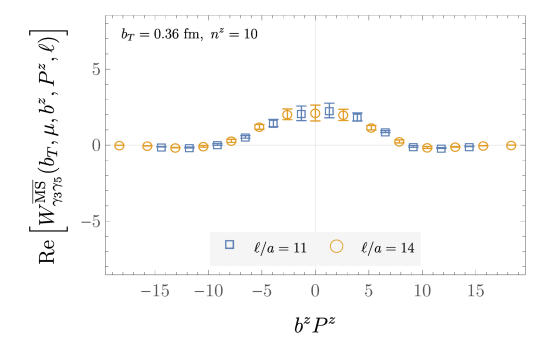} 
        \hspace{20pt}
         \includegraphics[width=0.46\textwidth]{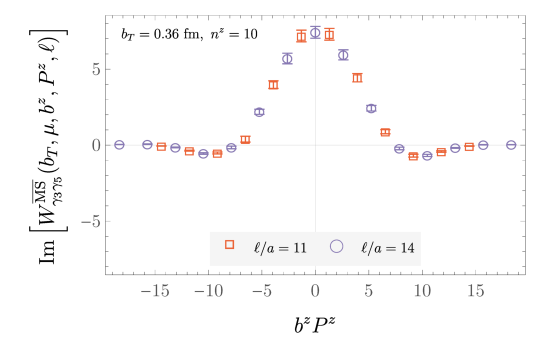} 
        \includegraphics[width=0.46\textwidth]{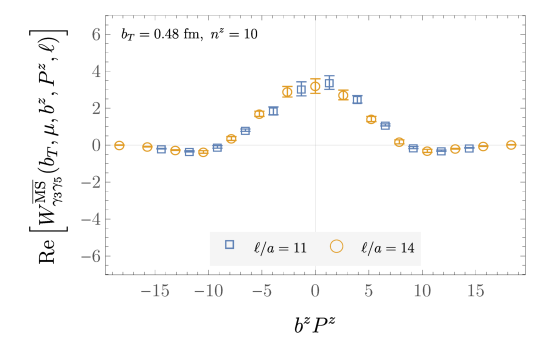} 
        \hspace{20pt}
        \includegraphics[width=0.46\textwidth]{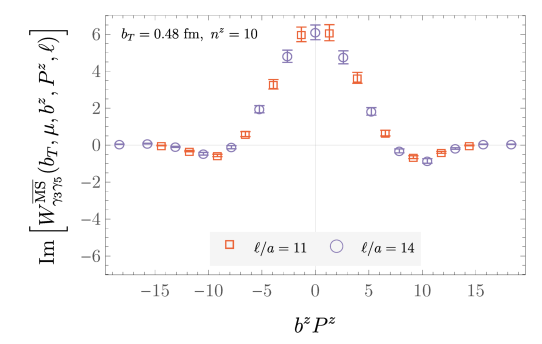} 
        \caption{As in \cref{fig:wf_ms_pz4_gamma7_a}, for $\Gamma = \gamma_3\gamma_5$ and $P^z = \SI{2.15}{\GeV}$.
        \label{fig:wf_ms_pz10_gamma11_a}}
\end{figure*}
\begin{figure*}[t]
    \centering
        \includegraphics[width=0.46\textwidth]{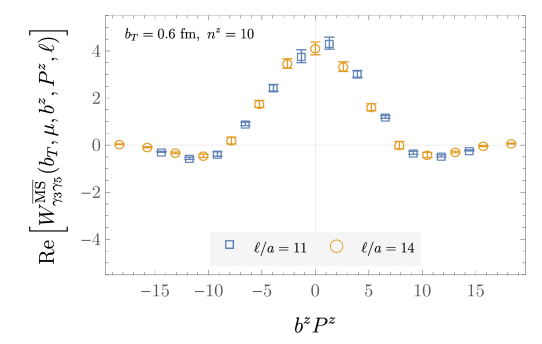} 
        \hspace{20pt}
        \includegraphics[width=0.46\textwidth]{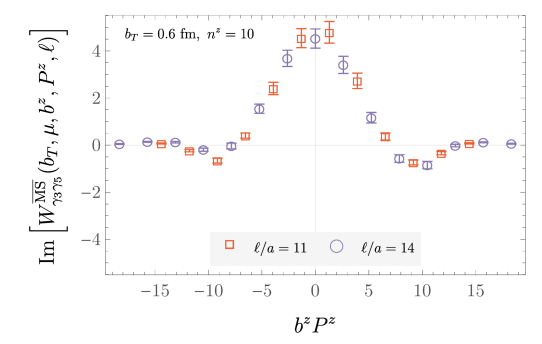} 
        \includegraphics[width=0.46\textwidth]{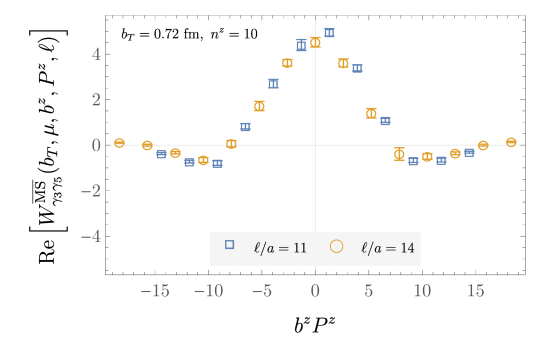} 
        \hspace{20pt}
        \includegraphics[width=0.46\textwidth]{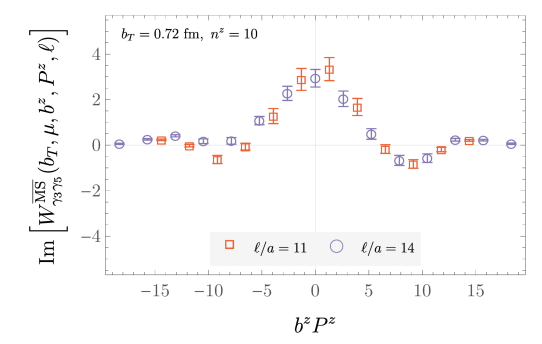} 
        \includegraphics[width=0.46\textwidth]{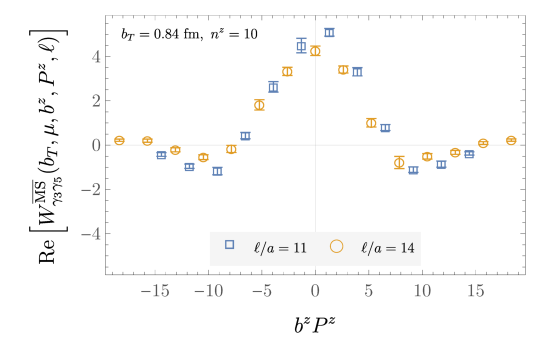}  
        \hspace{20pt}
        \includegraphics[width=0.46\textwidth]{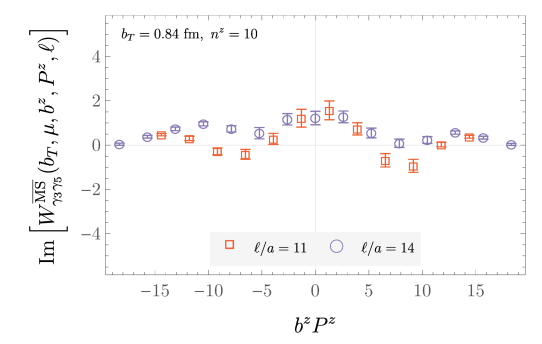}  
        %\includegraphics[width=0.46\textwidth]{WF_MS_pz10_gamma11_bT8_vs_bz_re} 
        %\hspace{20pt}
        %\includegraphics[width=0.46\textwidth]{WF_MS_pz10_gamma11_bT8_vs_bz_im} 
    \caption{As in \cref{fig:wf_ms_pz4_gamma7_a}, for $\Gamma = \gamma_3\gamma_5$, $P^z = \SI{2.15}{\GeV}$, and $\SI{0.60}{\femto\meter} \leq b_\tran \leq \SI{0.84}{\femto\meter}$.
    \label{fig:wf_ms_pz10_gamma11_b}}
\end{figure*}


\clearpage

%\bibliography{refs}

\end{document}